\documentclass{aa} 
\usepackage[english] {babel}
\usepackage{astron}
\usepackage[dvips]{graphicx}
\usepackage{ae}
\usepackage{color}
\usepackage{amsfonts}
\usepackage{tabularx}
\usepackage{adjustbox}
\usepackage{natbib}
\usepackage{amssymb,amsmath}
\usepackage{listliketab}
\usepackage[toc,page]{appendix}
\usepackage{txfonts}
\usepackage{placeins}
\usepackage{float}
\usepackage{afterpage}
\usepackage{longtable}
\usepackage{supertabular}
\usepackage{url}
\usepackage{rotating}
\usepackage{wasysym}
\usepackage{adjustbox,lipsum}
\usepackage{pdflscape}
\usepackage[breaklinks]{hyperref}
\usepackage[hyphenbreaks]{breakurl}
\usepackage{supertabular}
\begin{document}
\bibliographystyle{aa} 

   \title{The matter distribution in the local Universe as derived from galaxy groups in SDSS DR12 and 2MRS}   

   \author{Christoph Saulder
          \inst{1}
          \and
          Eelco van Kampen
          \inst{2}
          \and
          Igor V. Chilingarian
          \inst{3,4}
          \and
          Steffen Mieske
          \inst{5}
          \and
          Werner W. Zeilinger
          \inst{1}                
          }

   \institute{
   Department of Astrophysics, University of Vienna, T\"urkenschanzstra\ss e 17, 1180 Vienna, Austria\\
   \email{christoph.saulder@equinoxomega.net}\\
   \and
   European Southern Observatory,   Karl-Schwarzschild-Stra\ss e 2, 85748 Garching bei M\"unchen, Germany\\
   \and   
   Smithsonian Astrophysical Observatory, Harvard-Smithsonian Center for Astrophysics, 60 Garden St. MS09 Cambridge, MA 02138, USA\\
   \and
   Sternberg Astronomical Institute, Moscow State University, 13 Universitetski prospect, 119992 Moscow, Russia\\  
   \and
   European Southern Observatory, Alonso de C\'{o}rdova 3107, Vitacura, Casilla 19001, Santiago, Chile\\
             }

   \date{Received June 9 2015 ; accepted July 29, 2016}

\abstract
{
Friends-of-friends algorithms are a common tool to detect galaxy groups and clusters in large survey data. In order to be as precise as possible, they
have to be carefully calibrated using mock catalogues.
}
{
We create an accurate and robust description of the matter distribution in the local Universe using the most up-to-date available data. This will provide the input for a specific cosmological test planned as follow-up to this work, and will be useful for general extragalactic and cosmological research.
}
{
We created a set of galaxy group catalogues based on the 2MRS and SDSS DR12 galaxy samples using a friends-of-friends based group finder algorithm. The algorithm was carefully calibrated and optimised on a new set of wide-angle mock catalogues from the Millennium simulation, in order to provide accurate total mass estimates of the galaxy groups taking into account the relevant observational biases in 2MRS and SDSS.  
}
{
We provide four different catalogues\thanks{Tables \ref{sample_cat_2mrs} to \ref{sample_cat_fi} are only available in electronic form at the CDS via anonymous ftp to cdsarc.u-strasbg.fr (130.79.128.5) or via \url{http://cdsweb.u-strasbg.fr/cgi-bin/qcat?J/A+A/}.} (i) a 2MRS based group catalogue; (ii) an SDSS DR12 based group catalogue reaching out to a redshift $z=0.11$ with stellar mass estimates for 70\% of the galaxies; (iii) a catalogue providing additional fundamental plane distances for all groups of the SDSS catalogue that host elliptical galaxies; (iv) a catalogue of the mass distribution in the local Universe based on a combination of our 2MRS and SDSS catalogues.  
}
{
While motivated by a specific cosmological test, three of the four catalogues that we produced are well suited to act as reference databases for a variety of extragalactic and cosmological science cases. Our catalogue of fundamental plane distances for SDSS groups provides further added value to this paper.
}

\titlerunning{The matter distribution in the local Universe from SDSS DR12 and 2MRS}
   \maketitle

\section{Introduction}
Galaxy clusters and groups have been an important tool in extragalactic astronomy since the discovery of their nature.  \citet{Zwicky:1933} used the internal dynamics of nearby clusters to postulate dark matter for the first time.  Messier was the first to notice an overdensity of nebulae in the Virgo constellation \citep{Biviano:2000} and thereby discovered the first galaxy cluster being unaware of its nature or the nature of the nebulae (galaxies).  The investigation of galaxy clusters started shortly after the Great Debate, when it became established that the Universe contains other galaxies than our own.  The first milestone was the discovery of dark matter in galaxy clusters \citep{Zwicky:1933}.  The first significant cluster catalogues were produced by \citet{Abell:1958} and \citet{Zwicky:1961}.  Starting with the pioneering work of \citet{Turner:1976} and heavily applied in \citet{Huchra:1982}, \citet{Zeldovich:1982} and \citet{Press:1982}, the methods of finding clusters became more sophisticated and reproducible.  The most common algorithm even up to the present day is the friend-of-friends algorithm \citep{Press:1982}, although there are other techniques around \citep{Yang:2005,Gal:2006,Koester:2007,Hao:2010,Makarov:2011,Munoz:2012}. \citet{Knebe:2011} provide a comprehensive comparison between halo finder algorithms for simulated data.  A detailed study on the optimization of cluster and group finders with a focus on friend-of-friends (FoF) algorithms was performed by \citet{Eke:2004a}. \citet{Duarte:2014,Duarte:2015} discuss the quality of FoF group finders and the impact of different linking lengths on the recovery of various group parameters.  Efficient and reliable algorithms become crucially important, especially during the last decade and in the time of big data and surveys, such as 2MASS \citep{2MASS}, SDSS \citep{SDSS_early,SDSS_DR12}, 2dFGRS \citep{2dFGRS}, 6dF Galaxy Survey \citep{6dFGS_basics,6dFGS_DR3}, and GAMA \citep{GAMA_basics}.  Information on galaxy grouping and clustering is important because it provides a laboratory to study the dependence of galaxy morphology on the environment \citep{Einasto:1974,Oemler:1974,Davis:1976,Dressler:1980,Postman:1984,Dressler:1997,Goto:2003,vdWel:2010,Wilman:2011,Cappellari:2011} or environmental influence on different properties of galaxies and groups \citep{Huertas:2011,Luparello:2013,Hearin:2013,Hou:2013,Yang:2013,Wetzel:2013,Budzynski:2014,Einasto:2014}.  It also provides a way to study the halo mass-luminosity relationship \citep{Yang:2009,Wake:2011} and thereby helps us understand the dark matter distribution in the Universe.

Notable group and cluster catalogues besides those already mentioned are \citet{Turner:1976}, \citet{Moore:1993}, \citet{Eke:2004b}, \citet{Gerke:2005}, \citet{Yang:2007}, \citet{Berlind:2006}, \citet{Brough:2006}, \citet{Crook:2007}, \citet{Knobel:2009}, \citet{Tempel:2012}, \citet{Nurmi:2013}, \citet{Tempel:2014}, and \citet{Tully:2015b}.  In our study, we will refer to all groups and clusters as groups independent of their sizes.  This also includes individual galaxies to which we refer to as a group with just one member.

It is important to make a suitable choice for the linking length, which is the distance that defines which object is still a ``friend'' of others.  Most FoF algorithms differ in the choice of scaling the linking length \citep{Huchra:1982,Ramella:1989,Nolthenius:1987,Moore:1993,Robotham:2011,Tempel:2012}, which is an important modification of all non-volume limited samples.  In the way we implement the FoF algorithm of the group finder, we mainly follow \citet{Robotham:2011} and use the corrected luminosity function of our sample for scaling.

Finally, it is crucial to calibrate the group finder on a set of mock catalogues to test its reliability.  We created suitable mock catalogues using the Millennium Simulation \citep{Millennium}.  When calculating the group catalogue, we paid specific attention that the mass in the considered volume matches the mass predicted by the cosmology.  As discussed for various methods in \citet{Old:2014,Old:2015}, it is notoriously difficult to assign accurate masses for groups, especially if they are below $10^{14} M_{\astrosun}$. The group catalogues which we obtained provide valuable insights into the matter distribution of the local Universe.

One of the motivations for this work is to obtain the basic dataset for a cosmological test, which was outlined in \citet{Saulder:2012}.  In a follow-up paper we will elaborate the applications of the data for test and the full background (Saulder et al. in prep.)  Hence, here we provide only essential details on that theory, in order to explain the motivation for our work. The tested theory, ``timescape cosmology" \citep{Wiltshire:2007}, aims to explain the accelerated expansion of the Universe by backreactions due to General Relativity and the observed inhomogeneities in the Universe instead of introducing dark energy.  The theory makes use of so-called ``finite infinity" regions and their sizes are provided in this paper.  The term was coined by \citet{Ellis:1984} and describes a matter horizon \citep{Ellis:1987} of the particles that will eventually be bound.  In our approach, we approximate the finite infinity regions with spheres of a mean density equal to the renormalized critical density (``true critical density" in \citet{Wiltshire:2007}), which is slightly lower than the critical density in the $\Lambda$-CDM model.

This paper is structured as follows: In Section 2, we describe the samples, which we used for the group finder and its calibration.  These calibrations are explained in detail in Section 3.  The results of the group finder are provided in Section 4 and they are discussed and summarized in Section 5. In Section 6, we give brief conclusions.  The appendices provide additional information on calibrations used throughout the paper.

\section{Galaxy samples}
We used the the 12th data release of the Sloan Digital Sky Survey (SDSS DR12, \citealp{SDSS_DR12}) and the 2MASS Redshift Survey (2MRS, \citealp{2MRS}), which is a spectroscopic follow-up survey of the Two Micron All Sky Survey (2MASS \citep{2MASS}), as our input observational data.

From the SDSS database, we retrieved data for 432,038 galaxies, which fulfilled the following set of criteria: spectroscopic detection, photometric and spectroscopic classification as galaxy (by the automatic pipeline), observed redshift between 0 and 0.112\footnote{We set the upper limit of 0.112 in order to avoid an asymmetric cut-off due to the corrections for our motion relative to the CMB.  It was reduced to 0.11 later.}, and the flag \emph{SpecObj.zWarning} is zero.  Additional constraints were applied later to further filter this raw dataset.  We obtained the following parameters for all galaxies: photometric object ID, equatorial coordinates, galactic coordinates, spectroscopic redshift, Petrosian magnitudes in the $g$, $r$, and $i$ band, their measurement errors, and extinction values based on \citet{Schlegelmaps}.

We used all 44,599 galaxies of \emph{table3.dat} from 2MRS catalogue \citep{2MRS_catdis}.  We obtained the following parameters for these galaxies: 2MASS-ID, equatorial coordinates, galactic coordinates, extinction-corrected total extrapolated magnitudes in all three 2MASS bands ($K_{s}$, $H$, and $J$), their corresponding errors, and spectroscopic redshift (in km/s), with its uncertainty.

In order to derive stellar masses for our catalogues, we also used additonal data from the Reference Catalog of Galaxy Spectral Energy
Distributions (Chilingarian et al. submitted), the UKIDSS survey \citep{UKIDSS}, and the SIMBAD\footnote{\url{http://simbad.u-strasbg.fr/simbad/}} database.

We choose a large set of simulated data provided the Millennium simulation \citep{Millennium} in order to calibrate our group finder and properly assess potential observational biases.

Aside from the dark matter halos produced by its original run, the Millennium  simulation database contains  data from re-runs with updated cosmological
parameters, semi-analytic models of galaxies, and the full particle information of a smaller run (called \emph{millimil}). 

\begin{table*}
\begin{center}
\begin{tabular}{c|cccccccccc}
& $\Omega_{M}$ & $\Omega_{b}$ & $\Omega_{\Lambda}$ & $h_{100}$ & $n_{\textrm{s}}$ & $\sigma_{8}$ & $N_{\textrm{p}}$ & $m_{\textrm{p}}$ [$M_{\astrosun}/h_{100}$] & $L$ [Mpc/$h_{100}$] & $\epsilon$ [kpc/$h_{100}$]\\ \hline 
original & 0.25 &	0.045 &	0.75 &	0.73 &	1 &	0.9  & 2160$^3$ & $8.61 \cdot 10^{8}$ & 500 &	5 \\
 MM & 0.25 &	0.045 &	0.75 &	0.73 &	1 &	0.9  &	270$^3$ & $8.61 \cdot 10^{8}$ & 62.5 &	5 \\
\end{tabular}
\end{center}
\caption{Parameters of the Millennium simulation (original run and \emph{millimil}(MM)).  Column one: total matter density $\Omega_{M}$, column two: baryonic matter density $\Omega_{b}$, column three: dark energy density $\Omega_{\Lambda}$, column four: $h_{100}$ is the Hubble parameter $H_{0}$ per 100 km/s/Mpc, column five: spectral index of density perturbations $n_{\textrm{s}}$, column six: size of linear density fluctuation at 8 Mpc/$h_{100}$ $\sigma_{8}$, column seven: number of particles in the simulation $N_{\textrm{p}}$, column eight: mass of a particle $m_{\textrm{p}}$, column nine: side length of the simulation box $L$, column ten: force softening parameter $\epsilon$.}
\label{cosmopar}
\end{table*}

Given that the cosmological parameters (see Table \ref{cosmopar}) used by the Millennium simulation in its original run \citep{Millennium} substantially deviate from the recently adopted values based on CMB data from the Planck satellite \citep{Planck} than reruns \citep{Guo:2013}, which used WMAP7 cosmological parameters \citep{WMAP7}, it was appealing to use this rerun.  However, for the reruns \citep{Guo:2013}, the database of the Millennium simulation does not contain the friends-of-friends groups, which are required for a proper calibration of a group finder algorithm, but only the halos derived from them.  Hence, we had to use the data from the original run.

\begin{table*}
\begin{center}
\begin{tabular}{cc|ccc}
\emph{snapnum} & redshift & number of & number of & percentage of \\
& & FoF-groups & galaxies & particles in groups \\ \hline
63 & 0.000 & 7 913 369 & 6 981 224 & 46.8\\
62 & 0.020 & 7 933 951 & 7 032 122 & 46.5\\
61 & 0.041 & 7 955 548 & 7 124 656 & 46.2\\
60 & 0.064 & 7 979 530 & 7 226 286 & 45.8\\
59 & 0.089 & 8 003 794 & 7 337 200 & 45.4\\
58 & 0.116& 8 033 674 & 7 455 464 & 45.0\\
\end{tabular}
\end{center}
\caption{List of number of galaxies, number of FoF groups and the percentage of all particles in these groups (with at least 20 particles) by a snapshot used, and corresponding redshifts.}
\label{listnumbers}
\end{table*}

For our group catalogue, we intended to reach up to a redshift of 0.11, which corresponds to the co-moving distance of about 323 Mpc/$h_{100}$. Therefore, we had to obtain a cube of at least this side-length. Using our method to derive multiple mock catalogues (see the next section) from the dataset, we would ideally retrieve the full volume of the Millennium run (a cube with 500 Mpc/$h_{100}$ side-length), but due to limitations in our available computational facilities, we had to restrict ourselves to a cube with the 400~Mpc/$h_{100}$ side-length.  To properly consider evolutionary effects, we obtained not only data from the last (present day) snapshot of the Millennium simulation, but also from all snapshots up-to (actually slightly beyond) the limiting redshift of our group catalogue.  In the end, we received six cubes with a side-length of 400 Mpc/$h_{100}$ for the latest snapshots (\emph{snapnum} 63 (present day) to 58) of the Millennium simulation. The basic data is listed in Table \ref{listnumbers}.  For the FoF-groups of these cubes, we obtained the following parameters: FoF-group ID number, co-moving coordinates, and number of particles in the FoF-group.

Because the Millennium simulation is a dark matter only simulation, all data on the luminous part of galaxies was derived from semi-analytic galaxy models with which the dark matter halos were populated.  We used the semi-analytic models from \citet{Guo:2011} created using the L-galaxies galaxy formation algorithm \citep{Croton:2006,DeLucia:2006}, and obtained the following list of parameters for all galaxies brighter than $-15$~mag in the SDSS $r$ band in the previously defined cubes of the snapshots (see Table \ref{listnumbers}): galaxy ID, FoF-group ID number to which the galaxy belongs, co-moving coordinates, peculiar velocities, number of particles in the galaxy's halo, and ``dusty''\footnote{Considering internal dust extinction effects in the galaxies} absolute SDSS magnitudes for the $g$, $r$, and $i$ band.

Less than half of the particles of the simulation are in FoF-groups (see Table \ref{listnumbers}).  This is a cause for concern, because we aim to create a model of the local Universe, which provides a highly complete picture of the mass distribution.  Full particle information on the original run of the Millennium simulation was not available to us (and would exceed our computational capacities anyway).  However, for a smaller volume (see Table \ref{cosmopar}) of the \emph{millimil} run (MM), the complete particle information as well as the FoF-groups are available in the Millennium simulation database.  For the entire MM volume and the latest six snapshots (the same as for the original run) we obtained the following parameters for all FoF groups: FoF-group ID, Cartesian coordinates, number of particles in the FoF-group, and radius $R_{200,\textrm{NFW}}$, within which the FoF group has an overdensity 200 times the critical density of the simulation when fitted by a NFW-profile.  In addition to that, we received the same set of parameters of semi-analytical galaxy models as we did for the semi-analytical models in the original run using the same conditions as earlier for the entire MM volume.  We also obtained the Cartesian coordinates for all particles of the latest snapshots of MM.

\section{Description of the method}
\subsection{Processing observational data}
We had to further process the data retrieved from SDSS and 2MRS in order to construct a dataset suitable for our group finder algorithm.  The calibrations were largely the same for SDSS and 2MRS.  Methodological differences are explained below.

The first step was to correct for the Solar system motion relative to the comic microwave background (CMB).  We used the measurements from \citet{WMAP_5} and correct observed redshifts $z_{\textrm{obs}}$ applying a method explained in \citet{Saulder:2013}, Appendix A, but slightly updated using the redshift addition theorem \citep{Davis:2014}, to corrected redshifts $z_{\textrm{cor}}$.

Afterwards, we corrected the observed SDSS magnitudes $m_{\textrm{sdss}}$ for the Galactic foreground extinction according to \citet{Schlegelmaps}. 

The 2MRS magnitudes obtained from the database were already corrected for galactic extinction.  Hence, we could directly the observed 2MRS magnitudes as extinction-corrected magnitudes $m_{\textrm{cor}}$.

In the next step, we calculated the extinction and K-corrected apparent magnitudes $m_{\textrm{app}}$ using analytical polynomial approximations of K-corrections from \citet{Chilingarian:2010} updated in \citet{Chilingarian:2012} for SDSS bands:
\begin{equation}
K(z,m_{\textrm{cor},f_{1}}-m_{\textrm{cor},f_{2}})=\sum\limits_{i,j} B_{ij} z^{i} (m_{\textrm{cor},f_{1}}-m_{\textrm{cor},f_{2}})^{j}
\label{Kcorrection}
\end{equation}

We calculated the luminosity distance $D_{L}(z_{\textrm{cor}})$ following the equations \citep{Hogg:1999}:
\begin{equation}
D_{C}(z_{\textrm{cor}})=\frac{c}{H_{0}} \int_{0}^{z_{\textrm{cor}}} \textrm{d}z' \left( \Omega_{M} (1+z')^{3} + \Omega_{\Lambda} \right)^{-1/2}
\label{comovingdist}
\end{equation}
\begin{equation}
D_{L}(z_{\textrm{cor}})=D_{C} \left(1+z_{\textrm{cor}} \right).
\label{lumdistcom}
\end{equation} 
$D_{C}$ denotes for the co-moving distance.  For consistency reasons, we used the values of the original Millennium simulation run for the cosmological parameters (see Table \ref{cosmopar}). We then computed K-corrected absolute magnitudes $M_{\textrm{abs}}$.

After these calibrations, finer cuts were applied on the SDSS and 2MRS datasets to make them compatible with the mock catalogues built in the next sub-section and directly usable for our group finder algorithm.

From the SDSS data, we removed all galaxies which fulfilled all of the following conditions: a corrected redshift $z_{\textrm{cor}}$ higher than 0.11 or lower than zero, an apparent magnitude $m_{\textrm{app}}>18.27$~mag, 0.5~mag fainter than the official limiting magnitude in the $r$ band of 17.77~mag \citep{SDSS_spectarget} (to clean the sample from poorly identified or misclassified objects), a measured magnitude error greater than 1~mag in the $r$ or $g$ band, and $r$ band absolute magnitude $M_{\textrm{abs}}$ fainter than -15 mag.  Applying these cuts, our SDSS sample was reduced to 402,588 galaxies.

From the 2MRS data, we removed all objects for which no redshift information was provided.  We also removed all galaxies which fulfilled the following set of conditions: a corrected redshift $z_{\textrm{cor}}$ lower than zero, an apparent magnitude $m_{\textrm{app}}>12.25$~mag, 0.5 magnitude fainter than the official limiting magnitude in the $K_{s}$ band of 11.75~mag \citep{2MRS}, and $K_{s}$ band absolute magnitude $M_{\textrm{abs}}$ fainter than -18~mag.  After these cuts, we ended up with a 2MRS sample of 43,508 galaxies.

\subsection{Creating the mock catalogues}
We constructed sets of mock catalogues each corresponding to both observational datasets and the ``true'' dark matter distribution based on the simulated data.

\begin{figure*}[ht]
\begin{center}
\includegraphics[width=0.90\textwidth]{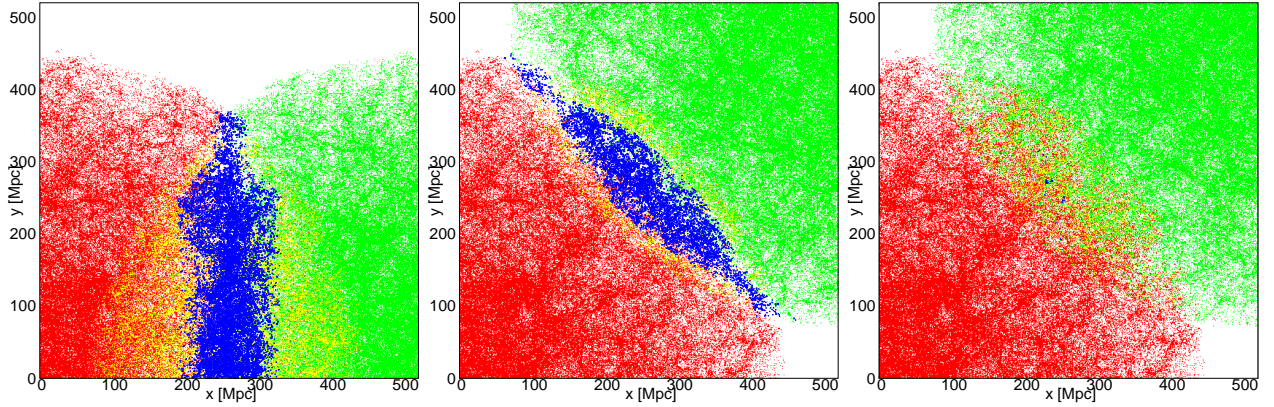}\\ 
\caption{Projections of the distribution of galaxies in the SDSS mock catalogues and their overlaps.  The fine green and red pixel in every plot show the projected (on the $xy$-plane) areas, where galaxies from two mock catalogues can be found, which belong to only one of the two catalogues.  Yellow pixels indicate galaxies from both catalogues.  The tiny blue crosses indicate galaxies that can be found in both catalogues in the same evolutionary stage (from the same redshift snapshot).  Left panel: overlap of two mock catalogues whose coordinate origins are located in the neighbouring corners.  Central panel: overlap of two mock catalogues whose origins are located in opposite corners, yet in the same plane (side of the cube).  Right panel: overlap of two mock catalogues whose origins are located diagonally, opposite across the entire cube.}
\label{SDSS_overlaps_map}
\end{center}
\end{figure*}
\begin{figure*}[ht]
\begin{center}
\includegraphics[width=0.90\textwidth]{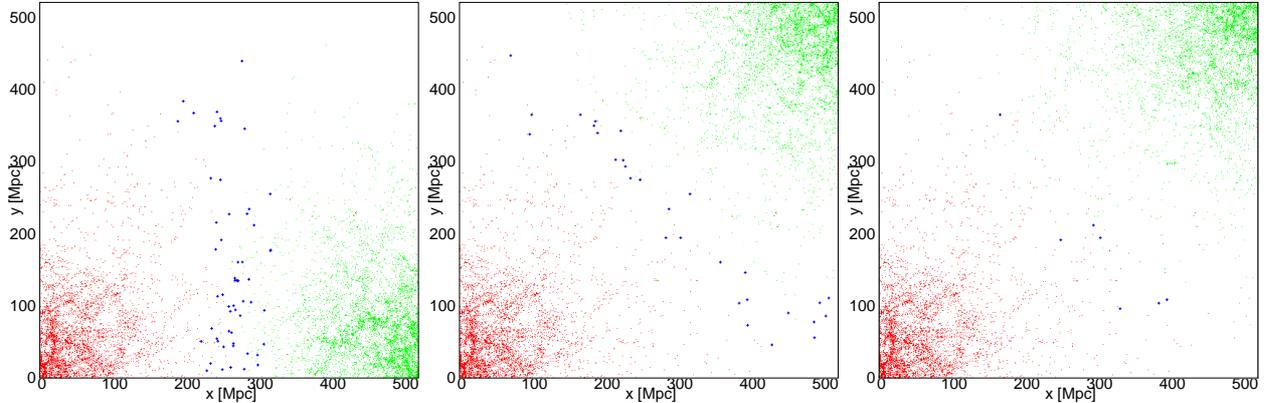}\\ 
\caption{Projections of the distribution of the galaxies in the 2MRS mock catalogues and their overlaps.  Symbols as in Figure~\ref{SDSS_overlaps_map}.}
\label{2MRS_overlaps_map}
\end{center}
\end{figure*}

We created eight (largely) independent mock catalogues for the data cubes obtained from original run of the Millennium simulation using the following method.  We put an observer into each of its 8 vertices\footnote{The origin shift was in fact handled by rotations in a way that all galaxies and halos are located in the first octant as seen from the origin.} and thereby obtained 8 different viewpoints, which became largely independent from each other once we included the Malmquist-bias into our calculations.  These mock catalogues are not completely independent, because the brightest galaxies can be seen across the entire cube and, therefore, some of them are included in every mock catalogue (yet at different distances/redshifts and consequently at slightly differently evolutionary states).  The overlaps of the different mock catalogues are illustrated in Figures \ref{SDSS_overlaps_map} and \ref{2MRS_overlaps_map} for different geometrical arrangements between a set of two mock catalogues for the same survey.  Depending on the distance between the corners, the overlap ranges from about nine percent to about a tenth of a percent for SDSS and always less than one percent for 2MRS\footnote{The survey is sufficiently shallow to only have a handful of galaxies in the overlap, resulting in a nearly empty Figure~\ref{2MRS_overlaps_map}.}.

Prior to constructing the mock catalogues, there was one important issue to be considered: the data from \citet{Guo:2011} does not contain 2MASS magnitudes.  Therefore, we derived them from SDSS magnitudes using a colour transformation inspired by \citet{Bilir:2008}, which is provided in Appendix \ref{2mass_sdss_transform}.

The (dark matter) FoF groups are important for the calibration of the group finder algorithm.  To avoid losing dark matter information on cut-off groups (edge-effect) in our sample, we moved the origin 10 Mpc/$h_{100}$ inwards in all directions and we restricted ``our view'' to the first octant.

The cosmological redshift $z_{\textrm{cosmo}}$ was calculated from the co-moving distance $D_{C}$ by inverting Equation~\ref{comovingdist}.  The co-moving distance itself was derived directly from the Cartesian coordinates of the Millennium simulation.  The luminosity distance $D_{L}$ was obtained from the co-moving distance and the cosmological redshift using Equation \ref{lumdistcom}.

However, one does not directly observe the cosmological redshift, but a parameter, which we call the ``observed'' redshift $z_{\textrm{obs}}$, which considers effects from peculiar motions and measurement uncertainties.

\begin{equation}
z_{\textrm{pec,rad}} = \frac{p_{x} v_{x} + p_{y} v_{y} + p_{z} v_{z} }{c \cdot D_{C}} 
\label{zpecrad}
\end{equation} 
\begin{equation}
z_{\textrm{obs,pec}} = ((1+z_{\textrm{cosmo}}) \cdot (1+z_{\textrm{pec,rad}})) - 1
\label{zprimeobs}
\end{equation} 
\begin{equation}
z_{\textrm{obs}} = z_{\textrm{obs,pec}} + \frac{\sigma_{z}}{c} \cdot \mathfrak{G}
\label{zobs}
\end{equation} 

The projection of the peculiar motion $v_{x}$, $v_{y}$, and $v_{z}$ on the line of sight from the coordinate origin (view point) to the galaxy at the Cartesian coordinates $p_{x}$, $p_{y}$, and $p_{z}$ yielded the redshift $z_{\textrm{pec,rad}}$ of the radial component of the peculiar motion.  We added it to the cosmological redshift using the theorem of \citet{Davis:2014}.  The redshift measurement uncertainty $\sigma_{z}$ is about 30~km/s for SDSS and about 32~km/s for 2MRS.  We use the symbol $\mathfrak{G}$ to indicate a random Gaussian noise with a standard deviation $\sigma$ of 1, which was implemented in our code using the function \emph{gasdev} (Normal (Gaussian) Deviates) from \citet{NumericalRecipes}.

The absolute magnitudes $M_{\textrm{abs}}$ from semi-analytic models were converted into apparent magnitudes $m_{\textrm{app}}$.

The magnitudes in the 2MRS catalogue were already corrected for extinction and the survey's limiting magnitude was applied on the extinction corrected values.  Consequently, we did not have to consider extinction for the 2MRS data.  

However, we had to take extinction into account for mock catalogues which we compared to SDSS data.  We used the reddening map of \citet{Schlegelmaps} and multiplied the reddening coefficients of that map with the conversion factors listed in \citet{SDSS_early} to obtain extinction values $A_{\textrm{model}}$.  To be more specific: only a part of the Schlegel map was used, which covers one eighth of the sky (the sky coverage of a single mock catalogue) and was at least 20 degrees away from the Galactic plane to reproduce a comparable extinction profile as the area covered SDSS.

\begin{equation}
m_{\textrm{obs}} = m_{\textrm{app}} + \bar{K}(z_{\textrm{app}},(m_{1,\textrm{app}}-m_{2,\textrm{app}})) + A_{\textrm{model}} + \bar{m}_{\textrm{err}} \cdot \mathfrak{G}.
\label{magobs}
\end{equation} 

Considering the extinction correction (for SDSS only), the mock K-correction $\bar{K}(z_{\textrm{app}},(m_{1,\textrm{app}}-m_{2,\textrm{app}}))$, and the photometric uncertainty to the apparent magnitude $\bar{m}_{\textrm{err}}$, we were able to calculate the observed magnitude $m_{\textrm{obs}}$.  The mock K-correction is explained in detail in Appendix \ref{mockK}.  The uncertainties in the Petrosian model magnitudes, which were used for the selection (limiting magnitude) of the spectroscopic sample of SDSS, were calculated using the uncertainties provided by the survey, which are 0.026 mag for the $g$ band and 0.024 mag for the $r$ band.  The photometric uncertainties of 2MRS are 0.037 mag in the $J$ band and 0.056 mag in the $K_{s}$ band.

Observations do not directly yield 3D positions as we have in the simulated data, but a 2D projection on the sky plus a redshift.  The equatorial coordinates $\alpha'$ and $\delta'$ were obtained by simple geometry:

\begin{align}
\alpha' &=\textrm{arctan}\left( \frac{p_{y}}{p_{x}} \right)\\
\delta' &=\textrm{arcsin}\left(\frac{p_{z}}{D_{C}} \right)\nonumber.
\label{radec}
\end{align} 

\begin{align}
\alpha &=\alpha'+\mathfrak{G}\cdot \sigma_{a} \textrm{sin}\left( 2 \pi \cdot \mathfrak{R} \right) \textrm{cos}\left( \delta \right)\\
\delta &=\delta'+\mathfrak{G}\cdot \sigma_{a} \textrm{cos}\left( 2 \pi \cdot \mathfrak{R} \right)\nonumber
\label{radecerr}
\end{align} 

Considering an astrometric uncertainty $\sigma_{a}$ of 0.1 arcseconds \footnote{\url{http://www.sdss3.org/dr10/scope.php\# opticalstats}}, we obtained the observed equatorial coordinates $\alpha$ and $\delta$.  The symbol $\mathfrak{R}$ indicates a uniformly distributed random variable between 0 and 1.

The evolutionary effects on galaxies and their distribution were taken into account by only using the galaxies from the snapshot (see Table \ref{listnumbers} for the redshifts of the snapshots) closest to their cosmological redshift $z_{\textrm{cosmo}}$.  This simplification is justified because passive evolution is sufficiently slow for nearby ($z_{\textrm{cosmo}} < \sim 0.1$) galaxies \citep{Kitzbichler:2007}.

The Malmquist bias was introduced into our mock catalogues by removing all galaxies with an `observed'' (apparent) magnitude $m_{\textrm{obs}}$ fainter than the limiting magnitude of the survey.  The limiting magnitude is 17.77 mag \citep{SDSS_spectarget} in the $r$ band for SDSS and 11.75 mag \citep{2MRS} in the $K_{s}$ for 2MRS.

All galaxies with an observed redshift $z_{obs}$ higher than 0.11 are removed from the SDSS mock catalogues.

We restricted our view of ``visible galaxies'' to the first octant of the coordinate system, which was necessary, because we had shifted the origin by 10 Mpc/$h_{100}$ inwards earlier to avoid potential problems with the groups of the mock catalogue contributing to the visible distribution being partially cut.  We still used the FoF group data from beyond these limits for later calibrations.

The completeness of 2MRS of 97.6$\%$ \citep{2MRS} was taken into account by randomly removing 2.4$\%$ of the galaxies from the mock catalogue.  The SDSS sample, before considering additional cuts due to fibre collisions (in SDSS only, 2MRS was obtained differently), is more than 99$\%$ complete \citep{Blanton:2003}, which was implemented in the same way as for 2MRS.

Taking the fibre collision into account correctly is very important, since this is more likely to happen in clusters or dense groups of galaxies, which are essential objects for our group finder.  The size of the fibre plugs of SDSS does not allow two spectra to be taken closer than 55 arcseconds of one another \citep{Blanton:2003}.  Consequently if we found any galaxy in our mock catalogue that was closer than this minimal separation to another galaxy we removed one of the two galaxies at random.  Foreground stars and background galaxies (at least as long they were in a luminosity range to be considered spectroscopic targets for SDSS) were not considered in our treatment of the fibre collisions, not only because this would be beyond the capabilities of our available simulated data, but also because they should not correlate with the distribution of galaxies in the range of our catalogue.  Thanks to the SDSS tiling algorithm \citep{Blanton:2003}, some areas are covered more than once.  This allows for spectra to be taken from galaxies that were blocked due to fibre collision the first time around.  An overall sampling rate of more than 92$\%$ was reached.  We implemented this by randomly re-including galaxies, which were previously removed due to fibre collision, until this overall sampling rate was reached.  Because we selected only SDSS objects, which were spectroscopically classified as galaxies by the automatic pipeline, we randomly exclude a number of galaxies from the mock catalogue corresponding to the number of QSO classifications in the same volume of SDSS (normalized from the spectroscopic sky coverage of SDSS 9,376 square degree \citep{SDSS_DR12} to the coverage of the mock catalogue (one octant = $\sim$5,157 square degree).

\begin{figure*}[ht]
\begin{center}
\includegraphics[width=0.90\textwidth]{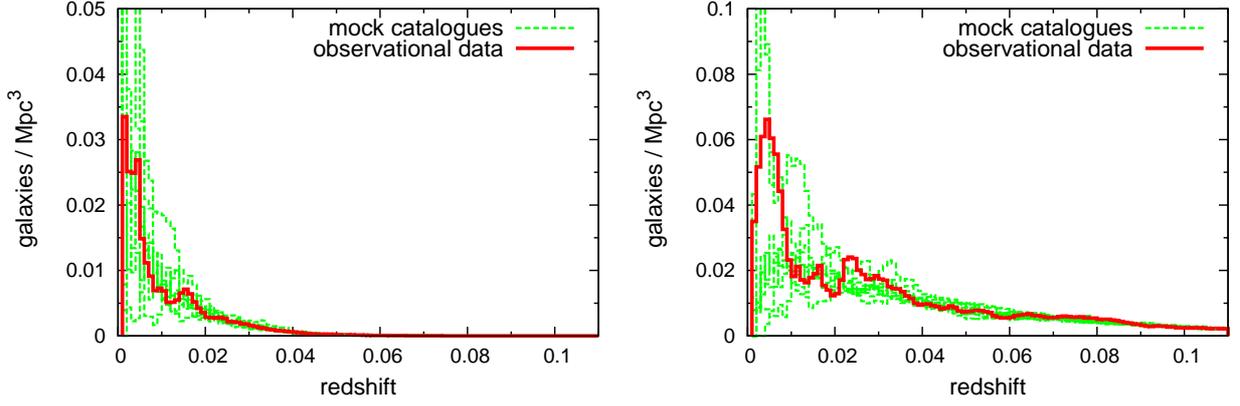}\\ 
\caption{Redshift dependence of the number densities of the observational data and the mock catalogues. Left panal: 2MRS data; right panel: SDSS data.}
\label{bin_ndes}
\end{center}
\end{figure*}

To complete the construction of our mock catalogues, we applied the same pipeline that we used to process the observational data on the mock catalogue data so far.  We ended up with eight mock catalogues for SDSS containing between 209,341 and 234,481 simulated galaxies.  These numbers correspond to a relative richness of our mock catalogues compared to SDSS of 94 to 104$\%$.  Also eight 2MRS mock catalogues consisting of 5,878 to 7,386 simulated galaxies (corresponding to a relative richness between 98 to 124$\%$ compared to 2MRS) were created.  The stronger variations in the number of galaxies per 2MRS mock catalogue is due to the survey's shallower depth, which makes the mock catalogues more affected by local variations in the matter distribution. As illustrated in Figure \ref{bin_ndes}, the number densities of the observational data are within the range and scatter of the various mock catalogues for both datasets. 

Additionally, we also calculated the distribution of the FoF groups corresponding to the mock catalogues by applying the before-mentioned origin shift and the method, which we used to consider the evolutionary effects on the FoF groups obtained from the Millennium simulation.  The masses of the FoF groups were obtained by multiplying the number of particles in them \emph{MField..FOF.np} with the mass per particle (see Table \ref{cosmopar}).  We produced FoF groups catalogues corresponding to the mock catalogues of simulated galaxies containing between 8,021,151 and 8,022,371 FoF groups.  Additonally, unbiased (no Malmquist bias, fibre collision correction, and sampling correction applied) mock catalogues of the simulated galaxies were created (containing between 7,403,914 and 7,407,193 simulated galaxies) for later use in our calibrations of the luminosity function for the Malmquist bias correction.

\subsection{Calibrating the group finder}

We used a friends-of-friends (FoF) algorithm to detect groups in our data. It recursively found all galaxies, which were separated by less than the linking length $b_{\textrm{link}}$ from any other galaxy and consolidated them in groups.

FoF algorithms are a straightforward procedure, if full 3D information is available (in fact the FoF-groups in the Millennium simulations were obtained that way from the particle distribution).  However, data from surveys is strongly affected by observational biases, which have to be considered for the group finder algorithm.  Therefore, we required some additional considerations and calibrations before we could get to the group catalogue.  We roughly followed \citet{Robotham:2011}, who created and applied a FoF-group finder algorithm on the GAMA survey \citep{GAMA_basics}.

We started by calculating a basic values for the linking length $b_{\textrm{link},0}$ for our surveys.  It was defined as the average distance from one galaxy to the nearest galaxy in the (unbiased) sample.  To estimate this distance, we used the present-day snapshot of MM and apply our limits for the absolute magnitudes (brighter than -15~mag in the SDSS $r$ band and -18~mag in 2MASS $K_{s}$).  We obtained the following values for $b_{\textrm{link},0}$: $\sim 0.64$ Mpc for 2MRS and $b_{\textrm{link},0}$ $\sim 0.67$ Mpc for SDSS (interestingly, very similar to the distance between the Milky Way and M31).  This was the baseline for the effective linking length $b_{\textrm{link}}$, which was used in the group finder and depends on several other parameters.

\begin{figure}[ht]
\begin{center}
\includegraphics[width=0.45\textwidth]{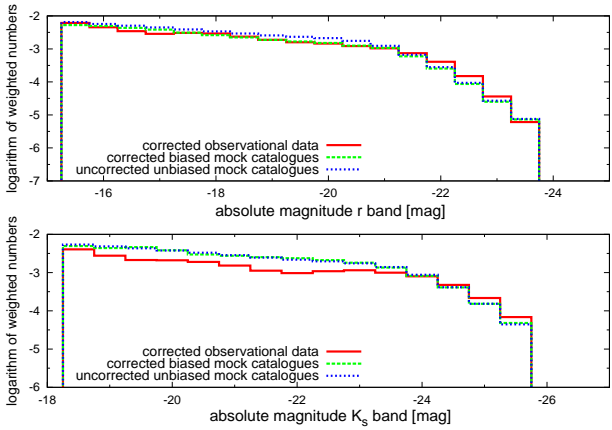}\\ 
\caption{Luminosity function for SDSS and 2MRS data are shown in the upper and lower panels correspondingly.  Green dashed line: true luminosity function derived using all galaxies from the unbiased mock catalogues.  Blue dashed line: corrected (reconstructed) luminosity function from the Malmquist biased mock catalogues.  Red solid line: corrected observational data.}
\label{lum_func}
\end{center}
\end{figure}

The dominant bias affecting our survey data is the Malmquist bias, which drastically removed the fainter end of the galaxy luminosity function from our sample at larger distances. To this end, we retrieved the luminosity function from the unbiased mock catalogues and compared it to the mock catalogues that consider all the biases (see Figure \ref{lum_func}). With some additional considerations, we proceeded to calcuate a correction that allowed us to retrieve the unbiased luminosity function from observational (biased) data.

\begin{equation}
\textrm{log}_{10}\left(D_{L,\textrm{limit}}\right)=\frac{-M_{\textrm{abs}} + m_{\textrm{limit}} + 5}{5}.
\label{limit_dist}
\end{equation}

Using the limiting magnitudes $m_{\textrm{limit}}$ of the surveys, we calculated the limiting luminosity distance $D_{L,\textrm{limit}}$ up to which a galaxy with an absolute magnitude $M_{\textrm{abs}}$ is still detected.

\begin{equation}
V_{C}(z_{\textrm{sat}},z_{\textrm{limit}})=\frac{4 \pi}{3} \frac{A_{\textrm{survey}}}{A_{\textrm{sky}}} \left(D_{C}^{3}(z_{\textrm{limit}})-D_{C}^{3}(z_{\textrm{sat}})\right)
\label{comvol}
\end{equation}

Using the inversion of Equation \ref{comovingdist}, we derived the limiting redshift $z_{\textrm{limit}}$ corresponding to the limiting luminosity distance $D_{L,\textrm{limit}}$ and using Equation \ref{lumdistcom} the corresponding co-moving distance $D_{C}$.  The saturation redshift $z_{\textrm{sat}}$ was defined in the same was as the limiting redshift (using Equation \ref{limit_dist}, but with the saturation limit of the survey $m_{\textrm{sat}}$ instead of the limiting magnitude $m_{\textrm{limit}}$).  The saturation magnitude of SDSS is 14 mag in $r$ band\footnote{\url{http://classic.sdss.org/dr7/instruments/technicalPaper/index.html}} and for 2MRS it does not apply, hence we set the limiting distance/redshift to zero.  We thus calculated the co-moving volume $V_{C}$ in which a galaxy with an absolute magnitude $M_{\textrm{abs}}$ can be detected in a survey covering $A_{\textrm{survey}}$ of the total sky area $A_{\textrm{sky}}$.

\begin{equation}
w_{\textrm{vol},i}=\frac{\left(V_{C,i}\right)^{-1}}{\sum\limits_{j} \left(V_{C,j}\right)^{-1}} .
\label{volweight}
\end{equation}

Using the co-moving volumes $V_{C,i}$ within which a galaxy with the index $i$ is detectable, we calculated the volume weights $w_{\textrm{vol},i}$. The volume weights allowed us to correct the observed luminosity function (either from observational data or from our mock catalogues) as it is illustrated in Figure \ref{lum_func}.  For both, the SDSS and 2MRS mock catalogues, the reconstruction works extremely well, however when comparing these results to the observational data, there are some small deviations, especially for 2MRS.  We attributed them to the additional uncertainty introduced into the mock data by using the SDSS-2MRS colour transformation (see Appendix \ref{SDSS2MASStransformation}).

Due to the Malmquist bias, the fainter members of a group are not visible anymore at higher redshifts.  With a constant linking length, this would cause groups to fragment at greater distances.  To avoid this, we adjusted the linking length of our FoF-group finder accordingly.

\begin{equation}
b_{\textrm{cor},\Phi}(z) = \left(\frac{\int_{-\infty}^{-5 \textrm{log}_{10}\left(D_{L}(z)\right) + m_{\textrm{limit}} + 5} \Phi \left(m \right) dm}{\int_{-\infty}^{M_{\textrm{abs,min}}} \Phi \left(m \right) dm} \right)^{-1/3}
\label{linking_lumfunccorr}
\end{equation}

The modification factor $b_{\textrm{cor},\Phi}(z)$ corrects for the Malmquist bias by calculating the fraction of the luminosity function $\Phi$ that is visible at a redshift $z$ for a luminosity limited survey with a limiting magnitude $m_{\textrm{limit}}$ and a minimal absolute magnitude $M_{\textrm{abs,min}}$ (-15~mag in the r band for SDSS and -18~mag in the $K_{s}$ band for 2MRS) for galaxies that are still included in dataset. $b_{\textrm{cor},\Phi}(z)$ was used to rescale the basic linking length $b_{\textrm{link,0}}$.

In physical space, the linking length is an isotropic quantity, it does not depend on the direction.  However, when observing galaxies projected on the sky, one can only directly obtain two coordinates, while the third dimension is derived from the redshift and is affected by so-called redshift-space distortions. This observed redshift cannot be exclusively attributed to the metric expansion of space-time, but also to the peculiar radial velocity of the galaxy.  The imprint on the observed redshift from these peculiar motions cannot be distinguished a priori from the cosmological redshift.  Hence, when using a redshift--distance relation to convert redshifts into distances, the estimated positions are smeared out in the radial direction with respect to the true positions.  In the case of galaxy groups/clusters, this effect is known as the ''Fingers of God'' effect \citep{Jackson:1972,Arp:1994,Cabre:2009}.

Although deforming the sphere with the linking length as radius to an ellipsoid appears to be the natural way to incorporate these effects into FoF-group finder algorithms, \citet{Eke:2004a} found that cylinders along the line of sight are more efficient.  Therefore, instead of one linking length we used two separate linking lengths: an angular linking length $\alpha_{\textrm{link}}$ and a radial linking length $R_{\textrm{link}}$.

\begin{equation}
\alpha_{\textrm{link}}=\textrm{tan}\left(\frac{b_{\textrm{link}}}{D_{A}} \right)
\label{angular_linking}
\end{equation}

The angular linking length is unaffected by the redshift-space distortion and directly relates to the linking length in real space $b_{\textrm{link}}$ by simple trigonometry.  The angular diameter distance $D_{A}$ is defined as:

\begin{equation}
D_{A}=D_{C} \left(1+z_{\textrm{cosmo}} \right)^{-1}.
\label{angulardist}
\end{equation} 

The radial linking length is larger than the linking length in real space because of the scatter in the redshift space due to the peculiar motions.  We transformed the $b_{\textrm{link}}$ distance into a corresponding redshift difference:

\begin{equation} 
R_{\textrm{link}}=b_{\textrm{link}} + 2 \sigma_{\textrm{rad}}
\label{radial_linking}
\end{equation}

\begin{figure}[ht]
\begin{center}
\includegraphics[width=0.45\textwidth]{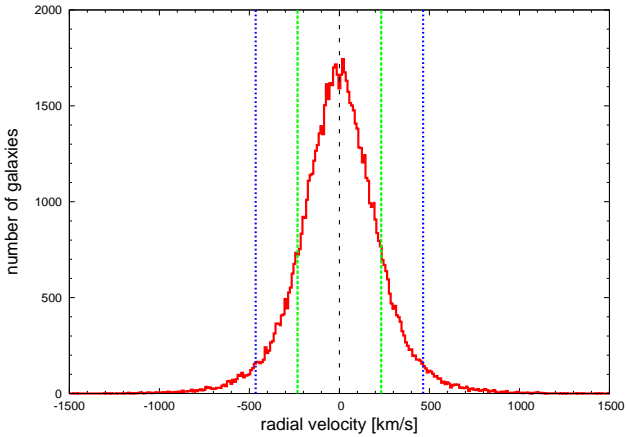}\\ 
\caption{The red solid line shows the distribution of radial proper motions in MM, which has a roughly Gaussian shape.  Black dashed line: zero radial velocity bin.  Green dashed lines: dispersion of the radial peculiar velocities $\sigma_{\textrm{rad}}$, which corresponds to the standard deviation of the plotted distribution.  Blue dashed lines: two-$\sigma$ interval, which is used to stretch the radial linking length.}
\label{velocities}
\end{center}
\end{figure}

By adding the dispersion of radial peculiar velocities $\sigma_{\textrm{rad}}$ to the linking length in real space, one gets a first estimate of the radial linking length $R_{\textrm{link}}$.  In MM, we found a $\sigma_{\textrm{rad}}$ of $\sim 232.3$ km/s for SDSS and $\sim 233.3$ km/s for 2MRS, hence about 95$\%$ of all possible radial velocity differences were included in an envelope of $\pm 464.7$ km/s, and $\pm 466.6$ km/s respectively, as illustrated in Figure \ref{velocities}.

We combined all these corrections and modifications to the linking length and obtained one set of equations:

\begin{equation}
\alpha_{\textrm{eff}}= \alpha_{\textrm{opt}} \cdot \alpha_{\textrm{link}}\cdot \left(b_{\textrm{cor},\Phi}(z)\right)^{\lambda_{\textrm{gal}}}
\label{alpha_final}
\end{equation} 
\begin{equation}
R_{\textrm{eff}}= R_{\textrm{opt}} \cdot R_{\textrm{link}} \cdot  \left(b_{\textrm{cor},\Phi}(z)\right)^{\lambda_{\textrm{opt}}}.
\label{R_final}
\end{equation} 

The effective angular linking length $\alpha_{\textrm{eff}}$ and the effective radial linking length $R_{\textrm{eff}}$ define the linking conditions of our group finder.  The exponent $\lambda_{\textrm{opt}}$ allows for a completeness correction (see Equation \ref{linking_lumfunccorr}).  The coefficients $\alpha_{\textrm{opt}}$, $R_{\textrm{opt}}$, and $\lambda_{\textrm{opt}}$ allow for fine-tuning of the group finder.  This was done using the mock catalogues.  For the optimization we followed \citet{Robotham:2011}, who provided a well-tested method that was successfully applied on GAMA survey data \citep{GAMA_basics}, and defined a group cost function in the following way:

\begin{equation}
E_{\textrm{fof}}=\frac{N_{\textrm{bij}}(n_{\textrm{limit}})}{N_{\textrm{fof}}(n_{\textrm{limit}})},  
\label{E_fof}
\end{equation} 

\begin{equation}
E_{\textrm{mock}}=\frac{N_{\textrm{bij}}(n_{\textrm{limit}})}{N_{\textrm{mock}}(n_{\textrm{limit}})},  
\label{E_mock}
\end{equation} 

\begin{equation}
E_{\textrm{tot}} = E_{\textrm{mock}} \cdot E_{\textrm{fof}}.
\label{E_tot}
\end{equation} 

The global halo finding efficiency measurement $E_{\textrm{tot}}$ is defined by the product of the halo finding efficiencies of the mock catalogue $E_{\textrm{mock}}$ and the FoF catalogue $E_{\textrm{fof}}$.  The parameters $N_{\textrm{mock}}(n_{\textrm{limit}})$ and $N_{\textrm{fof}}(n_{\textrm{limit}})$ are the number of groups with at least $n_{\textrm{limit}}$ members in the mock catalogue and in the results of our FoF-based group finder.  $N_{\textrm{bij}}(n_{\textrm{limit}})$ is the number of groups that are found bijectively in both samples (the mock catalogue, which consists of the ``true'' group information based purely on the simulation, and the FoF catalogue, which consists of the grouping found after applying the group finder on the mock catalogue).  This means that at least 50$\%$ of the members found in a group in one sample must make up at least 50$\%$ of a corresponding group in the other sample as well.

\begin{equation}
Q_{\textrm{fof}}=\frac{\sum_{i=1}^{N_{\textrm{fof}}} P_{\textrm{fof}}(i) \cdot N_{\textrm{members,fof}}(i)}{\sum_{i=1}^{N_{\textrm{fof}}} N_{\textrm{members,fof}}(i)}
\label{Q_fof}
\end{equation} 
\begin{equation}
Q_{\textrm{mock}}=\frac{\sum_{i=1}^{N_{\textrm{mock}}} P_{\textrm{mock}}(i) \cdot N_{\textrm{members,mock}}(i)}{\sum_{i=1}^{N_{\textrm{mock}}} N_{\textrm{members,mock}}(i)}
\label{Q_mock}
\end{equation} 
\begin{equation}
Q_{\textrm{tot}} = Q_{\textrm{mock}} \cdot Q_{\textrm{fof}}
\label{Q_tot}
\end{equation} 

The global grouping purity $Q_{\textrm{tot}}$ is defined using the grouping purity of the mock catalogue $Q_{\textrm{mock}}$ and the grouping purity of the FoF catalogue $Q_{\textrm{fof}}$.  The variables $N_{\textrm{members,mock}}(i)$ and $N_{\textrm{members,fof}}(i)$ are the numbers of galaxies in individual groups $i$ of the mock catalogue and the FoF catalogue respectively.  The purity products $P_{\textrm{mock}}(i)$ and $P_{\textrm{fof}}(i)$ are defined as the maximal product of the ratio of shared galaxies to all galaxies within a group of one catalogue and the ratio of the same shared galaxies within the other catalogue.  An illustrative example was provided in \citet{Robotham:2011}.

\begin{equation}
S_{\textrm{tot}} = E_{\textrm{tot}} \cdot Q_{\textrm{tot}}
\label{S_tot}
\end{equation} 

The group cost function $S_{\textrm{tot}}$ is defined by the product of the global halo finding efficiency measurement $E_{\textrm{tot}}$ and the global grouping purity $Q_{\textrm{tot}}$.  Following the definitions, $S_{\textrm{tot}}$ can take values between 0 (total mismatch) and 1 (perfect match).

\begin{table*}
\begin{center}
\begin{tabular}{c|ccccc}
sample & $b_{\textrm{link},0}$  & $\alpha_{\textrm{opt}}$ & $R_{\textrm{opt}}$ & $\lambda_{\textrm{opt}}$ & $S_{\textrm{tot}}$ \\ \hline \hline
2MRS & 0.64 Mpc & $0.609 \pm 0.004$ & $0.707 \pm 0.004$ & $0.582 \pm 0.008$ & $0.321$ \\
SDSS & 0.67 Mpc & $0.522 \pm 0.006$ & $0.750 \pm 0.026$ & $0.822 \pm 0.033$ & $0.188$ 
\end{tabular}
\end{center}
\caption{Optimal coefficients for the group finders for 2MRS and SDSS.  The parameter $b_{\textrm{link},0}$ is the basic linking length, which corresponds to the distance to nearest visible neighbour at redshift zero.  The coefficient $\alpha_{\textrm{opt}}$ allows for proper scaling of the angular linking length, while the coefficient $R_{\textrm{opt}}$ does the same for the radial linking length (in the redshift space).  The coefficient $\lambda_{\textrm{opt}}$ provides the optimized dependence on the scaling of the Malmquist bias correction.  $S_{\textrm{tot}}$ is the median value of the group cost function calculated using all mock catalogues and the optimal coefficients.}
\label{optimal_coefficients}
\end{table*}

We started our optimization by performing a coarse parameter scan for the three coefficients $\alpha_{\textrm{opt}}$, $R_{\textrm{opt}}$, and $\lambda_{\textrm{opt}}$ in one of our mock catalogues to get an initial guess for the order of magnitude of optimal coefficients.  We then used a Simplex algorithm \citep{Simplex} to maximise the mean group cost function $S_{\textrm{tot}}$ of all of our 8 mock catalogues.  The optimal coefficients for both samples, SDSS and 2MRS, are listed in Table \ref{optimal_coefficients}.  For the calculation of $S_{\textrm{tot}}$ and the optimisation, we used $n_{\textrm{limit}}=2$.  We also repeated it with different values for $n_{\textrm{limit}}$ and found very similar optimal coefficients (a few percent difference).

The coefficients $\alpha_{\textrm{opt}}$ and $R_{\textrm{opt}}$ are well within an order of magnitude of unity, indicating that our initial definitions of the effective linking lengths are reasonable.  The coefficient $\lambda_{\textrm{opt}}$ is clearly below the naive expected value of 1, which shows that it was important to consider this parameter in the optimization.  The distribution of the median group cost function depending on the coefficient $\alpha_{\textrm{opt}}$ and $R_{\textrm{opt}}$ is illustrated in Figure \ref{optimal_2MRS_SDSS}.

\subsection{Obtaining group parameters}
After detecting groups using our optimized group finder algorithm, we calculated various parameters for these groups, such as their positions, sizes, masses and luminosities.  For most of these parameters, we used methods, which were tested and found to be efficient and robust by \citet{Robotham:2011}.  For some, we had to calibrate them using our own mock catalogues.

\subsubsection{Group velocity dispersion}
\label{sec_veldisp}
We calculated group velocity dispersions using the gapper estimator of \citet{Beers:1990} including the modification of \citet{Eke:2004a}. Naturally, this could only be done for groups with at least two members.  This well-tested method requires the following calculations:

\begin{equation}
\sigma_{\textrm{gap}}=\frac{\pi}{N_{\textrm{fof}}(N_{\textrm{fof}}-1)}\sum\limits_{i=1}^{N_{\textrm{fof}}-1} w_{i} g_{i},
\label{gapper_basic}
\end{equation} 
\begin{equation}
w_{i}=i \cdot (N_{\textrm{fof}}-i),
\label{gapper_weight}
\end{equation}
\begin{equation}
g_{i}=v_{i+1}-v_{i},
\label{gapper_vgap}
\end{equation}
\begin{equation}
\frac{v_{i}}{c} =\frac{(1+z_{\textrm{obs},i})^{2}-1}{(1+z_{\textrm{obs},i})^{2}+1},
\label{vrad_z}
\end{equation}
\begin{equation}
\sigma_{\textrm{group}}=\sqrt{\frac{N_{\textrm{fof}}}{N_{\textrm{fof}}-1}\sigma_{\textrm{gap}}^{2}-\sigma_{\textrm{err}}^{2}}.
\label{gapper_mod}
\end{equation}

The gapper velocity dispersion $\sigma_{\textrm{gap}}$ of a group with $N_{\textrm{fof}}$ member was calculated by summing up the product of the weights $w_{i}$ and the radial velocity gaps $g_{i}$ for all all its members.  It was essential that the radial velocities $v_{i}$ were ordered for this approach, which we assured by applying a simple sorting algorithm for each group.  The radial velocities $v_{i}$ were calculated using the observed redshifts $z_{\textrm{obs},i}$.  The group velocity dispersion $\sigma_{\textrm{group}}$ also took into account the measurement errors of the redshift determination $\sigma_{\textrm{err}}$, which were 30 km/s for SDSS and $\sim 32$ km/s for 2MRS.  In the case that the obtained group velocity dispersion was lower than the measurement errors of the redshift determination, we set them to $\sigma_{\textrm{err}}$.

\subsubsection{Total group luminosity}
The observed group luminosity $L_{\textrm{obs}}$ was calculated by adding up the emitted light, in the SDSS r band or the 2MASS K$_{\textrm{s}}$ band respectively, of the group members.

\begin{equation}
L_{\textrm{obs}}=\sum\limits_{i=1}^{N_{\textrm{fof}}} L_{i}
\label{lum_obs}
\end{equation}
\begin{equation}
L_{i} = 10^{-0.4\cdot (M_{\textrm{abs},i} - M_{\textrm{abs},\astrosun})}
\label{lum_individual}
\end{equation}

The calculation of the luminosity of an individual galaxy $L_{i}$ required the absolute r band/K$_{\textrm{s}}$ band magnitudes $M_{\textrm{abs},i}$ and the solar absolute magnitude $M_{\textrm{abs},\astrosun}$ in the $r$ band of 4.76~mag or in the $K_{s}$ band of 3.28~mag respectively.

\begin{equation}
L_{\textrm{tot}}=L_{\textrm{obs}} \frac{\int_{-\infty}^{M_{\textrm{abs,min}}} \Phi \left(m \right) dm}{\int_{-\infty}^{-5 \textrm{log}_{10}\left(D_{L}(z)\right) + m_{\textrm{limit}} + 5} \Phi \left(m \right) dm}
\label{lum_tot}
\end{equation}

We obtained the total group luminosity $L_{\textrm{tot}}$ by rescaling the observed group luminosity $D_{L}$ with the fraction of the luminosity function $\Phi \left(m \right)$ visible at the group's luminosity distance.  $m_{\textrm{limit}}$ is the limiting magnitude of the survey and the parameter $M_{\textrm{abs,min}}$ denotes the minimal absolute magnitude to which the luminosity function is still considered in our sample.  This method only corrects for the Malmquist bias, but not the saturation bias, because there is no suitable way to predict the group luminosity if the brightest object of the group is missing (or maybe not).

\subsubsection{Group centre}
The radial and the projected group centre were calculated differently, because the former one was obtained using redshifts, while the latter one from astrometry.  \citet{Robotham:2011} compared various approaches and we applied those they found to be the most efficient.  In the case of the radial group centre, this turned to be simply taking the median of the redshifts of all detected group members.  The most efficient method to find the projected group centre was an iterative approach using the centre of light of the group members as explained in \citet{Robotham:2011}.  At first, the coordinates of the centre of light, which is the luminosity-weighted (using $L_{i}$ as weights), were calculated using all group members.  Then the group member had the largest angular separation from the centre of light was rejected and the new centre of light was calculated with the remaining members.  This process was repeated iteratively until only one galaxy remained and its coordinates were used as the coordinates of the projected group centre.

\subsubsection{Group radius}
We calculated a characteristic projected radius for groups with two or more members.  Again, following \citet{Robotham:2011}, who tested this method extensively, we define our group radius $R_{\textrm{group}}$ as the radius around the projected group centre in which 50$\%$ of the group members are located.  To illustrate this simple definition, we provide two examples: for a group with five members the radius corresponds to the distance of the third most distant member from the group centre and for a group with four members, the radius is the mean between the distance from the group centre of the second and third most distant members.

\subsubsection{Dynamical mass}
Using the previously defined group radii and group velocity dispersions, we calculated approximate dynamical masses $M_{\textrm{dyn}}$ for our groups using the following equation \citep{Spitzer:1969,Robotham:2011,Chilingarian:2008}:

\begin{equation}
M_{\textrm{dyn}} \sim \frac{10}{G} \cdot (\sqrt{3}\sigma_{\textrm{group}})^2  \cdot R_{\textrm{group}}.
\label{dyn_mass}
\end{equation}
\subsubsection{Stellar mass}
\label{sec_mstar}

As a part of the 2MRS and SDSS DR12 group catalogues, we provide stellar masses of galaxies included in the two samples.  The stellar masses were computed as follows.

For the objects included in the SDSS DR12 sample, we used the Reference Catalog of Galaxy Spectral Energy Distributions (RCSED, Chilingarian et al. submitted) in order to obtain stellar population ages and metallicities derived from the full spectrum fitting of SDSS DR7 spectra using the \emph{nbursts} code \citep{Chilingarian:2007a,Chilingarian:2007b} and a grid of simple stellar population (SSP) models computed with the \emph{pegase.hr} code \citep{LeBorgne:2004}.  371,627 objects from the SDSS DR12 sample matched the RCSED sample.  We derived their stellar mass-to-light ratios using the \emph{pegase.hr} models at the corresponding ages and metallicities, the \citet{Kroupa:2002} stellar initial mass function, and then converted them into stellar masses using the K-corrected integrated photometry of all galaxies in the SDSS $r$ and $g$ bands converted into Johnson $V$.  Here we assumed that the SDSS aperture spectra reflected 
global stellar population properties of galaxies.

\begin{figure}
\includegraphics[width=0.45\textwidth]{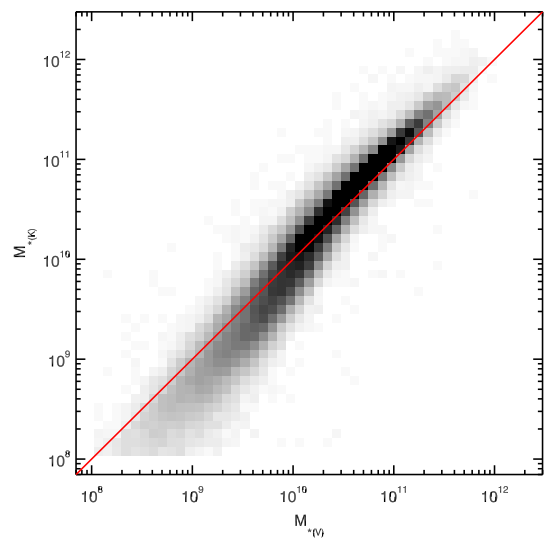}
\caption{Comparison between stellar masses computed using \emph{pegase.hr} stellar population models for the $V$ band photometry (converted from SDSS $g$ and $r$) with those calculated using the near-infrared UKIDSS $K$ band magnitudes and a grid of \citet{Maraston:2005} SSP models. RCSED SSP ages in metallicities and the Kroupa IMF were used in both stellar mass estimates.
\label{fig_ml_k_r}}
\end{figure}

125,383 galaxies of our SDSS DR12 sample possess fully corrected integrated photometry in the near-infrared $K$ band from the UKIDSS survey \citep{UKIDSS} available in RCSED, which is known to be a good proxy for a stellar mass.  We used the \citet{Maraston:2005} $K$ band mass-to-light ratios computed for the Kroupa IMF and SSP ages and metallicities obtained from the full spectrum fitting and computed stellar masses for the subsample of 125,383 SDSS DR12 galaxies.  The $K$ band mass-to-light have much weaker dependence on stellar population ages compared to the optical $V$ band mass-to-light ratios, therefore possible stellar population gradients in galaxies should introduce weaker bias in $K$ band stellar masses.  The comparison between the optical and $K$ band stellar masses is provided in Figure \ref{fig_ml_k_r}.

\begin{figure}
\includegraphics[width=0.45\textwidth]{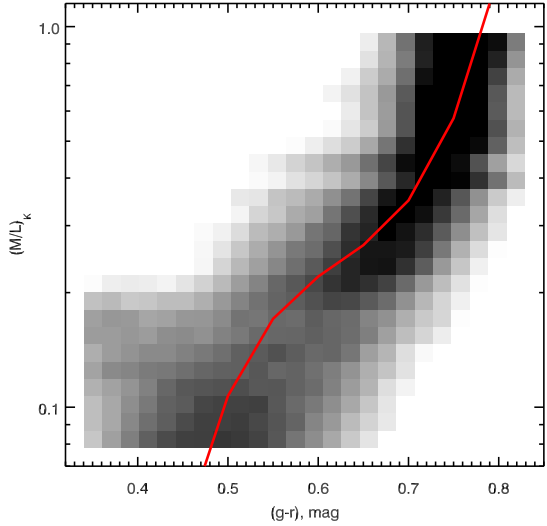}
\caption{An polynomial approximation of a $K$ band mass-to-light ratio as a function of the optical $(g-r)$ colour shown in red overplotted on a two-dimensional histogram presenting $K$ mass-to-light ratios estimated from the spectroscopic stellar population parameters for a sample of 424 163 galaxies. \label{fig_mlk_gr}}
\end{figure}

We also see a correlation between the derived $K$ band mass-to-light ratios and the integrated $g-r$ and $g-K$ colours (see Figure \ref{fig_mlk_gr}). We approximated the $K$ band mass-to-light ratio as a polynomial function of the integrated optical $g-r$ colour.  Because the distribution of galaxies in this parameter space is very non-uniform, we used the following procedure to derive the approximation:

\noindent (i) We restricted the sample of galaxies to objects at redshifts $0.01<z<0.15$ having Petrosian radii $R_{50} < 10$~arcsec, good quality of the full spectrum fitting ($\chi^2/DOF<0.8$), and SSP age uncertainties better than 30\%\ of the age values.

\noindent (ii) We selected 20 logarithmically spaced bins in $(M/L)_K$ between 0.07 and 1.15 in the Solar units, that is  $\sim0.061$ in $\log(M/L)_K$ and computed the median value of the $(g-r)$ colour in every bin.

\noindent (iii) We fit $\log(M/L)_K$ as a third order polynomial function of $(g-r)-0.50$ in the range $0.47<(g-r)<0.79$~mag and obtained the following approximation:

\begin{align}
&\log (M/L)_K = -0.97 + 5.39 (g-r)_{-0.5} \nonumber\\
&- 31.02 (g-r)_{-0.5}^2 + 84.46 (g-r)_{-0.5}^3
\end{align}

\noindent (iv) For $(g-r)<0.47$~mag we used the lower limit $(M/L)=0.07$, and for $(g-r)>0.79$~mag the upper limit $(M/L)=1.15$ which corresponds to the top of red sequence.  The uncertainty of the approximation is about 0.2~dex at $(g-r)<0.73$~mag and 0.3~dex at the red end, where we see no correlation between the colour and the $(M/L)_K$ value.  Therefore, we could estimate photometric stellar masses to a factor of 2 provided only broadband optical and $K$ band magnitudes.

7,361 galaxies out of 43,425 in the 2MRS sample matched with RCSED objects. For those galaxies we computed stellar masses using the same approach as for SDSS galaxies with available NIR magnitudes but using the 2MASS $K_{s}$ band photometry instead of UKIDSS.

For the remaining objects, we followed a different strategy.  We cross-matched the entire sample of galaxies with the SIMBAD\footnote{\url{http://simbad.u-strasbg.fr/simbad/}} database and extract integrated photometry in the $B$, $V$, and $R$ bands.  In total, we found $B-R$ colours for 11,210 galaxies.  We correct the magnitudes for the Galactic extinction, compute and apply K-corrections using the analytical approximations from \citep{Chilingarian:2010}, and convert them into the SDSS photometric system using the transformations from Lupton\footnote{\url{http://classic.sdss.org/dr7/algorithms/sdssUBVRITransform.html\# Lupton2005}} as: $(g-r) = 0.667 (B-Rc) - 0.216$.  Then we estimate the $K$-band stellar mass-to-light ratios using integrated $g-r$ and the calibration provided above and convert them into stellar masses.

In this fashion, we computed stellar masses for about 400,000 galaxies included into the 2MRS and SDSS DR12 catalogues.  We used them to provide an estimate (lower limit) on the stellar masses contained in the galaxy groups of our catalogues.

\subsubsection{Group mass}
\label{sec_mass}
\begin{table}
\begin{center}
\begin{tabular}{c|cc}
coefficients & 2MRS & SDSS \\ \hline \hline
$a_{1}$ & $-9.8 \pm 0.3$ & $-6.3 \pm 0.3$\\
$a_{2}$ & $0.89 \pm 0.03$ & $0.51 \pm 0.03$\\
$a_{3}$ &  $-0.0244 \pm 0.0007$ & $-0.0103 \pm 0.0010$\\
$a_{4}$ & $1.2 \pm 0.3$ & $-2.4 \pm 0.1$\\
$a_{5}$ & $-0.42 \pm 0.17$ & $1.36 \pm 0.06$\\
$a_{6}$ & $0.017 \pm 0.035$ & $-0.254 \pm 0.009$\\
$a_{7}$ & $44.8 \pm 1.2$ & $35.5 \pm 0.9$\\ \hline
$s_{\textrm{rms}}$ & $0.3401$ & $0.2606$
\end{tabular}
\end{center}
\caption{Coefficients of the mass dependence on observed parameters of isolated galaxies.  They were obtained by a least-square fit of mock catalogue data using Equation~\ref{fit_mass_single}.}
\label{fit_mass_single_coeff}
\end{table}
\begin{figure*}[ht]
\begin{center}
\includegraphics[width=0.90\textwidth]{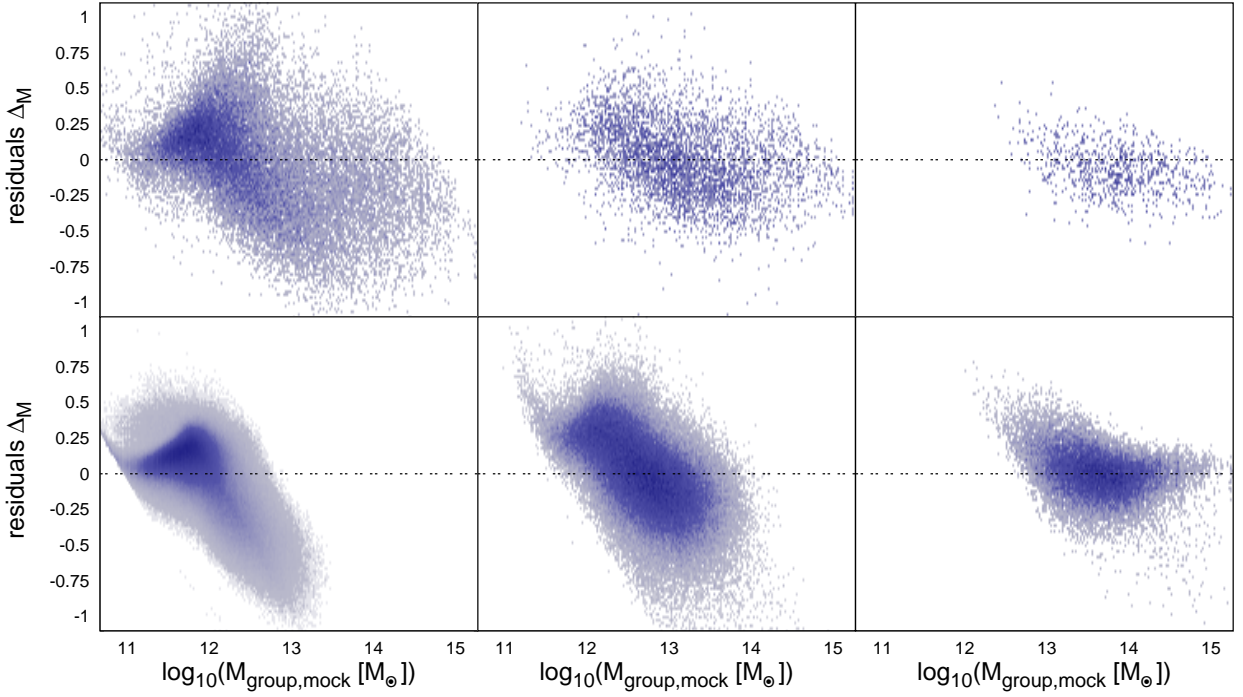}\\ 
\caption{Residuals of the fit for the mass determination of groups using 2MRS and SDSS mock data.  Top-left panel: residuals of our fit using isolated galaxies (groups with one visible member only) for the 2MRS data. Top-central panel: residuals of groups with two to four members using 2MRS data.  Top-right panel: residuals of groups with five or more members using 2MRS data.  Bottom-left panel: residuals of our fit using isolated galaxies (groups with one visible member only) for the SDSS data.  Bottom-central panel: residuals of groups with two to four members using SDSS data. Bottom-right panel: residuals of groups with five or more members using SDSS data.}
\label{all_residuals}
\end{center}
\end{figure*}
\begin{figure}[ht]
\begin{center}
\includegraphics[width=0.45\textwidth]{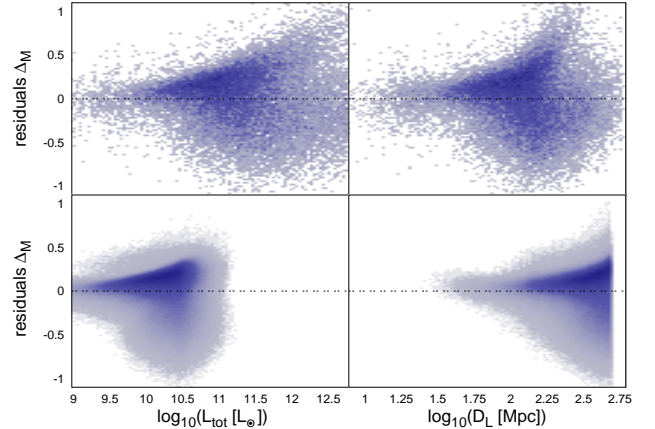}\\ 
\caption{Residuals of the fit for the mass determination of groups with only one visible member depending on the fitting parameter.  Top-left panel: residuals as a function of the total $K_{s}$ band group luminosity for the 2MRS data.  The top-right panel: residuals as a function of the luminosity distance for the 2MRS data.  Bottom-left panel: residuals as a function of the total $r$ band group luminosity for the SDSS data.  The Bottom-right panel: residuals as a function of the luminosity distance for the SDSS data.}
\label{res_single}
\end{center}
\end{figure}
In order to get robust mass estimates for the detected groups, we calibrated a set of mass functions depending on several parameters using our mock catalogues.  We split the sample into three sub-samples: isolated galaxies (groups with one visible member only) and all other groups with two to four members, and groups with more than four members.  In the case of the isolated galaxies, we had just two quantities at our disposal to derive their masses: luminosity and distance.  We fit the following function to these parameters:

\begin{align}
&\textrm{log}_{10}(M_{\textrm{group}})=\nonumber \\
&\sum\limits_{i=1}^{3} \left( a_{i} \left(\textrm{log}_{10}(L_{\textrm{tot}})\right)^{i} \right) 
&+ \sum\limits_{i=1}^{3} \left( a_{i+3} \left(\textrm{log}_{10}(D_{L})\right)^{i} \right) + a_{7}
\label{fit_mass_single}
\end{align}

The group mass $M_{\textrm{group}}$ depends on the group luminosity $L_{\textrm{tot}}$ and the luminosity distance $D_{L}$.  The coefficients $a_{1}$, $a_{2}$, $a_{3}$, $a_{4}$, $a_{5}$, $a_{6}$, and $a_{7}$ were estimated using a least-square fit to our mock catalogues.  The results are listed in Table \ref{fit_mass_single_coeff} and the residuals of the fitted group mass $\Delta_{M}=\textrm{log}_{10}(M_{\textrm{group,fit}})-\textrm{log}_{10}(M_{\textrm{group,mock}})$ depending on the true group mass of mock catalogue are illustrated in Figure \ref{all_residuals}.  Apparently, there is a trend in our fit to underestimate the true masses of groups with just one visible member in the case of high mass groups that is most prominent in the tail of the distribution for SDSS galaxies (see Figure \ref{all_residuals}). However, the dependence of the residuals on the fitted parameters (see Figure \ref{res_single}) does not exhibit any obvious trends, which indicates that considering higher order terms in the fit would not improve our mass function.  Not considering the luminosity distance would not visibly change any residuals (in Figures \ref{all_residuals} and \ref{res_single}), but it would increase the root mean square $s_{\textrm{rms}}$ of the fit by several percent.  Although for Malmquist-bias corrected data the distance should not have a significant impact on the fit, we chose to keep this parameter in, because otherwise it would unnecessarily increase the uncertainty in our mass estimates.  We also considered using the colour as an additional parameter, but we found that it did not have any notable impact on the quality of the fit and dropped it for reasons of simplicity and consistency with the other fits.  The apparent bi-modality in Figure \ref{res_single} is not a consequence of different galaxy populations, but the difference between truly isolated galaxies and groups with just one visible member.

\begin{table}
\begin{center}
\begin{tabular}{c|cc}
coefficients & 2MRS & SDSS \\ \hline \hline
$a_{1}$ & $-16.6 \pm 1.0$ & $-15.6 \pm 1.2$\\ 
$a_{2}$ & $1.43 \pm 0.09$ & $1.30 \pm  0.12$\\ 
$a_{3}$ & $-0.039 \pm 0.002$ & $-0.033 \pm  0.004$\\ 
$a_{4}$ & $-0.7 \pm 0.5$ & $-1.5 \pm 0.5$ \\ 
$a_{5}$ & $0.9 \pm 0.3$ & $0.6 \pm 0.2$\\ 
$a_{6}$ & $-0.27 \pm 0.06$ & $-0.09 \pm 0.04$\\ 
$a_{7}$ & $0.175 \pm 0.010$ & $0.203 \pm 0.002$\\ 
$a_{8}$ & $0.049 \pm 0.005$ & $0.160 \pm 0.002$\\ 
$a_{9}$ & $73.1 \pm 4.2$ & $71.4 \pm 4.1$\\ \hline
$s_{\textrm{rms}}$ & $0.2592$ & $0.2812$
\end{tabular}
\end{center}
\caption{Coefficients of the mass dependence on observed parameters of groups with two to four members.  They were obtained by a least-square fit on mock catalogue data using Equation~\ref{fit_mass_multi_low}.}
\label{fit_mass_multi_low_coeff}
\end{table}

\begin{figure*}[ht]
\begin{center}
\includegraphics[width=0.90\textwidth]{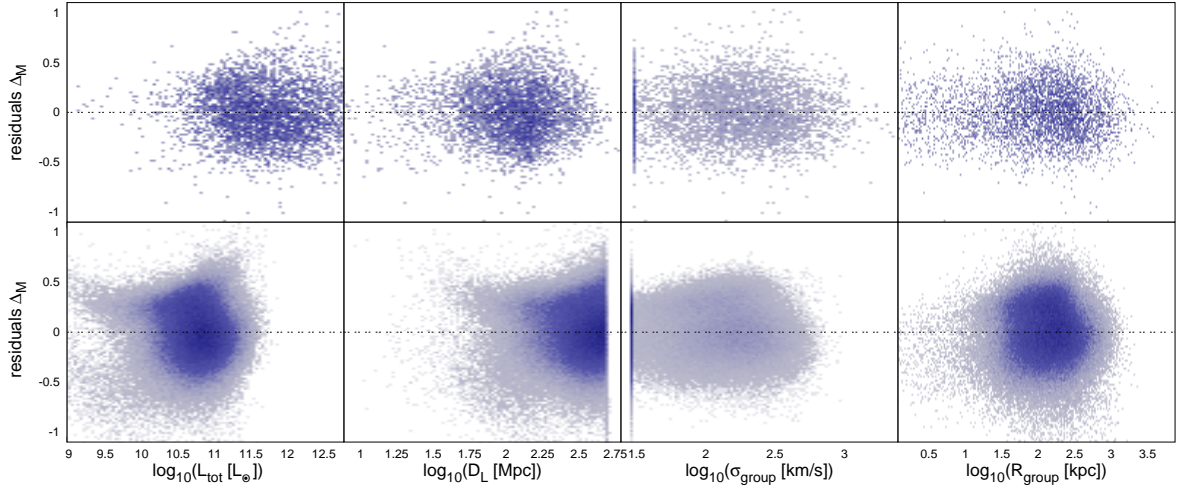}\\ 
\caption{Residuals of the fit for the mass determination of groups with two to four members depending on the fitting parameter.  Top row: residuals for 2MRS data.  Bottom row: residuals for SDSS data.  First column: residuals as a function of the luminosity ($K_{s}$ band for 2MRS, $r$ band for SDSS). Second column: residuals as a function of the luminosity distance.  Third column: residuals as a function of the group velocity dispersion.  Forth column: residuals as a function of the group radius.}
\label{res_multi_low}
\end{center}
\end{figure*}

For groups with two to four members, we included a dependence on the group velocity dispersion and group radius into the fitting function, which is defined as follows:

\begin{align}
&\textrm{log}_{10}(M_{\textrm{group}})= \label{fit_mass_multi_low}\\
&\sum\limits_{i=1}^{3} \left( a_{i} \left(\textrm{log}_{10}(L_{\textrm{tot}})\right)^{i} \right) + \nonumber  \\ 
&\sum\limits_{i=1}^{3} \left( a_{i+3} \left(\textrm{log}_{10}(D_{L})\right)^{i} \right) +  \nonumber \\ 
& a_{7} \textrm{log}_{10}(\sigma_{\textrm{group}})  + a_{8}  \textrm{log}_{10}(R_{\textrm{group}})   + a_{9} \nonumber.
\end{align}

The group velocity dispersion $\sigma_{\textrm{group}}$ and the group radius $R_{\textrm{group}}$ were used in addition to the parameters of the fit above in order to obtain the coefficients $a_{1}$ to $a_{9}$.  We also considered using the dynamical group mass $\sigma_{\textrm{group}}$, which is based on the previous two parameters.  However, we found that fitting both parameters separately instead of the dynamical mass improves the quality by up-to one percent.  The results of the fit are listed in Table \ref{fit_mass_multi_low_coeff}, whereas the residuals of these fits depending on the group mass are shown in Figure \ref{all_residuals}.  The residuals of the fits depending on the fit parameters, which are shown in Figure \ref{res_multi_low}, do not indicate any strong trends.

For the richest groups in our catalogue (with five or more members), we can define the following fitting function:

\begin{align}
&\textrm{log}_{10}(M_{\textrm{group}})= \label{fmmh}\\
&\sum\limits_{i=1}^{3} \left( a_{i} \left(\textrm{log}_{10}(L_{\textrm{tot}})\right)^{i} \right) + \nonumber  \\ 
&\sum\limits_{i=1}^{3} \left( a_{i+3} \left(\textrm{log}_{10}(D_{L})\right)^{i} \right) +  \nonumber \\ 
&a_{7} \textrm{log}_{10}(\sigma_{\textrm{group}})  + a_{8}  \textrm{log}_{10}(R_{\textrm{group}})  + a_{9} \textrm{log}_{10}(N_{\textrm{fof}}) + a_{10} \nonumber.
\end{align}

\begin{figure*}[ht]
\begin{center}
\includegraphics[width=0.90\textwidth]{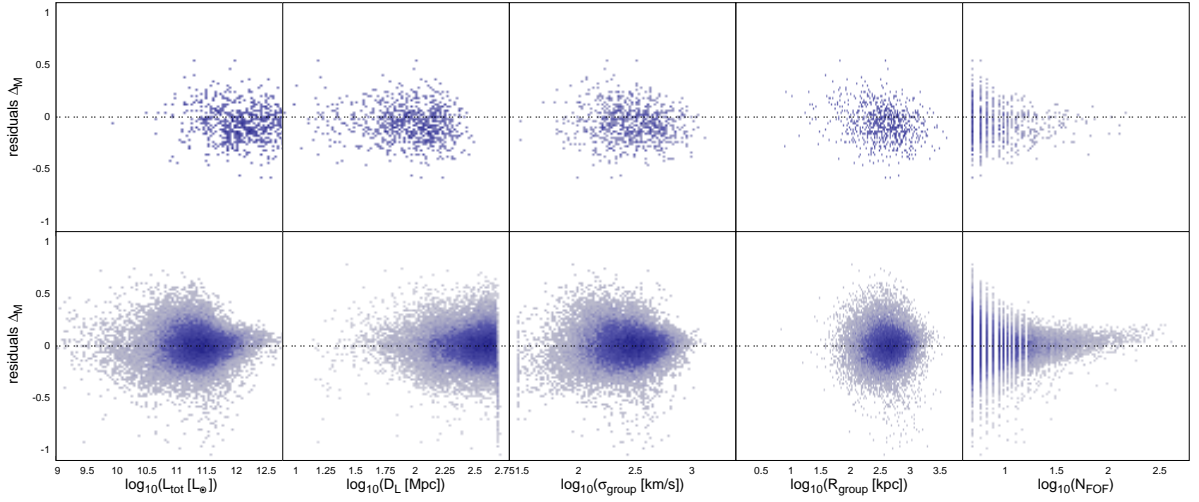}\\ 
\caption{Residuals of the fit for the mass determination of groups with five or more members depending on the fitting parameter.  Top row: residuals for 2MRS data.  Bottom row: residuals for SDSS data.  First column: residuals as a function of the luminosity ($K_{s}$ band for 2MRS, $r$ band for SDSS).  Second column: residuals as a function of the luminosity distance.  Third column: residuals as a function of the group velocity dispersion.  Forth column: residuals as a function of the group radius.  Fifth column: residuals as a function of the number of galaxies detected inside a group.}
\label{res_multi_high}
\end{center}
\end{figure*}

The number of detected galaxies within a group $N_{\textrm{fof}}$ was used in addition to the parameters of the previous fit to obtain the coefficients $a_{1}$ to $a_{10}$.  The results of the fit are listed in Table \ref{fit_mass_multi_high_coeff}, whereas the residuals of these fits depending on the group mass are shown in Figure \ref{all_residuals}.  The residuals of the fits depending on the fit parameters, which are shown in Figure \ref{res_multi_high}, do not indicate any strong trends.

\begin{table}
\begin{center}
\begin{tabular}{c|cc}
coefficients & 2MRS & SDSS \\ \hline \hline
$a_{1}$ & $-19.5 \pm 4.2$ & $-15.1 \pm 1.2$\\ 
$a_{2}$ & $1.6 \pm 0.3$ & $1.3 \pm 0.1$\\ 
$a_{3}$ & $-0.043 \pm 0.009$ & $-0.039 \pm  0.003$\\ 
$a_{4}$ & $2.8 \pm 0.7$ & $5.5 \pm 0.5$ \\ 
$a_{5}$ & $-1.8 \pm 0.4$ & $-3.2 \pm 0.2$\\ 
$a_{6}$ & $0.43 \pm 0.09$ & $0.60 \pm 0.04$\\ 
$a_{7}$ & $0.371 \pm  0.027$ & $0.382 \pm 0.006$\\ 
$a_{8}$ & $0.090 \pm 0.015$ & $0.108 \pm 0.005$\\ 
$a_{8}$ & $0.604 \pm 0.040$ & $0.626 \pm 0.009$\\ 
$a_{9}$ & $87.5 \pm 16.9$ & $63.7 \pm 4.5$\\ \hline
$s_{\textrm{rms}}$ & $0.1404$ & $0.1603$
\end{tabular}
\end{center}
\caption{Coefficients of the mass dependence on observed parameters of groups with more than four members.  They were obtained by a least-square fit on mock catalogue data using Equation \ref{fmmh}.}
\label{fit_mass_multi_high_coeff}
\end{table}

The majority of our groups have masses below $10^{14} \textrm{M}_{\astrosun}$ (see Figure \ref{all_residuals}).  At such low masses, the uncertainty in the mass determination is notoriously high \citep{Old:2015}.  Therefore, our overall uncertainties for the fits are higher than those of the mass determination methods presented in \citet{Old:2015}, because they primarily focused on groups more massive than $10^{14} \textrm{M}_{\astrosun}$.

\subsection{Calculating the finite infinity regions}
\label{sec_fi_rescale}
In order to derive a catalogue of finite infinity regions \citep{Ellis:1984}, which will be used in our upcoming paper (Saulder et al., submitted) to perform a cosmological test that was outlined in \citet{Saulder:2012} comparing timescape cosmology \citep{Wiltshire:2007} and $\Lambda$-CDM cosmology, we used a simple approximation for the finite infinity regions. They were calculated using spherical regions with an average density that corresponds to the critical density of the timescape cosmology. More complex geometrical shapes would be possible, however with our available computional power as well as the general uncertainties of the mass estimates in mind, we found that this model practical and sufficient for our planned test (Saulder et. al, submitted). In the timescape model of \citet{Wiltshire:2007}, the Universe is approximated by a two-phase model of empty voids and walls with an average density of the true critical density.  This renormalized critical density is about 61$\%$ of the critical density in the $\Lambda$-CDM model according to the best fits for this alternative cosmology (the Hubble parameter\footnote{We are aware that the absolute values of the Hubble parameter are relatively low compared to values in standard cosmology.  We also point out that we only use relative values for our test.} for the wall environment (inside finite infinity regions) is expected to be 48.2 km $\textrm{s}^{-1}$ $\textrm{Mpc}^{-1}$ and void environment (outside finite infinity regions) to be 61.7 km $\textrm{s}^{-1}$ $ \textrm{Mpc}^{-1}$ \citep{Wiltshire:2007,Leith:2008} according to the best fit on supernovae Type Ia \citep{Riess:2007}, CMB \citep{WMAP_pre,WMAP3} and Baryonic acoustic oscillations \citep{Cole:2005,Eisenstein:2005} data within the framework of the simple two phase model presented in \citet{Wiltshire:2007}).  Because our preliminary results \citep{Saulder:2012} had already shown that this basic approximation required all our available computational resources, more complex models for the finite infinity regions are not possible to consider at this time.

In practice, calculating finite infinity regions requires a solid knowledge of the matter distribution.  In Table \ref{listnumbers}, we found that less than half of the simulation's particles are bound in FoF-groups of 20 or more particles.  Because the timescape model that we want to test is a two-phase model and distribution of the matter is very important for it, we used the MM data to locate the missing particles.  We calculated the radii $R_{f}$ of homogeneous spheres around the FoF-groups in MM in the following way:

\begin{equation}
R_{f} = \left( \frac{3 M_{\textrm{FoF}}}{4 \pi \rho_{\textrm{crit}} f} \right)^{1/3}.
\label{radcrit}
\end{equation} 

$M_{\textrm{FoF}}$ stands for the mass bound in a FoF-group according to the Millennium simulation.  $\rho_{\textrm{crit}}$ is the critical density and $f$ is a factor to modify the critical density.  It was set to 0.61 for timescape cosmology.

We found that about 77 percent of all particles are within $R_{0.61}$ around the FoF-groups.  Expanding the spherical regions iteratively using the masses of the particles, we found that about 82 percent of all particles are located within finite infinity regions, which occupy about 23 percent of the simulations volume, around the FoF-groups.  The remaining particles can be assumed to be either uniformly distributed all across the voids or arranged in tendrils (fine filaments in voids using the terminology of \citet{Alpaslan:2014}) of small halos (less than 20 particles) far outside main clusters and groups.

\begin{table*}[ht]
\begin{center}
\begin{tabular}{c|ccccc|ccc}
rescaling & redshift & $f_{0}$ & $f_{1}$ & $f_{2}$ & $s_{\textrm{rms}}$ & mass within & volume within & groups \\ 
 &  &  &  &  &  & fi-regions & fi-regions & detected/used \\
\hline \hline
first & 0 & $0.008 \pm 0.002$ & $0.77 \pm 0.06$ & $1.7 \pm 0.3$ & $0.1247$ & 74.68$\%$ & 13.27$\%$ & 15413 \\ 
final & 0 & $0.028 \pm 0.004$ & $0.26 \pm 0.11$ & $5.1 \pm 0.6$ & $0.2187$ & 80.28$\%$ & 22.76$\%$ & 9187 \\
first & 0.020 & $0.005 \pm 0.003$ & $0.85 \pm 0.08$ & $1.3 \pm 0.5$ & $0.1230$ & 72.16$\%$ & 12.80$\%$ & 8952 \\ 
final & 0.020 & $0.031 \pm 0.006$ & $0.17 \pm 0.15$ & $5.7 \pm 0.9$ & $0.2188$ & 78.24$\%$ & 22.27$\%$ & 5676 \\
first & 0.041 & $0.007 \pm 0.005$ & $0.80 \pm 0.12$ & $1.7 \pm 0.7$ & $0.1198$ & 69.15$\%$ & 12.10$\%$ & 4993 \\ 
final & 0.041 & $0.030 \pm 0.009$ & $0.16 \pm 0.22$ & $5.9 \pm 1.3$ & $0.2197$ & 75.61$\%$ & 21.35$\%$ & 3415 \\
first & 0.064 & $0.001 \pm 0.006$ & $0.95 \pm 0.15$ & $0.8 \pm 0.9$ & $0.1158$ & 66.35$\%$ & 11.59$\%$ & 3165 \\ 
final & 0.064 & $0.026 \pm 0.012$ & $0.24 \pm 0.30$ & $5.6 \pm 1.8$ & $0.2210$ & 73.20$\%$ & 20.63$\%$ & 2336 \\
first & 0.089 & $0.003 \pm 0.008$ & $0.89 \pm 0.21$ & $1.2 \pm 1.3$ & $0.1165$ & 62.45$\%$ & 10.82$\%$ & 1941 \\ 
final & 0.089 & $0.031 \pm 0.016$ & $0.10 \pm 0.41$ & $6.7 \pm 2.6$ & $0.2237$ & 69.93$\%$ & 19.85$\%$ & 1530 \\
first & 0.116 & $0.001 \pm 0.012$ & $0.93 \pm 0.31$ & $1.0 \pm 1.9$ & $0.1184$ & 55.60$\%$ & 9.44$\%$ & 1002 \\ 
final & 0.116 & $0.043 \pm 0.024$ & $-0.24 \pm 0.61$ & $9.1 \pm 3.9$ & $0.2313$ & 64.75$\%$ & 18.75$\%$ & 826 
\end{tabular}
\end{center}
\caption{Coefficients of the mass rescaling for the finite infinity regions and the mass as well as the volume covered by them.  Column one: indicator if first of final rescaling of FI regions, column two: cosmological redshift of that snapshot, column three to five: coefficients of the fit (see Equation \ref{fi_rescale}),  column six: root mean square of that fit, column seven: percentage of mass within FI-regions compared to the total mass in the simulation, column eight: percentage of volume within FI-regions compared to the total volume in the simulation, column nine: number FI regions/groups used for the fit and remaining after iteration.}
\label{fit_fi_rescale}
\end{table*}

We developed a method to derive the finite infinity regions from observational data.  Assuming that the last snapshots of the Millennium simulation provide a reasonable model of the present-day large-scale matter distribution of the universe, we took the six snapshots of the redshift range of our project from the MM and introduced magnitude limits into them corresponding to the Malmquist bias of SDSS.  Because the observational data, which we used in Section \ref{merge_cat} was a combined catalogue of SDSS and 2MRS groups, it compensated for the selection effect due to the SDSS saturation limit.  Then we assigned radii using Equation \ref{radcrit} to every group that has at least one visible/detectable member.  We used the total mass of those groups to derive the sizes of these ``proto''-finite infinity regions.  Afterwards we merged all groups that were fully within the radii of other groups by adding their masses to their host groups and shift their centre of mass accordingly.  In the next step, we counted the particles inside the recalculated radii of the remaining groups.  To correctly handle the overlapping regions, particles that are located within more than one finite infinity region were assigned weights corresponding to the reciprocal values of the number of finite infinity region they were shared with.  Then we adjusted the masses of groups according to the total (weighted) mass of particles inside them.  This step was used to calibrate the first (of the two) re-scaling of the masses.

\begin{equation}
\textrm{log}_{10}\left(M_{\textrm{fi}}\right)=f_{0} \left(\textrm{log}_{10}\left(\sum\limits_{i=1}^{n} M_{\textrm{group},i}\right)\right)^{2}+f_{1} \textrm{log}_{10}\left(\sum\limits_{i=1}^{n} M_{\textrm{group},i}\right)+f_{2}
\label{fi_rescale}
\end{equation}

We performed a least square fit to calibrate the mass of the finite infinity region $M_{\textrm{fi}}$ derived from the particles within it at this step depending on the sum of (original) group masses it is composed of $M_{\textrm{group}}$.  The values of coefficients $f_{0}$, $f_{1}$, and $f_{2}$ are listed in Table \ref{fit_fi_rescale}.  We performed this fit for each of the six MM snapshots.  The distribution of the parameters and our fit on them is illustrated in Figure \ref{map_fi_first}.  In the next step, we iteratively merged and expanded (using masses of the particles within the group's finite infinity radii) the groups in the same way as before until the change of total mass within all finite infinity regions was less than $0.1\%$\footnote{Below this value, they algoritm converges very slowly and the total change in mass becomes insignificant.} from one step to the next.  Once this was reached, we used the final masses (derived from the particles) of the finite infinity regions to calibrate another re-scaling relation in the same way as before.  We fitted the final masses of the finite infinity regions to the sum of the inital masses of the FoF-groups using Equation \ref{fi_rescale}.  The results of the fit for the final rescaling are listed in Table \ref{fit_fi_rescale} and illustrated in Figure \ref{map_fi_final}.

After the calibrations of the rescaling were completed, we applied our method to real data, a combined catalogue of SDSS and 2MRS, in the following way: The masses of groups were obtained using the calibrations described in Section~\ref{sec_mass}.  Afterwards we calculated the radii of the proto-finite infinity regions using Equation \ref{radcrit}.  In the next step, we iteratively merged all groups that were fully within the radii of other proto-finite infinity regions and shifted the centre of mass of the remaining groups accordingly.  Then we applied the first rescaling on the data using the coefficients from Table \ref{fit_fi_rescale}, which belong to the redshift closest to the actual redshift of each group.  Afterwards we recalculated the radii using the new masses and repeated the merging procedure from before.  Then we applied the final rescaling using the sum of the initial masses of the groups that merged as a basis in the same way as we did for the first rescaling, but with the coefficients of the final rescaling (see Table \ref{fit_fi_rescale}).  We used the initial masses for the final rescaling instead of the masses obtained after the first rescaling, because the overall uncertainty was slightly lower this way.  The reason to do the first rescaling at all, and not to proceed directly to the final rescaling without it, is that the merging process, which is done after it, slightly shifts the distribution of the mass function and we would introduce an unnecessary source of error by skipping it.

As illustrated in Table \ref{fit_fi_rescale}, the total mass within finite infinity regions decreases at higher redshifts (from $\sim 80\%$ at $z=0$ to $\sim 65\%$ at $z=0.116$), but not as strikingly as does the number of detectable groups.  This means, that although we miss some of the smaller structures, the biggest contributors to the mass are still detected at higher redshifts.  Furthermore, many of the smaller masses are close to the bigger ones, hence we still obtained a good representation of the finite infinity regions.  Even after the final rescaling there are about $20-30\%$ of the mass missing (over the redshift range of our projects) and not within the defined finite infinity regions.  Because the model \citep{Wiltshire:2007}, which we intend to test with this data, is only a two-phase model consisting of completely empty voids and finite infinity regions (walls with an average density of the renormalized critical density), we had two options on how to proceed: (i) we could either add the $20-30\%$ missing mass to the detected finite infinity regions and adjust their sizes accordingly, or we could (ii) assume that the $20-30\%$ missing mass is distributed homogeneously throughout the rest of the simulations volume (the voids) and define them as not completely empty.  In this paper we provided data for both options.  In our follow-up paper (Saulder et al., submitted), we will test them and we will see, which one is the best-suited.

\section{Results}
We provide four group catalogues, which are made available on VizieR: \url{http://cdsarc.u-strasbg.fr/viz-bin/qcat?J/A+A/}.

\subsection{The 2MRS group catalogue}
\begin{figure}[ht]
\begin{center}
\includegraphics[width=0.45\textwidth]{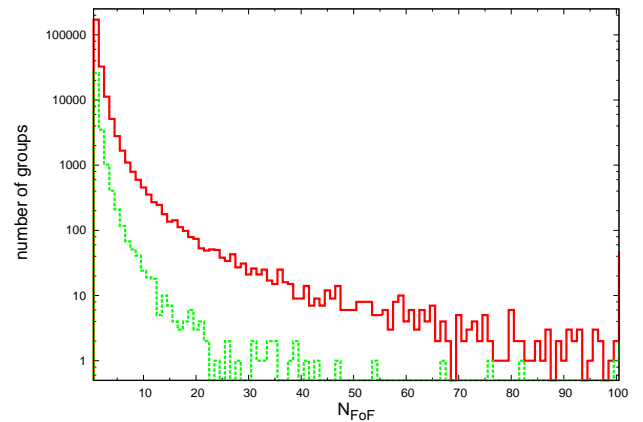}\\ 
\caption{Distribution of the multiplicity $N_{FoF}$ of the groups detected in 2MRS (green dotted line) and in SDSS (red solid line).}
\label{bin_clusters_combo}
\end{center}
\end{figure}

Our 2MRS based group catalogue is composed of 43,425 galaxies from \citet{2MRS} covering $91\%$ of the sky.  Using our group finder with the optimal coefficients from Table \ref{optimal_coefficients}, we detected  31,506 groups in the 2MRS data.  As illustrated in Figure \ref{bin_clusters_combo}, the majority (25,944 to be precise) of the galaxies can be found in groups with only one visible member.  This does not necessarily mean that all of them are isolated objects, but that there is only one galaxy sufficiently bright to be included in the 2MRS.  We identify 5,452 groups within the multiplicity range from two to ten and only 110 with higher multiplicities (two of them with more than 100 members each).  Figure \ref{bin_clusters_combo} clearly shows that the number of groups rapidly decreases with increasing number of multiplicity.

\begin{figure*}[ht]
\begin{center}
\includegraphics[width=0.90\textwidth]{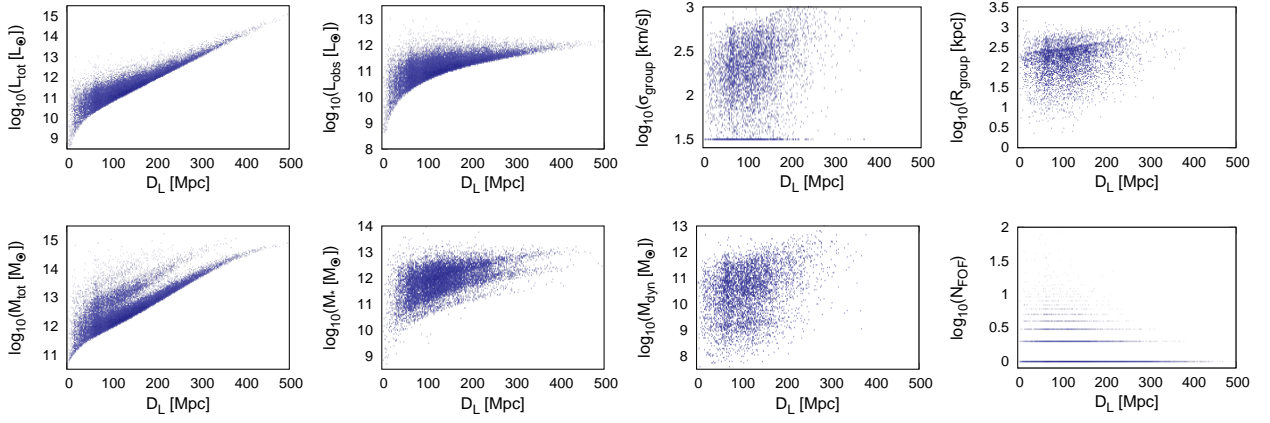}\\ 
\caption{Dependence of various group parameters of the 2MRS catalogue on the luminosity distance. First panel upper row: dependence on the total group luminosity $L_{\textrm{tot}}$; second panel upper row: dependence on the observed group luminosity $L_{\textrm{obs}}$; third panel upper row: dependence on the group velocity dispersion $\sigma_{\textrm{group}}$; fourth panel upper row: dependence on the group radius $R_{\textrm{group}}$; first panel lower row: dependence on the total group mass $M_{\textrm{group}}$; second panel lower row: dependence on the stellar mass $M_{*}$; third panel lower row: dependence on the dynamical mass $M_{\textrm{dyn}}$; fourth panel lower row: dependence on the number of detected galaxies in the group $N_{\textrm{FoF}}$.}
\label{resdist_2MRS}
\end{center}
\end{figure*}

As illustrated in Figure \ref{resdist_2MRS}, the group parameters show different dependences on the luminosity distance.  The group luminosity is strongly affected by the Malmquist bias, because fainter groups fall below the detection threshold at larger distances and only the brightest groups are detected. A comparison to the observed luminosity indicates that our corrections might slightly overestimate the total group luminosity at greater distances. The total group mass shows a very similar distribution as does the stellar mass.  However, the dynamical mass, as well as the group velocity dispersion and the group radius, show much weaker dependences on the distance.  There is an accumulation at the minimum value of the group velocity dispersion, which is due to the way it was calculated (see Section \ref{sec_veldisp}).  The distribution of group radii shows a step, which corresponds to half of the effective angular linking length.  In the case of groups with two members, this is the maximum value of the group radius for which the group is still recognised as bound by the FoF-algorithm.  Richer groups containing larger numbers of detected members, become rarer at larger distances, because the fainter group members are not detected in 2MRS any more.
         
Appendix \ref{cat_descrip} contains a detailed description of the catalogue.  We identified some of the richest clusters \footnote{We use the NASA/IPAC Extragalactic Database (\url{http://ned.ipac.caltech.edu/}) for a manual search by coordinates to identify this and all other other groups in this section.}.  A list is provided Appendix \ref{cat_descrip}. The fact, that we were able to locate many well known clusters in our catalogue suggests that our group finding algorithm worked as expected.

\subsection{The SDSS DR12 group catalogue}

Our SDSS based group catalogue is composed of 402,588 galaxies from \citet{SDSS_DR12} covering 9,274 square degrees on the sky.  Using our group finder with the optimal coefficients from Table \ref{optimal_coefficients}, we detected 229,893 groups in the SDSS DR12 data up to a redshift of 0.11.  As illustrated in Figure \ref{bin_clusters_combo}, a large fraction (170,983 to be precise) of the galaxies are found in groups with only one visible member.  Similar to the results of the 2MRS catalogue, the number of groups rapidly decreases with increasing number of multiplicity, but since the SDSS sample is much deeper, there are more groups with higher multiplicity than for the 2MRS sample.  We detected 48 groups with more than 100 visible members each.

\begin{figure*}[ht]
\begin{center}
\includegraphics[width=0.90\textwidth]{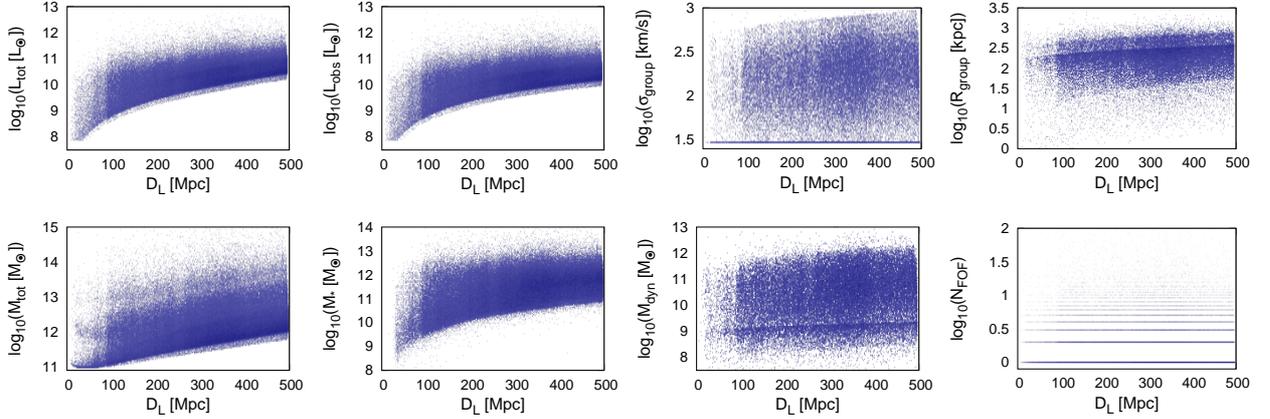}\\ 
\caption{Dependence of various group parameters of the SDSS catalogue on the luminosity distance. Panels: as in Figure \ref{resdist_2MRS}.}
\label{resdist_SDSS}
\end{center}
\end{figure*}

As illustrated in Figure \ref{resdist_SDSS}, all group parameters show dependences on the luminosity distance to a varying degree.  Similar to the 2MRS catalogue, the SDSS group catalogue is also affected by the Malmquist bias, which removes fainter groups from the sample at large distances.  However, SDSS is significantly deeper than 2MRS and there are no indications of any over-corrections within the limits of the catalogue.  Therefore, the effect of the bias on the dependence of the total group luminosity and the total group mass is not as striking as for 2MRS, but it is still clearly visible in the plots. Despite the impact of the Malmquist bias, the distribution of the stellar mass is nearly constant over the redshift range of our catalogue. The dependence of the dynamical mass, the group velocity dispersion, and the group radius is barely noticeable.  Again there is an accumulations at the minimum value of the group velocity dispersion and a step in the distribution of the group radii corresponding to half the effective linking length for the same reasons as for the other group catalogue.  The groups with largest numbers of detected members are found at intermediate distances, where the saturation bias effect becomes insignificant, while the Malmquest bias is not dominant yet.

A full description of the catalogue data and a list of prominent clusters can be found in Appendix \ref{cat_descrip}.  SDSS covers a much smaller area of the sky than 2MRS.  Furthermore, due to the saturation limits of SDSS spectroscopy, there is a dearth of galaxies in the SDSS survey at very low redshifts.  Hence, we did not detect as many rich nearby groups as in the 2MRS.  However, we were able to identify some of the same groups and clusters in both surveys independently.

\subsection{The fundamental plane distance group catalogue}
\begin{figure}[ht]
\begin{center}
\includegraphics[width=0.45\textwidth]{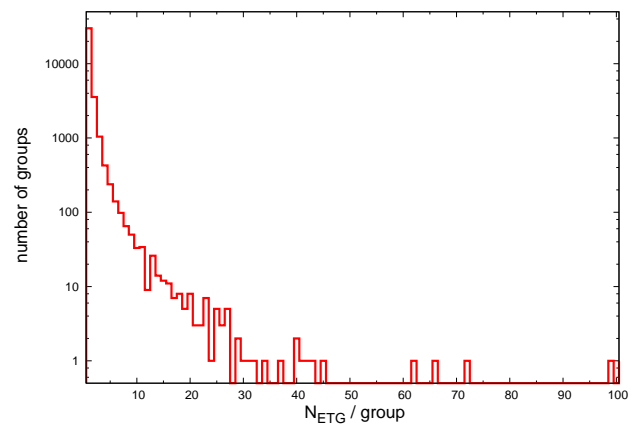}\\ 
\caption{Distribution of the number of early-type galaxies found in our SDSS groups.}
\label{bin_earlytypes_percluster}
\end{center}
\end{figure}

We took advantage of our previous work \citep{Saulder:2013} on the fundamental plane of elliptical galaxies to provide additional information for a subset of groups of our SDSS catalogue.  We provide redshift independent fundamental plane distances for all groups that contain at least one early-type galaxy based on our extended and up-dated fundamental plane calibrations in the Appendix of our recent paper \citep{Saulder:2015}.  We found 49,404 early-type galaxies distributed over 35,849 groups of our SDSS group catalogue.  These groups themselves host over 145,000 galaxies of various morphological types.  As illustrated in Figure \ref{bin_earlytypes_percluster}, the majority (30,017 to be exact) of the early-type galaxies are the only detected early-type galaxy in their group. We also found 5,832 groups hosting two or more early-type galaxies and 803 of these groups even contain five or more early-type galaxies.

\begin{figure}[ht]
\begin{center}
\includegraphics[width=0.45\textwidth]{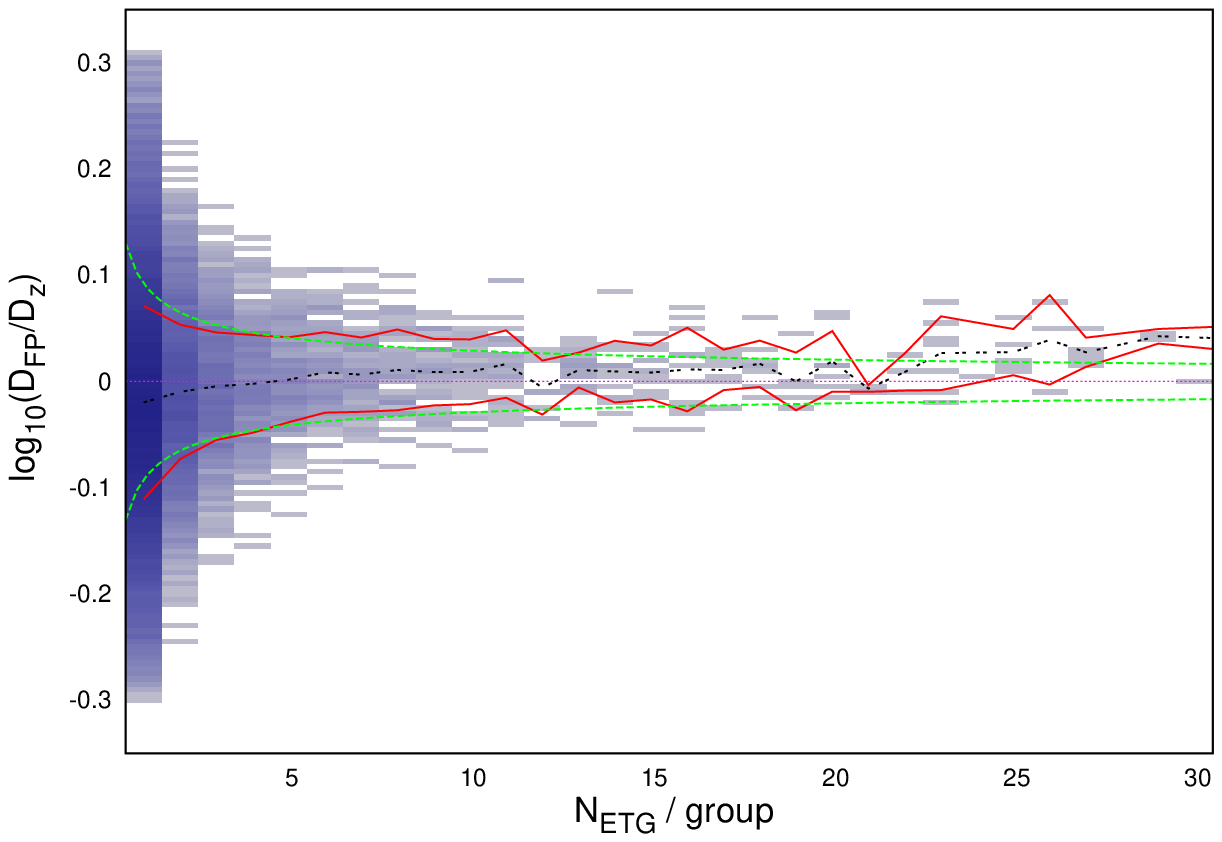}\\ 
\caption{The difference in the distance measurements for our clusters by comparing the fundamental plane distances with the redshift distances depending on the number of early-type galaxies $N_{\textrm{ETG}}$ per group.  Black dashed line: average ratio per early-type galaxies multiplicity bin.  Red solid line: 1-$\sigma$ intervals.  Green dashed line: expected progression of the 1-$\sigma$ intervals around one based on the root mean square of our fundamental plane distance error of 0.0920.  Magenta dotted line: reference for where both distance measurements yield exactly the same values.}
\label{fp_distance_err}
\end{center}
\end{figure}

In Figure \ref{fp_distance_err}, we show that the difference between the co-moving fundamental plane distance and the co-moving redshift distance decreases with the increasing number of elliptical galaxies per group.  For higher multiplicities, the statistics is affected by the small number of groups hosting so many detected elliptical galaxies.  We compared this to the expected decrease based on the root mean square (of 0.0920 in the $z$ band) of our fundamental plane calibration in our recent paper \citep{Saulder:2015} and found that the measured decrease is comparable to the expected one, although there is a trend for the mean value to rise with higher early-type multiplicity per cluster.

A detailed description of the data included in the catalogue is provided in Appendix \ref{cat_descrip}.

\subsection{The finite infinity regions catalogues - merging 2MRS and SDSS}
\label{merge_cat}
While the three previous catalogues were kept relatively general allowing for a wide range of applications, the catalogue of finite infinity regions has been exclusively created as a preparatory work for our next paper (Saulder et al., submitted).

\begin{figure}[ht]
\begin{center}
\includegraphics[width=0.45\textwidth]{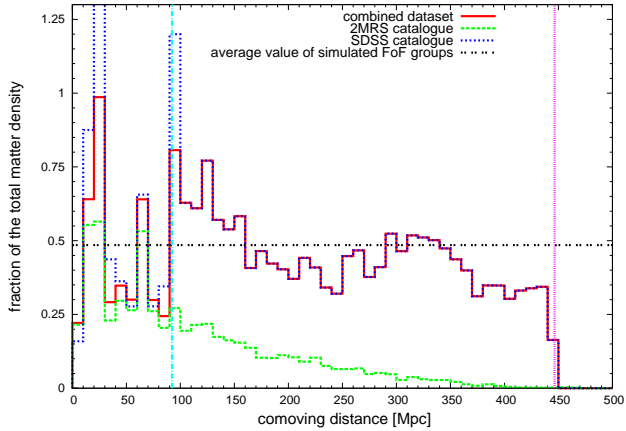}\\ 
\caption{Distribution of the matter density in dependence of the distance for our catalogues.  Dotted black line: expected average value based on the mock catalogue used to calibrate the group masses.  Dotted blue line: density distribution of our SDSS catalogue.  Dashed green line: density distribution of our 2MRS catalogue.  Solid red line: density distribution of our combined dataset, which is a mixture of the SDSS and 2MRS cataloguesbelow the distance at which the SDSS saturation limit becomes negligible.  Dashed-dotted cyan line: only the SDSS catalogue is used above this distance.  Dotted magenta line: maximum depth of catalogue ($z$=0.11).}
\label{density_function}
\end{center}
\end{figure}

\begin{figure}[ht]
\begin{center}
\includegraphics[width=0.45\textwidth]{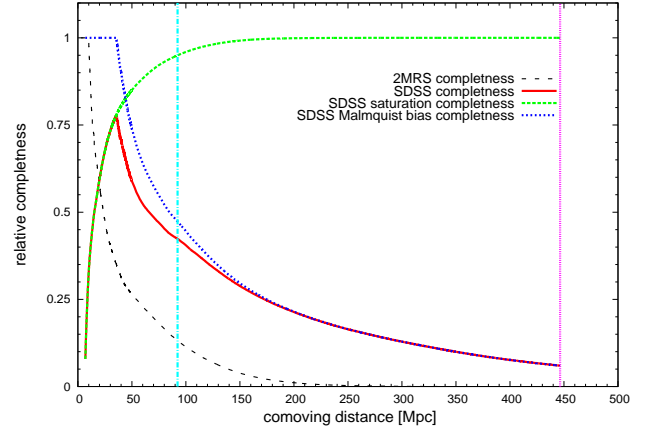}\\ 
\caption{Completeness of our catalogues based on the luminosity function of their galaxies depending on the co-moving distance.  Dashed black line: completeness function of the 2MRS catalogue, which is only affected by the Malmquist bias.  Solid red line: completeness function of SDSS, which is affected by the Malmquist bias and a saturation limit.  Dotted blue line: Malmquist bias for SDSS.  Dashed green line: saturation bias for SDSS.  Dashed-dotted cyan line denotes the point, when the impact of saturation limit on the completeness function of SDSS becomes negligible (the saturation limit completeness rises about $95\%$).  Dotted magenta line: maximum depth of catalogue ($z$=0.11).}
\label{completness_function}
\end{center}
\end{figure}

The main reason for using the 2MRS catalogue in addition the the SDSS sample is the fact that the SDSS suffers from incompleteness at very low redshifts due to the saturation of their spectroscopic data.  The 2MRS catalogue has no saturation limit and allows us to fill in at least some gaps.  The merging of the two catalogues is a delicate procedure, which required some deliberations beforehand.  When plotting the matter density as a function of distance (see Figure \ref{density_function}), the density of the 2MRS catalogue fluctuates (due to large local structures) around the average value expected from the mock catalogues in the inner mass shells, but then drops drastically in the outer shells.  The density of SDSS catalogue tends to be higher than the density of the 2MRS catalogue in the innermost shells and also exhibits much bigger fluctuations in these shell due to the smaller area of the sky covered by it.

We used a combination of the 2MRS and SDSS catalogue up to a certain distance and beyond that the SDSS catalogue only.  We defined this limit as the distance at which the effect of saturation limit on SDSS becomes negligible. We placed it at a co-moving distance of 92.3 Mpc, which corresponds to where $95\%$ of the luminosity function (down to an absolute magnitude of -15~mag in the $r$ band, which was our limit on SDSS data) is unaffected by the saturation limit as illustrated in Figure \ref{completness_function}. This is also where the mass density of 2MRS starts dropping significantly below the expected value (see Figure \ref{density_function}).  The 2MRS catalogue is already strongly effected by the Malmquist bias at this distance.  As illustrated in Figure \ref{completness_function} only the brightest $\sim 15\%$ of the luminosity function (down to an absolute magnitude of -18 mag in the $K_{s}$ band, which was our limit on 2MRS data) are still visible.

The first step in merging the 2MRS and SDSS catalogue was to remove all 22,190 2MRS groups beyond a co-moving distance of 92.3 Mpc.  From this distance outward, only SDSS data were used.  In the overlapping area we can encounter three cases, for which the third case required careful assessment: 

\noindent (i) a group is detected only in SDSS, because its galaxies are too faint for 2MRS; 

\noindent (ii) The group is only detected in 2MRS because they are too near and too bright for SDSS or simply because they are outside the area covered by SDSS -- in both those cases the group is fully included in the new merged catalogue; 

\noindent (iii) The same group (or parts of it) is detected in both catalogues.  In this case, both detections need to be merged in a meaningful manner.

For this group merging, we took our truncated 2MRS catalogue and verified how many SDSS groups were within one linking length of our 2MRS groups.  We use the definition of the linking length of Equations \ref{alpha_final} and \ref{R_final} with the optimised parameters for 2MRS from Table~\ref{optimal_coefficients} but with $\lambda_{\textrm{opt}}$ set to zero, because the corresponding term was calibrated using the galaxy luminosities and not the group luminosities and it was a minor correction at that distance anyway.  If one of the linking lengths $\alpha_{\textrm{eff}}$ or $R_{\textrm{eff}}$ was smaller than the group angular radius or the group velocity dispersion respectively, it was scaled up accordingly.  We found that 1,990 of the 10,085 SDSS groups in the overlapping volume had to be merged with 1,654 2MRS groups.  There were obviously several cases in which we found more than one SDSS group within a 2MRS group.  The parameters of the newly merged groups were the weighted averages of the parameters of their predecessors.  One part of the weights was the completeness function:

\begin{equation}
f_{c}(z) = \frac{\int_{-5 \textrm{log}_{10}\left(D_{L}(z)\right) + m_{\textrm{sat}} + 5}^{-5 \textrm{log}_{10}\left(D_{L}(z)\right) + m_{\textrm{limit}} + 5} \Phi \left(m \right) dm}{\int_{-\infty}^{M_{\textrm{abs,min}}} \Phi \left(m \right) dm}. 
\label{completeness_int}
\end{equation}

It depends on the luminosity distance $D_{L}$ derived from the redshift $z$, the saturation limit $m_{\textrm{sat}}$, which is 14~mag in the SDSS $r$ band and none for 2MRS, the limiting magnitude $m_{\textrm{limit}}$ of the survey, which is 17.77~mag in the $r$ band for SDSS and 11.75~mag in the $K_{s}$ band for 2MRS.  We used the corrected observed luminosity function $\Phi \left(m \right)$ as illustrated in Figure \ref{lum_func}.  The 2MRS and SDSS completeness functions are plotted in Figure \ref{completness_function}.  The other part of the weights $w_{\textrm{merge}} = f_{c} \cdot N_{FoF}$ used for merging the groups is the number of members per group $N_{FoF}$.  This ensured that the masses of galaxy clusters were not biased by a single galaxy from the other survey. In the merging process the centre of mass of the new groups was adjusted according to the weights as well.  Our combined catalogue consists of 237,219 groups.  They contain in total $39.9\%$ of mass expected for the  volume using the Millennium simulation cosmology, which is slightly below the 43.4$\%$ of matter in visible groups in the latest snapshot of the \emph{millimil} simulation.

We applied the method described in Section~\ref{sec_fi_rescale} to our combined catalogue and obtained the finite infinity regions.  We assigned radii for the first estimate of the finite infinity regions based on the masses in the combined catalogue and merged all groups that were fully within these regions of other groups into their hosts.  We ended up with 189,712 groups after performing this procedure.  The masses of these groups were rescaled using Equation~\ref{fi_rescale} with the coefficients for the first rescaling from Table~\ref{fit_fi_rescale} of the snapshot nearest in redshift to our groups.  We calculated the new finite infinity radii and repeated the merging procedures to find 171,801 groups containing $67.4\%$ of the mass expected for the volume and the cosmology used.  The masses of 171,801 groups are rescaled again using the sum of the original masses of member groups as a basis and the coefficients for the final rescaling from Table \ref{fit_fi_rescale} of the snapshot nearest in redshift to our groups.  The total mass of the groups adds up to $70.5\%$ and the finite infinity regions occupy $19.5\%$ of the volume covered by our combined catalogue. These values are within the expected range (see Table \ref{fit_fi_rescale}) for a combined dataset of all snapshots.  We note that based on the theory of the two-phase model of \citet{Wiltshire:2007}, we expect $\sim 25\%$ of the volume to be inside finite infinity regions. 

A full description of the format of the catalogue is given in Appendix~\ref{cat_descrip}.  Additionally, we provide another list in the same format, in which the group masses were rescaled so that their sum covers the full mass expected for the catalogue's volume and the Millennium simulation cosmology.  We also applied the merging procedure on the rescaled groups, which left us with 154,903 groups whose finite infinity regions cover $27.1\%$ of the catalogue's volume.
  
\section{Summary and discussion}
In order to provide a robust model of the matter distribution in the local Universe, we used data from the SDSS DR12 \citep{SDSS_DR12} and the 2MRS \citep{2MRS}.  After we preformed some filtering and calibrated the data, we ended up using 43,508 of 44,599 galaxies from the entire 2MRS catalogue and 402,588 of 432,038 galaxies from the SDSS below a redshift of 0.112\footnote{This value was reduced during the filtering after applying a correction for Solar system motion relative to the CMB to 0.11.}. We created several mock catalogues from the Millennium simulation in order to carefully calibrate our techniques.

For the SDSS sample and the 2MRS sample, we provide eight independent mock catalogues each.  Every one of them covers one eighth of the sky and the distribution of the luminous matter in them is based on semi-analytic galaxy models from \citep{Guo:2011}.  The 2MRS mock catalogues consider the Malmquist bias, peculiar motions and all possible measurement uncertainties.  The SDSS mock catalogues takes the same effects into account as for the 2MRS mock catalogues, but also includes the saturation limits and fibre collisions bias of SDSS.  We also provide a corresponding set of dark catalogues, which were used to optimise the group finder and calibrate the masses derived from the groups.  Although our mock catalogues were primarily designed as a calibration tool for our group finder algorithm presented in this paper, they are kept sufficiently general to be used in future work as well.

The core result of our study is the group finder that we developed.  It was strongly inspired by the one presented in \citep{Robotham:2011}.  We considered several effects for our group finder algorithm, which we had to calibrate independently for the 2MRS and the SDSS sample.  First, we calculated the basic linking length $b_{\textrm{link},0}$, which we defined as the average co-moving distances between the two nearest sufficiently luminous (this requirement excludes (most) dwarf galaxies) neighbour galaxies in our unbiased mock catalogue.  The derived parameters provided a first basic estimate of the adaptive linking length used in our algorithm.  The linking length was split into a radial and angular (transversal) components and the latter was modified to account for stretching effect in the redshift space due to peculiar motions inside  groups (illustrated in Figure \ref{velocities}).  We also considered a correction that rescaled the linking length depending on the fraction of the galaxy luminosity function (see Figure \ref{lum_func}) that is visible at a certain distance.  This rescaling effectively corrected for the incompleteness of our data due to the Malmquist bias.  The final linking length, which was defined in Equations \ref{alpha_final} and \ref{R_final}, depends on three free coefficients, that are $\alpha_{\textrm{opt}}$, $R_{\textrm{opt}}$, and $\lambda_{\textrm{opt}}$.  Following \citet{Robotham:2011}, they were optimized by maximizing the group cost function (see Equations \ref{E_fof} to \ref{S_tot}) using a Simplex algorithm \citep{Simplex}.  The results of the optimization are provided in Table \ref{optimal_coefficients}.  All coefficients have values between 0.5 and 1.  The optimized group finder was applied on our datasets in the next step.
 
The 2MRS group catalogue includes 31,506 groups, which host a total of 43,425 galaxies.  We identified many well-known structures of the local Universe in this catalogue.  Aside from basic parameters such as coordinates and redshifts, the most important quantity for our groups are their masses.  Therefore, we used our mock catalogues to carefully calibrate the mass function depending on several other parameters of the groups.  The advantages of the 2MRS catalogue are its large sky coverage and its high completeness at very low redshifts.

The SDSS group catalogue comprises 229,893 groups, which host 402,588 galaxies.  It is restricted to a smaller area of the sky than the 2MRS group catalogue, but it is also deeper providing a clearly more complete sample at higher (up to our limiting redshift of 0.11) redshifts. At very low redshifts the SDSS group catalogue suffers from some additional incompleteness due to the saturation limits of SDSS spectroscopy.  This was the main reason why we also provided the 2MRS catalogue to complement the SDSS catalogue a very low redshifts.  The two catalogues are not completely disjunct, there is some overlap between them and we were able to identify a few prominent galaxy clusters in both catalogues.

We also studied the dependence of the group parameters on the luminosity distance for the 2MRS and SDSS group catalogues (see Figure \ref{resdist_2MRS} and \ref{resdist_SDSS}).  The total group luminosity and the total group mass strongly depend on the luminosity distance.  This is mainly due to the impact of the Malmquist bias.  For group catalogues based on volume limited data, all group parameters should be independent of the distance, however 2MRS and SDSS are magnitude limited.  Despite some corrections that are applied (for example to compensate the Malmquist bias for the detected groups), the Malmquist bias leaves a distinctive imprint on the distribution of the group parameters, because fainter groups are not detected at larger distances.  However, the impact of the bias on parameters such as the group velocity dispersion and the group radius are minimal.

\begin{figure}[ht]
\begin{center}
\includegraphics[width=0.45\textwidth]{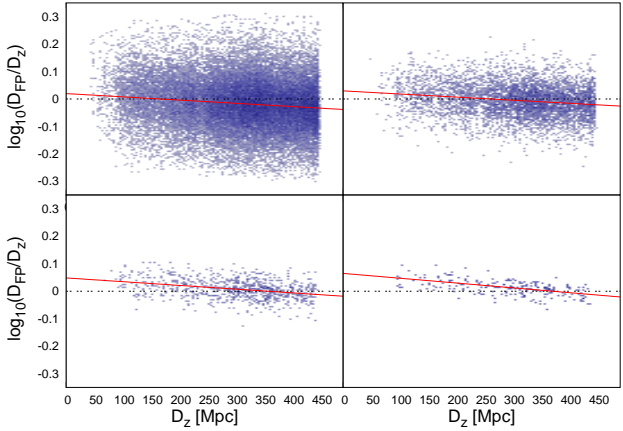}\\ 
\caption{Dependencies of the difference in the distance measurements for our clusters by comparing the co-moving fundamental plane distances with the co-moving redshift distances on the co-moving redshift distance itself and the number of elliptical galaxies per group.  Top-left panel: all groups that contain at least one elliptical galaxy.  Top-right panel: all groups containing at least two elliptical galaxies.  Bottom-left panel: all groups that host five or more elliptical galaxies.  Bottom-right panel: all groups hosting at least 10 elliptical galaxies.  The solid red lines indicate fits on the displayed data, which show an increasing dependence of the ratio between the fundamental plane distance and the redshift distance on the redshift distance with an increasing number of elliptical galaxies per group.  The dashed black line only provides a reference, if this trend was not detected.}
\label{fp_distance_multiplicity_dependence}
\end{center}
\end{figure}

The fundamental plane distance catalogue was obtained by combining the SDSS group catalogue with our previous work.  We provided very detailed calibration of the fundamental plane in \citet{Saulder:2013} and listed updated coefficients based on an extended sample from a more recent paper \citep{Saulder:2015}. We calibrated the fundamental plane of early-type galaxies fitting the fundamental plane parameters, which are the physical scale radius, the central velocity dispersion, and the surface brightness using a Malmquist-bias corrected modified least-square technique.  The physical scale radii were obtained from the angular radii of the galaxies and the redshift distances.  We used 119,085 early-type galaxies from SDSS DR10 \citep{SDSS_DR10}, which were classified using GalaxyZoo \citep{GalaxyZoo,GalaxyZoo_data} and additional criteria.  The aim of the fundamental plane distance catalogue is to provide redshift independent distance measurements for groups hosting elliptical galaxies.  As illustrated in Figure \ref{fp_distance_err}, the accuracy of our alternative (to redshift) distance measurements improves for groups with higher elliptical galaxy multiplicities, but there are some residual trends visible.  The fundamental plane distances tend to be on average larger than the redshift distances for groups with a higher number of early-type galaxies detected in them.

When we examine our sample closer, we discover a trend in the ratio between the fundamental plane distance and the redshift distance.  As illustrated in Figure~\ref{fp_distance_multiplicity_dependence}, there is a trend that the ratio between the co-moving fundamental plane distance and the co-moving redshift distance slightly decreases with a growing co-moving redshift distance. An interesting fact is that this trend becomes steeper for groups hosting more and more elliptical galaxies, as already suggested in Figure~\ref{fp_distance_err}.  We considered a selection effect on the elliptical galaxies in the nearer groups, a general dependence of elliptical galaxies on their environment, or some selection effect on measurement of the median redshift of the groups.  Understanding this systematic trend will be especially relevant for of our planned cosmological test (Saulder et al., submitted).  We notice a dearth of early-type galaxies, and consequently groups hosting them, at low distances, which we attribute to the saturation limit of SDSS of 14~mag in the $r$ band.  The average absolute $r$ band magnitude of an elliptical galaxy (not counting dwarf galaxies) of roughly -21~mag \citep{Saulder:2013}, which means that a significant part of the elliptical galaxies at a luminosity distance of about 100~Mpc is still not included in the sample.  A comparison with our mock catalogues allows us to rule out that this selection effect is a major contributor to the observed trend, hence we conclude that it is primarily due to effects of the environment on early type galaxies, which would agree with recent finding on the dependences of fundamental plane residuals by \citet{Joachimi:2015}.

We provide catalogues of fundamental plane distances for all cluster hosting early-type galaxies for three different cosmological: the Millennium simulation cosmology, the cosmology used in \citet{Saulder:2013} and the recent Planck cosmology \citep{Planck_cosmopara}.

The final catalogue of this paper contains the finite infinity regions, which are a necessary part of the foreground model for our planned cosmological test \citep{Saulder:2012}.  Although we will be testing the a cosmology that is notably different from the $\Lambda$-CDM cosmology used in the Millennium simulation, the assumption that the large-scale matter distribution in the last snapshot of that simulations provides a reasonable model of the matter distribution in the present-day universe allowed us to use that data. A more detailed discussion of this assumption and its implications will be provided in our recently follow-up paper (Saulder et. al, submitted). We found that for the mock catalogues and the Millimil simulation slightly less than half of the simulation's particles are actually bound to the FoF groups.  However, when calculating the finite infinity regions around the groups we saw that up to $80\%$ of the particles of the simulations are within them.  We used this to develop and calibrate a rescaling method, which allowed us to calculate the finite infinity regions from the masses of groups as explained in Section \ref{sec_fi_rescale}.  We applied this method on a combined dataset of the 2MRS and SDSS catalogue.  The two catalogues were merged using weights based on their completeness function (see Figure \ref{completness_function}) and multiplicities of the groups for the groups, which were in both catalogues. Furthermore, we did not use any 2MRS data beyond 92.3 Mpc.  In the end, we provided a catalogue of 171,801 groups and the sizes of the finite infinity regions surrounding them.  Since the distribution of the remaining mass influences the cosmological test, we prepared data for two different scenarios.  In one, we assumed that the rest is distributed sufficiently homogeneous in the voids between the finite infinity regions.  In the other one, we added the missing mass to detected groups and rescaled their masses in a way that their sum accounts for all of the expected mass in the catalogue's volume, yielding a catalogue of 154,903 finite infinity regions. In our follow-up paper (Saulder et. al, submitted), we study in detail which one is a better description of the real data. The finite infinity regions cover approximiately the same fraction of the total volume ($\sim 25\%$ \citep{Wiltshire:2007}) as expected for timescape cosmology. They also partially overlap. We calculated that there are between $2.1\%$ and $2.7\%$ (depending on the catalogue) of the total volume are within more than one finite infinity region. In our calibrations, we ensured that the masses of these regions were correctly estimated.

Several catalogues with similar premises, but also notable differences, can be found in the literature.  The SDSS DR4-based\footnote{with an updated version based on SDSS DR7 \citep{SDSS_DR7}} \citep{SDSS_DR4} catalogue by \citet{Yang:2007,Yang:2009} does not contain any galaxies below a redshift of 0.01 in contrast to our SDSS DR12 \citep{SDSS_DR12}, which does contain galaxies below that redshift and even compensates for the saturation bias at low redshifts with the provided 2MRS based catalogue.  The catalogue of \citet{Nurmi:2013} provides an SDSS DR7 \citep{SDSS_DR7} based sample of groups.  However, they use sub-samples of different magnitude range to construct it and they mainly concern themselves with rich groups and clusters and leave isolated galaxies aside.  The more recent catalogue by \citet{Tempel:2014} is based on SDSS DR10 \citep{SDSS_DR10} and supplemented with data from 2dFGRS \citep{2dFGRS,2dFGRS_final}, 2MASS \citep{2MASS}, 2MRS \citep{2MRS}, and the Third Reference Catalogue of Bright Galaxies \citep{deVaucouleurs:1991,Corwin:1994}.  The main difference to our catalogues is that \citet{Tempel:2014} created a set of volume limited samples, which is in total less complete than our magnitude limited (up to the selected redshift limit) catalogue.  Furthermore, we put a special emphasis on mass completeness, which is essential for the planned cosmological test.  The new 2MRS \citep{2MRS} based group catalogue of \citet{Tully:2015b} has much in common with our own 2MRS based group catalogue.  They used a slightly limited 2MRS sample (clusters between 3,000 and 10,000 km/s redshift velocity), but investigated their group data into greater detail.  For reasons of comparison, we created a sub-sample of our 2MRS group catalogue with the same limits as the catalogue of \citet{Tully:2015b}.  We found 25,396 galaxies with our calibrations in that range, while \citet{Tully:2015b} found 24,044 galaxies.  There is a notable difference in the number of group, which we detected in the sample: we identified 17,191 groups compared to 13,607 groups in the Tully catalogue.  3,630 of these groups have two or more members in our catalogue and \citet{Tully:2015b} found a similar number, 3,461, for theirs. Furthermore, there are several prominent clusters, which could be identified in both catalogues.

\section{Conclusions}

We created a set of group catalogues with the intention to use them for a cosmological test in our next paper.  In the process, we devised a set of wide-angle mock catalogues for 2MRS and SDSS, which consider all possible biases and uncertainties.  We use them to calibrate and optimize our FoF-based group finder algorithm and a mass function.  After applying the group finder on 2MRS and SDSS data, we obtain a set of group catalogues, which can be used for various future investigation, even beyond the initial motivation of this paper.  As matter of fact, three of the four catalogues, which we create in the process of this paper, are fully independent of the intended cosmological test.  The 2MRS and the SDSS group catalogue complement the existing group catalogues based on these surveys, such as \citet{Crook:2007}, \citet{Yang:2007}, \citet{Nurmi:2013}, \citet{Tempel:2014}, \citet{Tully:2015b}, and others.  The advantages of our catalogues are that they consider all groups of all sizes ranging from individual galaxy to massive clusters and that they provide well-calibrated group parameters with a special focus put on total masses and stellar masses.  The fundamental plane distance catalogue adds a unique dataset to results of this paper.  The final catalogue was obtained by merging or 2MRS and SDSS group catalogue and calculating of the finite infinity regions around the groups.  In our upcoming paper (Saulder et al., submitted), we will perform our test of timescape cosmology by using the fundamental distance catalogue and the finite infinity region catalogue.

\section*{Acknowledgments}
CS is grateful to Aaron Robotham for helpful advice and also acknowledges the support from an ESO studentship.  

IC is supported by the Telescope Data Center, Smithsonian Astrophysical Observatory.  His activities related to catalogues and databases are supported by the grant MD-7355.2015.2 and Russian Foundation for Basic Research projects 15-52-15050 and 15-32-21062. IC also acknowledges the ESO Visiting Scientist programme.

Funding for SDSS-III has been provided by the Alfred P. Sloan Foundation, the Participating Institutions, the National Science Foundation, and the U.S. Department of Energy Office of Science. The SDSS-III web site is \url{http://www.sdss3.org/}.

SDSS-III is managed by the Astrophysical Research Consortium for the Participating Institutions of the SDSS-III Collaboration including the University of Arizona, the Brazilian Participation Group, Brookhaven National Laboratory, University of Cambridge, Carnegie Mellon University, University of Florida, the French Participation Group, the German Participation Group, Harvard University, the Instituto de Astrofisica de Canarias, the Michigan State/Notre Dame/JINA Participation Group, Johns Hopkins University, Lawrence Berkeley National Laboratory, Max Planck Institute for Astrophysics, Max Planck Institute for Extraterrestrial Physics, New Mexico State University, New York University, Ohio State University, Pennsylvania State University, University of Portsmouth, Princeton University, the Spanish Participation Group, University of Tokyo, University of Utah, Vanderbilt University, University of Virginia, University of Washington, and Yale University. 

This publication makes use of data products from the Two Micron All Sky Survey, which is a joint project of the University of Massachusetts and the Infrared Processing and Analysis Center/California Institute of Technology, funded by the National Aeronautics and Space Administration and the National Science Foundation.

\FloatBarrier

\addcontentsline{toc}{section}{References}
\bibliography{paper}\label{bib}

\appendix
\onecolumn 
\begin{landscape}
\section{Catalogue description}
\label{cat_descrip}
\centering
\begin{table*}[ht]
\begin{center}
\begin{adjustbox}{max width=\linewidth}
\begin{tabular}{ccccccccccccccccc}
group ID & $\alpha_{\textrm{group}}$ & $\delta_{\textrm{group}}$ & $z_{\textrm{group}}$ & $\textrm{log}_{10} \left( L_{\textrm{tot,K}_{\textrm{s}}} \right) $ & $\textrm{log}_{10} \left( \Delta L_{\textrm{tot,K}_{\textrm{s}}} \right) $ & $\textrm{log}_{10} \left( L_{\textrm{obs,K}_{\textrm{s}}} \right)$ & $\textrm{log}_{10} \left( M_{\textrm{group}} \right)$ & $\textrm{log}_{10} \left( \Delta M_{\textrm{group}} \right)$ & $\textrm{log}_{10} \left( M_{*} \right)$ & $N_{M_{*}}$ &  $\textrm{log}_{10} \left( M_{\textrm{dyn}} \right)$ & $\sigma_{\textrm{group}}$ & $R_{\textrm{group}}$ & $a_{\textrm{group}}$ & $D_{L}$  & $N_{\textrm{group}}$  \\ 
 & [$^{\circ}$] &  [$^{\circ}$] & & [$\textrm{log}_{10} \left( L_{\astrosun} \right)$] & [$\textrm{log}_{10} \left( L_{\astrosun} \right)$] & 
 [$\textrm{log}_{10} \left( L_{\astrosun} \right)$] & [$\textrm{log}_{10} \left( M_{\astrosun} \right)$] & [$\textrm{log}_{10} \left( M_{\astrosun} \right)$] & [$\textrm{log}_{10} \left( M_{\astrosun} \right)$] &  &  [$\textrm{log}_{10} \left( M_{\astrosun} \right)$] & [km/s] & [kpc] & [$^{\circ}$] & [Mpc] & \\ \hline  
8 & 187.9967 & 14.4204 &  0.0049 &  13.00 & 0.21 &  12.99 &   14.80 & 0.14 & 11.04 & 15 & 12.63 & 668.9 & 1376 &  3.9046 & 20.4 & 205 \\
10 & 54.7163 & -35.5944 &  0.0046 &   12.37 & 0.21 & 12.36 &   13.95 &  0.14 & 12.11 & 27 &  11.46 &  308.5 & 438 & 1.3365 & 19.9 & 41 \\
91 & 192.5164 & -41.3820 & 0.0121 & 12.78 &  0.21 &  12.74 & 14.59 &   0.14 & 12.53 & 45 &  12.36 & 847.2 & 457 & 0.5334 & 50.3 & 67 \\
279 & 157.6104 & -35.3595 & 0.0101 & 12.37 & 0.21 &  12.33 &  14.15 & 0.14 & 12.26 & 27 & 11.80 &  522.6 & 333 & 0.4635 &  42.0 & 39 \\
298 & 243.7661 & -60.9072 &  0.0167 &  13.04 &  0.21 &   12.97 &  14.88 &  0.14 & 11.87 & 3 &   12.46 & 751.2 & 738 & 0.6271 & 69.7 & 104 \\
344 & 49.9510 & 41.5117 &   0.0171 & 12.97 & 0.21 &  12.90 & 14.89 & 0.14 & 12.15 & 9 &  12.56 &  1064.3 &  465 &  0.3874 & 71.1 & 100 \\
354  & 159.1784 & -27.5283 & 0.0131 & 12.68 & 0.21 &  12.63 & 14.56 & 0.14 & 12.53 & 47 &  12.25 & 634.8 &  627 & 0.6778 & 54.5 & 76 \\
400 & 21.5025 & -1.3451 &  0.0171 & 12.52 &  0.21 & 12.45 &  14.34 &  0.14 & 11.49 & 6 & 11.93 &  456.5 &  581 &  0.4841 & 71.1 & 39 \\
442 & 195.0338 &  27.9770 & 0.0241 &   13.16 &  0.21 & 13.00  & 15.02 &  0.14 & 12.96 & 78 &  12.62 &  945.4&  667 &  0.3975 & 100.9 & 82 \\
539 & 28.1936 &  36.1518 &  0.0154 & 12.66 & 0.21 & 12.59 & 14.45 &  0.14 & 11.85 & 9 & 12.00 & 442.6 & 728 & 0.6712 &  64.1 & 54 \\
1007 & 176.0090 & 19.9498 &   0.0223 & 12.85 & 0.21 & 12.72 & 14.66 &  0.14 & 12.69 & 40 & 12.20 &  591.0 & 654  & 0.4183 & 93.4 & 47 \\
2246 & 201.9870 & -31.4955 & 0.0509 & 13.71 &  0.21 & 12.90 &  15.22 &  0.14 & 12.77 & 11 &  12.67 & 788.8 & 1086 & 0.3150 & 218.2 & 22 
\end{tabular}
\end{adjustbox}
\end{center}
\caption{Parameters of a selected sample of groups as they appear in our 2MRS group catalogue.  Column 1: 2MRS group ID, column 2 and 3: equatorial coordinates of the group centre, column 4: redshift of the group centre, column 5: total group luminosity in the $K_{s}$ band, column 6: uncertainty of the total group luminosity, column 7: observed group luminosity in the $K_{s}$ band, column 8: group mass, column 9: uncertainty of the group mass, column 10: stellar mass, column 11: number of galaxies in the group with stellar masses available, column 12: dynamical mass of the group, column 13: group velocity dispersion, column 14: physical group radius, column 15: angular group radius, column 16: luminosity distance to the group centre, and column 17: detected group members.}
\label{sample_cat_2mrs}
\end{table*}

\begin{table*}[ht]
\begin{center}
\begin{adjustbox}{max width=\linewidth}
\begin{tabular}{ccccccccccccccccc}
group ID & $\alpha_{\textrm{group}}$ & $\delta_{\textrm{group}}$ & $z_{\textrm{group}}$ & $\textrm{log}_{10} \left( L_{\textrm{tot,r}} \right) $ & $\textrm{log}_{10} \left( \Delta L_{\textrm{tot,r}} \right) $ & $\textrm{log}_{10} \left( L_{\textrm{obs,r}} \right)$ & $\textrm{log}_{10} \left( M_{\textrm{group}} \right)$ & $\textrm{log}_{10} \left( \Delta M_{\textrm{group}} \right)$ &  $\textrm{log}_{10} \left( M_{*} \right)$ & $N_{M_{*}}$  & $\textrm{log}_{10} \left( M_{\textrm{dyn}} \right)$ & $\sigma_{\textrm{group}}$ & $R_{\textrm{group}}$ & $a_{\textrm{group}}$ & $D_{L}$  & $N_{\textrm{group}}$  \\ 
 & [$^{\circ}$] &  [$^{\circ}$] & & [$\textrm{log}_{10} \left( L_{\astrosun} \right)$] & [$\textrm{log}_{10} \left( L_{\astrosun} \right)$] & 
 [$\textrm{log}_{10} \left( L_{\astrosun} \right)$] & [$\textrm{log}_{10} \left( M_{\astrosun} \right)$] & [$\textrm{log}_{10} \left( M_{\astrosun} \right)$] & [$\textrm{log}_{10} \left( M_{\astrosun} \right)$] &    &  [$\textrm{log}_{10} \left( M_{\astrosun} \right)$] & [km/s] & [kpc] & [$^{\circ}$] & [Mpc] & \\ \hline 
13370 & 220.1785 &  3.4654 & 0.0285 & 12.25 &  0.22 &  12.23 & 14.92 &  0.16 & 12.65 & 215  & 12.41 &  547.9 & 1219 &  0.6167 & 119.8 & 220 \\
20266 & 351.0837 &  14.6471 & 0.0408 & 12.43 & 0.22 & 12.38 &  14.98 &  0.16 & 12.79 & 197  &  12.32 &  525.3 &  1074.1 &  0.3847 & 173.3 & 214 \\
30885 & 184.8427 &  6.0987 &  0.0629 &  11.70 & 0.22 &  11.70 &  14.82 & 0.16 & 11.25 & 63  &  12.60 & 692.3 &  1195.8 & 2.6698 & 26.0 & 252\\
39456 & 247.4371 & 40.8116 & 0.0303 &  12.74 &  0.22 & 12.71 &  15.38 &  0.16 & 13.11 & 550 &   12.77 & 620.0 &  2219  & 1.0558 & 127.8 & 567 \\
42643 & 227.7808 &  5.3173 &  0.0795 &  12.92 &  0.22 & 12.79 & 15.48 &  0.16 & 13.25 & 259 &  13.15 &   864.2 &   2712 &   0.5192&  348.7 & 277 \\
82182 & 49.1791 &  41.3249 &  0.0162 &  12.01 & 0.22 & 12.00 &  14.77 &  0.16 & -99.99 & 0 & 12.59 &  834.8 & 799 &  0.7015 & 67.4 &  141 \\
112271 & 239.5833 & 27.2334 &  0.0900 & 12.97 &  0.22 & 12.80 &  15.50 &  0.16 & 13.30 & 249 & 12.88 &  884.2 & 1405 & 0.2401 & 398.5 & 250 \\
127065 & 14.0672 & -1.2554 & 0.0431 & 12.53 &  0.22 & 12.49 & 15.09 & 0.16 & 12.82 & 162 & 12.67 &   626.9 & 1688 &   0.5742 & 183.3 & 227 \\
149473 & 240.5827 & 16.3460 & 0.0372 & 13.02 & 0.22 & 12.98 &  15.66 &  0.16 & 13.40 & 891 &  13.24 & 1011.4 &  2413 &  0.9428 & 157.7 & 896 \\
153084 & 195.0339 & 27.9769 & 0.0243 & 12.76 & 0.22 & 12.74 & 15.49 & 0.16 & 13.22 & 761 & 12.81 &  876.8 &  1214 & 0.7175 & 101.7 & 765 \\
160410 & 167.6615 & 28.7689 &  0.0342 &  12.39 & 0.22 & 12.36 & 15.01 & 0.16 & 12.77 & 224 & 12.52 & 634.4 & 1175.8 &  0.4980 & 144.7 & 229 \\
176804 & 176.0090 & 19.9498 &  0.0234 &  12.52 &  0.22 & 12.50 &  15.22 & 0.16 & 12.95 & 367 &  12.71 &  758.1 &  1272 &  0.7790 & 98.0 & 396
\end{tabular}
\end{adjustbox}
\end{center}
\caption{Parameters of a selected sample of groups as they appear in our SDSS group catalogue.  Column 1: SDSS group ID, column 2 and 3: equatorial coordinates of the group centre, column 4: redshift of the group centre, column 5: total group luminosity in the $r$ band, column 6: uncertainty of the total group luminosity, column 7: observed group luminosity in the $r$ band, column 8: group mass, column 9: uncertainty of the group mass, column 10: stellar mass, column 11: number of galaxies in the group with stellar masses available, column 12: dynamical mass of the group, column 13: group velocity dispersion, column 14: physical group radius, column 15: angular group radius, column 16: luminosity distance to the group centre, and column 17: detected group members.}
\label{sample_cat_sdss}
\end{table*}

\begin{table*}[ht]
\begin{center}
\begin{adjustbox}{max width=\linewidth}
\begin{tabular}{ccccccccccccccccccc} 
group ID & $\alpha_{\textrm{group}}$ & $\delta_{\textrm{group}}$ & $z_{\textrm{group}}$ & $\Delta z_{\textrm{group}}$ & $D_{A,\textrm{FP}}$ & $\Delta D_{A,\textrm{FP}}$ & $D_{C,\textrm{FP}}$ & $\Delta D_{C,\textrm{FP}}$ & $D_{L,\textrm{FP}}$ & $\Delta D_{L,\textrm{FP}}$ & $D_{A,z}$  & $\Delta D_{A,z}$  & $D_{C,z}$ & $\Delta D_{C,z}$ & $D_{L,z}$ & $\Delta D_{L,z}$ & $N_{\textrm{ETG}}$ & $N_{\textrm{group}}$  \\ 
& [$^{\circ}$] &  [$^{\circ}$] & & & [Mpc] & [Mpc] & [Mpc] & [Mpc] & [Mpc] & [Mpc] & [Mpc] & [Mpc] & [Mpc] & [Mpc] & [Mpc] & [Mpc] &  & \\ \hline 
2 & 131.9540 & 52.4033 & 0.09456 & 0.00007 & 345.9 & 82.1 & 378.1 & 89.8 & 413.3 & 98.1 & 350.9 & 0.3 & 384.1 & 0.3 & 420.4 & 0.3 & 1 & 2 \\
3 & 119.9849 & 41.5277 & 0.09424 & 0.00007 & 428.2 & 101.7 & 478.8 & 113.7 & 535.3 & 127.1 & 349.8 & 0.3 & 382.7 & 0.3 & 418.8 & 0.3 & 1 & 2 \\
8 & 177.0247 & -1.6887 & 0.09601 & 0.00005 & 288.8 & 68.6 & 310.9 & 73.8 & 334.7 & 79.5 & 355.7 & 0.2 & 389.9 & 0.2 & 427.3 & 0.2 & 1 & 4 \\
13 & 128.2924 & 49.6682 & 0.05304 & 0.00002 & 214.7 & 36.0 & 226.7 & 38.1 & 239.4 & 40.2 & 205.3 & 0.1 & 216.2 & 0.1 & 227.7 & 0.1 & 2 & 25 \\
16 & 149.7203 & 59.2838 & 0.07569 & 0.00010 & 349.3 & 82.9 & 382.2 & 90.7 & 418.2 & 99.3 & 286.2 & 0.4 & 307.8 & 0.4 & 331.1 & 0.4 & 1 & 1 \\
22 & 190.9071 & 64.5601 & 0.07536 & 0.00004 & 270.7 & 64.3 & 290.0 & 68.8 & 310.7 & 73.8 & 285.0 & 0.2 & 306.5 & 0.2 & 329.6 & 0.2 & 1 & 5 \\
25 & 133.2216 & 1.0330 & 0.05247 & 0.00010 & 312.7 & 74.2 & 338.7 & 80.4 & 367.0 & 87.1 & 203.3 & 0.4 & 213.9 & 0.4 & 225.2 & 0.4 & 1 & 1 \\
31 & 171.6762 & -1.4590 & 0.07625 & 0.00006 & 303.0 & 71.9 & 327.4 & 77.7 & 353.8 & 84.0 & 288.1 & 0.2 & 310.1 & 0.2 & 333.7 & 0.3 & 1 & 3 \\
39 & 185.9856 & -3.1172 & 0.09479 & 0.00006 & 269.6 & 64.0 & 288.7 & 68.5 & 309.2 & 73.4 & 351.6 & 0.2 & 385.0 & 0.2 & 421.5 & 0.3 & 1 & 3 \\
41 & 183.7026 & -2.9358 & 0.07649 & 0.00010 & 258.3 & 61.3 & 275.9 & 65.5 & 294.6 & 69.9 & 289.0 & 0.4 & 311.2 & 0.4 & 334.9 & 0.4 & 1 & 1 \\
51 & 258.8457 & 57.4112 & 0.02923 & 0.00001 & 112.2 & 10.9 & 115.4 & 11.2 & 118.7 & 11.5 & 116.14 & 0.05 & 119.54 & 0.05 & 123.0 & 0.1 & 6 & 67
\end{tabular}
\end{adjustbox}
\end{center}
\caption{Parameters of a selected sample of groups as they appear in our fundamental plane distance catalogue.  Column 1: SDSS group ID, column 2 and 3: equatorial coordinates of the group centre; column 4: redshift of the group centre; column 5: redshift uncertainty of the group centre; columns 6, 8, and 10: the angular diameter distance, the co-moving distance, and the luminosity distance all calculated using the fundamental plane; columns 7, 9, and 11: the uncertainties for the corresponding fundamental plane distances; columns 12, 14, and 16: the angular diameter distance, the co-moving distance, and the luminosity distance all calculated using the redshift; columns 13, 15, and 17: the statistical uncertainties for the corresponding redshifts distances, additional systematic uncertainties from peculiar motions are of the order of $\pm$3 Mpc; column 18: number of detected early-type galaxies; and column 19: all detected group members.}
\label{sample_cat_fp}
\end{table*} 

\begin{table*}[ht]
\begin{center}
\begin{tabular}{ccccccccccccc} 
galaxy ID & group ID & SDSS ObjID & $\alpha_{\textrm{gal}}$ & $\delta_{\textrm{gal}}$ & $z_{\textrm{group}}$ & $D_{L,\textrm{FP}}$ & $\Delta D_{L,\textrm{FP}}$ & $D_{C,\textrm{FP}}$ & $\Delta D_{C,\textrm{FP}}$ & $D_{A,\textrm{FP}}$ & $\Delta D_{A,\textrm{FP}}$ & ETG \\
 & & & [$^{\circ}$] &  [$^{\circ}$] & & [Mpc] & [Mpc] & [Mpc] & [Mpc] & [Mpc] & [Mpc] & \\ \hline 
 2 & 2 & 1237651191358292129 &  131.9540 & 52.4033 & 0.09456 &   429.0  &  101.2 &  392.4 & 92.6 & 359.0 & 84.7 &  1 \\
 3 &  3 & 1237651190815457668 &  119.9849 &  41.5277 &  0.09424 &  556.7 & 131.3 & 497.8 &   117.4 &  445.1 &  105.0 &   0 \\
 8 &  8 &  1237650371554377861 &  177.0247 &  -1.6887 & 0.09601 &  347.4 &   82.0 & 322.8 &  76.2 &  299.9 &  70.8 &   1 \\
 13 &  13 & 1237651191356719236 &  128.1192 & 49.6723 & 0.05304 & 248.6 &  40.2 & 235.5 &  38.1 &  223.0 &  36.0 &  0 \\
 16 &  16 &  1237651190289793215 & 149.7203 & 59.2838 & 0.07569 &   435.3 &   102.7 &   397.8 &   93.8 &   363.5 &   85.8 &  1 
\end{tabular}
\end{center}
\caption{Parameters of the first five galaxies as they appear in our fundamental plane distance catalogue.  Column 1: internal SDSS galaxy ID, column 2: SDSS cluster ID, column 3: SDSS DR12 photometric Object ID, column 4 and 5: equatorial coordinates of the galaxy; column 6: redshift of the group hosting the galaxy; columns 7, 9, and 11: the luminosity distance, the co-moving distance, and the angular diameter distance all calculated using the fundamental plane; columns 8, 10, and 12: the uncertainties for the corresponding fundamental plane distances; columns 13: early-type galaxy flag (set to one if it is an early-type galaxy used in the calibration of the fundamental plane).}
\label{galaxies_fp_groups}
\end{table*} 

\begin{table*}[ht]
\begin{center}
\begin{tabular}{cccccccccc} 
 SDSS ObjID & $\alpha_{\textrm{gal}}$ & $\delta_{\textrm{gal}}$ & $z_{\textrm{gal}}$ & $D_{L,\textrm{FP}}$ & $\Delta D_{L,\textrm{FP}}$ & $D_{C,\textrm{FP}}$ & $\Delta D_{C,\textrm{FP}}$ & $D_{A,\textrm{FP}}$ & $\Delta D_{A,\textrm{FP}}$  \\
 & [$^{\circ}$] &  [$^{\circ}$] & & [Mpc] & [Mpc] & [Mpc] & [Mpc] & [Mpc] & [Mpc] \\ \hline 
  1237648704066879657 & 244.3231 & -0.2069 &   0.09316 &  437.9 &  103.3 & 400.0  & 94.4 & 365.3 &  86.2 \\   
  1237648704605127522 & 247.5173 &  0.2144 &  0.16488 &  805.6 &  190.1 & 691.6 & 163.2 & 593.7  & 140.081589   \\ 
  1237648705130528881 & 221.3076 &  0.4984 &   0.11034 &   412.9 & 97.4 & 378.8 & 89.4 & 347.6 &  82.0170212    \\
  1237648705668120860 & 222.9668 &  1.0458 &  0.12011 & 453.0 & 106.9 & 412.6 & 97.3 & 375.8  & 88.6597137    \\
  1237648720687399047 & 165.2095 &  -0.5044 &  0.15186 &  596.6  & 140.8  & 529.8 & 125.0  & 470.5 & 111.007607    
  \end{tabular}
\end{center}
\caption{Parameters of the first five galaxies as they appear in our fundamental plane distance catalogue.  Column 1: SDSS DR12 photometric Object ID, column 2 and 3: equatorial coordinates of the galaxy; column 4: CMB-corrected redshift of the galaxy; columns 5, 7, and 9: the luminosity distance, the co-moving distance, and the angular diameter distance all calculated using the fundamental plane; columns 6, 8, and 10: the uncertainties for the corresponding fundamental plane distances.}
\label{galaxies_fp_beyond}
\end{table*} 

\twocolumn
\end{landscape} 

\begin{table*}[htpb]
\begin{center}
\begin{tabular}{ccccccc} 
galaxy ID & group ID & 2MASS ID & $\alpha_{\textrm{gal}}$ & $\delta_{\textrm{gal}}$ & $z_{\textrm{gal}}$ &  $\textrm{log}_{10} \left( M_{*} \right)$  \\
 & & & [$^{\circ}$] &  [$^{\circ}$] & &  [$\textrm{log}_{10} \left( M_{\astrosun} \right)$] \\ \hline 
  1  & 1 & 09553318+6903549 & 148.8882 &  69.0653 &  0.00016 & -99.99 \\
  2  &  2 & 13252775-4301073 &  201.3656 &  -43.0187 &  0.00267 & -99.99 \\
  3 & 3 & 13052727-4928044 &196.3637 &  -49.4679 & 0.00269 & -99.99 \\
 4 & 1 & 09555243+6940469 & 148.9685 & 69.6797 &  0.00094 & -99.99 \\
 5 & 4 & 13370091-2951567 & 204.2538 & -29.8658 &  0.00264 & -99.99 
 \end{tabular}
\end{center}
\caption{Identification numbers and parameters of the first five galaxies as they appear in the galaxy list for our 2MRS group catalogue.  Column 1:
internal 2MRS galaxy ID, column 2: 2MRS cluster ID, column 3: 2MASS ID, column 4 and 5: equatorial coordinates of the galaxy; column 6: CMB-corrected redshift of the galaxy, column 7: stellar mass (if available).}
\label{galaxies_2mrs_list}
\end{table*}  

\begin{table*}[htpb]
\begin{center}
\begin{tabular}{ccccccc} 
galaxy ID & group ID & SDSS ObjID & $\alpha_{\textrm{gal}}$ & $\delta_{\textrm{gal}}$ & $z_{\textrm{gal}}$ &  $\textrm{log}_{10} \left( M_{*} \right)$  \\
 & & & [$^{\circ}$] &  [$^{\circ}$] & &  [$\textrm{log}_{10} \left( M_{\astrosun} \right)$] \\  \hline 
 1 & 1  & 1237651213361217849  &   259.8924 &   58.2375 &  0.05106 & 10.17 \\
 2 & 2  & 1237651191358292129  &   131.9540 &  52.4033 &  0.09433 & 11.07\\
 3 & 3  & 1237651190815457668  &   119.9849 &  41.5277 &  0.09416 & 10.50 \\ 
 4 & 4  & 1237651067891220642  &   191.3432 & 65.9404 &   0.07532 & 10.42\\
 5 & 5  & 1237650796220842186  &   148.0828 &  0.2037 &  0.07611 & 11.16
 \end{tabular}
\end{center}
\caption{Identification numbers and parameters of the first five galaxies as they appear in the galaxy list for our SDSS group catalogue.  Column 1: internal SDSS galaxy ID, column 2: SDSS cluster ID, column 3: SDSS DR12 photometric Object ID, column 4 and 5: equatorial coordinates of the galaxy; column 6: CMB-corrected redshift of the galaxy, column 7: stellar mass (if available)..}
\label{galaxies_sdss_list}
\end{table*}   

\begin{table*}[htpb]
\begin{center}
\begin{tabular}{ccccccc}
$c_{x}$ & $c_{y}$ & $c_{z}$ & $\textrm{log}_{10} \left( M_{\textrm{fi}} \right)$ & $\textrm{log}_{10} \left(\Delta M_{\textrm{fi}} \right)$ & $R_{\textrm{fi}}$ & $ \Delta R_{\textrm{fi}}$\\
 $[\textrm{Mpc}]$ & $[\textrm{Mpc}]$ & $[\textrm{Mpc}]$ & [$\textrm{log}_{10} \left( M_{\astrosun} \right)$] & [$\textrm{log}_{10} \left( M_{\astrosun} \right)$] & $[\textrm{Mpc}]$ & $[\textrm{Mpc}]$\\\hline 
-0.91 & 0.55 & 2.87 & 12.51 & 0.18 & 2.0 & 0.6  \\
-6.60 & -2.59 & -6.56 & 13.26 & 0.18 & 3.6 & 1.1 \\  
-6.88 & -2.02 & -8.39 & 12.83 & 0.18 & 2.6 & 0.8 \\  
-8.14 & -3.67 & -5.13 & 12.90 & 0.18 & 2.7 & 0.8 \\
-8.14 & 0.13 & 10.51 & 13.97 & 0.18 & 6.3 & 1.8 \\ 
-19.15 & -1.51 & 5.13 & 15.37 & 0.18 & 18.4 & 5.3 \\ 
9.03 & 12.45 & -10.32 & 14.32 & 0.18 & 8.2 & 2.4 \\  
10.24 & 8.87 & -0.03 & 12.84 & 0.18 & 2.6 & 0.8 \\
5.88 & 6.85 & -7.88 & 12.61 & 0.18 & 2.2 & 0.6 \\ 
-2.69 & -1.61 & 4.37 & 11.99 & 0.18 & 1.4 & 0.4 
\end{tabular}
\end{center}
\caption{Parameters of the first ten groups as they appear in our finite infinite regions catalogue.  Column 1 to 3: the Cartesian co-moving coordinates of the centre of the finite infinity region; column 4: the mass within the finite infinity region; column 5: the uncertainty for mass within the finite infinity regions; column 6: the radius of the finite infinity region; and column 7: the uncertainty in the radius of the finite infinity region.}
\label{sample_cat_fi}
\end{table*} 

\FloatBarrier

In the section, we provide a detailed description of all catalogues presented in this paper and their internal structure.

\emph{2MRS group catalogue:} We provide a list of all detected groups that includes their 2MRS group ID, the coordinates of the group centre (right ascension and declination are both given in degrees), the median redshifts, the total group luminosity in the $K_{s}$ band and its uncertainty, the observed group luminosity in the $K_{s}$ band, the calculated group mass (using the method explained in Section \ref{sec_mass}) and its uncertainty, the stellar mass using the method of Section \ref{sec_mstar} and the number of galaxies within that group for which we had stellar masses available\footnote{For some galaxies, there was not sufficient data available to derive solid stellar masses.}, the dynamical mass of the group, the group velocity dispersion, the physical group (in kpc), the angular group radius (in degrees), the luminosity distance to the group centre (in Mpc), and the number of detected group members.  An excerpt of this list is provided in Table \ref{sample_cat_2mrs}.  In addition to that list, a list of all the galaxies used is provided in Table \ref{galaxies_2mrs_list}, containing the our internal galaxy IDs, the 2MRS group IDs of the group the galaxy belongs to, the 2MASS IDs of the galaxies, as well as its coordinates and redshift.

The cluster with the most detected members and the 2MRS group ID~8 in our catalogue is the well-known Virgo cluster.  Furthermore, we identified the following clusters with their parameters listed in Table \ref{sample_cat_2mrs}: the Fornax Cluster with the ID~10, the Centaurus Cluster with the ID~91, the Antlia Cluster with the ID~279, the Norma cluster with the ID~298, the Perseus Cluster with the ID~344, the Hydra Cluster with the ID~354, Abell 194 with the ID~400, the Coma Cluster with the ID~442, Abell~262 with the ID~539, the Leo Cluster with the ID~1007, and the cluster with ID~2246 is associated with the Shapley Supercluster.

\emph{SDSS group catalogue:} We provide a list of all detected groups containing their SDSS group ID, the coordinates of the group centre (right ascension and declination are both given in degrees), the median redshifts, the total group luminosity in the $r$ band and its uncertainty, the observed group luminosity in the $r$ band, the calculated group mass (using the method explained in Section \ref{sec_mass}) and its uncertainty, the stellar mass (using the method of Section \ref{sec_mstar}) and the number of galaxies within that group for which we had stellar masses available, the dynamical mass of the group, the group velocity dispersion, the physical group (in kpc), the angular group radius (in degrees), the luminosity distance to the group centre (in Mpc), and the number of detected group members.  An excerpt of this list is provided in Table \ref{sample_cat_sdss}.  In addition to that list, a list of all the galaxies used is provided in Table \ref{galaxies_sdss_list}, containing the our internal galaxy IDs, the SDSS group IDs of the group the galaxy belongs to, the SDSS object IDs of the galaxies, as well as its coordinates and redshift.

We identified some of the richest clusters in our catalogue: the Coma Cluster with the ID~153084, the Perseus Cluster with the ID~82182, the Leo Cluster with the ID~176804, Abell~2142 with the ID~112271 and the cluster with the ID~149473 is associated with structures of the Hercules Supercluster.  A list of these clusters and others with more than two hundred detected members is provided in Table \ref{sample_cat_sdss}.

\emph{Fundamental plane distance group catalogue:} We provide three lists (one of them is shown in Table \ref{sample_cat_fp} as an example) of all our SDSS groups hosting elliptical galaxies for three slightly different sets of cosmological parameters.  Naturally, since this paper uses the cosmology of the Millennium simulation (see Table \ref{cosmopar} for the parameters), one of our lists is based on it.  Another list is based on the cosmological parameters of our previous papers \citep{Saulder:2013,Saulder:2015}, which are $H_{0}=70$ km/s/Mpc, $\Omega_{M} = 0.3$, and $\Omega_{\Lambda} = 0.7$, while the last list uses the cosmological parameters of the Planck mission \citep{Planck}, $H_{0}=67.3$ km/s/Mpc, $\Omega_{M} = 0.315$, and $\Omega_{\Lambda} = 0.685$ \citep{Planck_cosmopara}.  The lists contain their SDSS group ID, the coordinates of the group centre (right ascension and declination are both given in degrees), the median redshifts and their uncertainties, the angular, co-moving, and luminosity fundamental plane distances and their respective uncertainties, the angular, co-moving, and luminosity redshift distances and their respective statistical uncertainties, the number of elliptical galaxies hosted in that group andthe total number of detected group members.  Additional $\sim 3$ Mpc of systematic uncertainties have to be considered for the redshift distances due to peculiar motions (see Figure \ref{velocities}).  As an example of our catalogue, the top ten lines of our catalogue is provided in Table \ref{sample_cat_fp}.  Additional, we suppliment our catalogue with a list of all 145,359 galaxies within groups that host early-type galaxies for all cosmologies mentioned earlier.  As illustrative example we provide that list containing the corresponding IDs, coordinates, and fundamental plane distances in Table \ref{galaxies_fp_groups}.  The galaxies used in the fundemtantal plane calibration of \citet{Saulder:2015}, which lie beyond the limits of our group catalogue, are listed in a seperate table.  Table \ref{galaxies_fp_beyond} displays a tiny sample of these almost 70,000 galaxies to illustrate the structure of that list.

\emph{Finite infinity region catalogue:} We provide a list of all 164,509 remaining groups containing their Cartesian co-moving coordinates $c_{x}$, $c_{y}$, and $c_{z}$, their final masses, and their radii for the finite infinity regions.  A sample of the first ten lines of our catalogue is given in Table \ref{sample_cat_fi}.  Additionally, we provide another list in the same format, in which the masses of the groups have been rescaled so thattheir sum covers the full mass expected for the catalogue's volume and the Millennium simulation cosmology.  We also applied the merging procedure on the rescaled groups, which leaves us with 152,442 groups whose finite infinity regions cover $28.3\%$ of the catalogue's volume.

\FloatBarrier
\section{SDSS-2MASS transformation}
\label{SDSS2MASStransformation}
The data from \citet{Guo:2011} does not contain any 2MASS magnitudes for the semi-analytic galaxy models in contrast to models in previous runs \citep{DeLucia:2006}.  Since we want to use the new models from the Millennium simulation for our mock catalogues, we had to extrapolate from the SDSS magnitudes provided in the \citet{Guo:2011} data  to 2MASS magnitudes.  In addition to the data from the Millennium simulation's first run \citep{Millennium} with its semi-analytic galaxy models \citep{DeLucia:2006}, we have the full data from the 2MASS Redshift Survey \citep{2MRS} and all galaxies from the twelfth data release of SDSS \citep{SDSS_DR12} at hand.  We found 8,499 galaxies which are in both datasets (2MRS and SDSS)\footnote{Tolerance of angular separation: 5 arcseconds; tolerance for separation in redshift space: 300 km/s.} and we therefore have 2MASS and SDSS magnitudes for them.  We adopted the functional form of the SDSS-2MASS colour transformation proposed by \citet{Bilir:2008}:

\begin{equation}
(m_{g}-m_{X}) = d_{X} (m_{g}-m_{r}) + e_{X} (m_{r}-m_{i}) + f_{X} .
\label{colourcolourcolour_relation}
\end{equation} 
\begin{figure*}[ht]
\begin{center}
\includegraphics[width=0.90\textwidth]{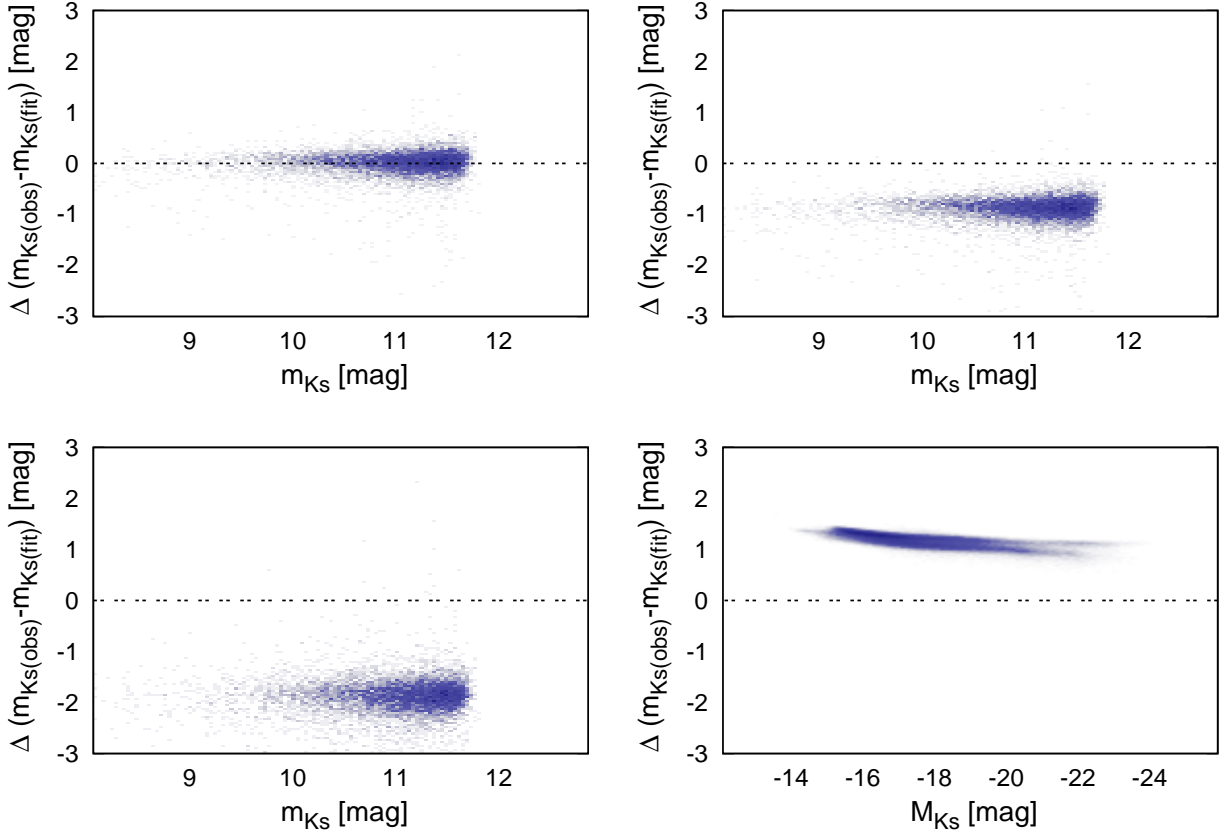}\\ 
\caption{Comparison of the correlations between SDSS and 2MASS magnitudes. Top-left panel: our fit on observational data.  Top-right panel: correlation from \citet{Bilir:2008} as the straight dashed line, which has a clear off-set from our observational data.  Bottom-left panel: performance of the relation derived from the one used in the Millennium simulation, which also deviates clearly from our observational data.  Bottom-right panel: correlation from \citet{Bilir:2008} applied on the Millennium-Simulation data, which does not fit either.}
\label{comparision_fits}
\end{center}
\end{figure*}

The wild card $X$ stands for any of the 3 2MASS bands ($J$, $H$, and $K_{s}$).  The relation connects the 3 SDSS magnitudes $m_{g}$, $m_{r}$, and $m_{i}$ with one 2MASS magnitude $m_{X}$.  We found that the values given for the coefficients $d_{X}$, $e_{X}$, and $f_{X}$ in \citet{Bilir:2008} were of no good use for galaxies.  The coefficients of that paper were derived for a general populations of stars and by applying them on galaxies one gets a clear offset and tilt with respect to the values from observations (see Figure \ref{comparision_fits} top-right panel) as well as to the values from the simulations (see Figure \ref{comparision_fits} bottom-right panel).  At first, we tried to obtain the coefficients of the colour transformation by fitting it to the SDSS and 2MASS magnitudes of the semi-analytic galaxy models \citep{DeLucia:2006} from the Millennium-Simulation.  However, we found that the fit derived from the simulated galaxies did not agree with the data from observed galaxies (see Figure \ref{comparision_fits} bottom-left panel).  Hence, we finally derived the coefficients of the colour transformation by directly fitting it to the observational data of 8,499 galaxies for which we had SDSS and 2MASS magnitudes both.  After a 3-$\sigma$ clipping to remove some clear outliers, 8,384 galaxies remained and we obtained the coefficients listed in Table \ref{sdsstwomass_coeff}.

\begin{table*}
\begin{center}
\begin{tabular}{c|ccc}
 coefficients & $J$ & $H$ & $K_{s}$ \\ \hline \hline
 $d$&  1.4925 $\pm$ 0.0013 &   1.5193 $\pm$ 0.0013  &  1.4731 $\pm$ 0.0013 \\ 
 $e$&  1.2305 $\pm$ 0.0023 &   1.4972 $\pm$ 0.0023 &  1.6164 $\pm$ 0.0023 \\ 
 $f$&  1.0948 $\pm$ 0.0008 &  1.6664 $\pm$ 0.0008 &  1.9080 $\pm$ 0.0008  \\ \hline
 rms& 0.1986  &  0.2130  &  0.2336  \\ 

\end{tabular}
\end{center}
\caption{Linear correlation coefficients of the fundamental-plane residuals for all possible combinations of the five SDSS filters.}
\label{sdsstwomass_coeff}
\end{table*}
\begin{figure*}[ht]
\begin{center}
\includegraphics[width=0.90\textwidth]{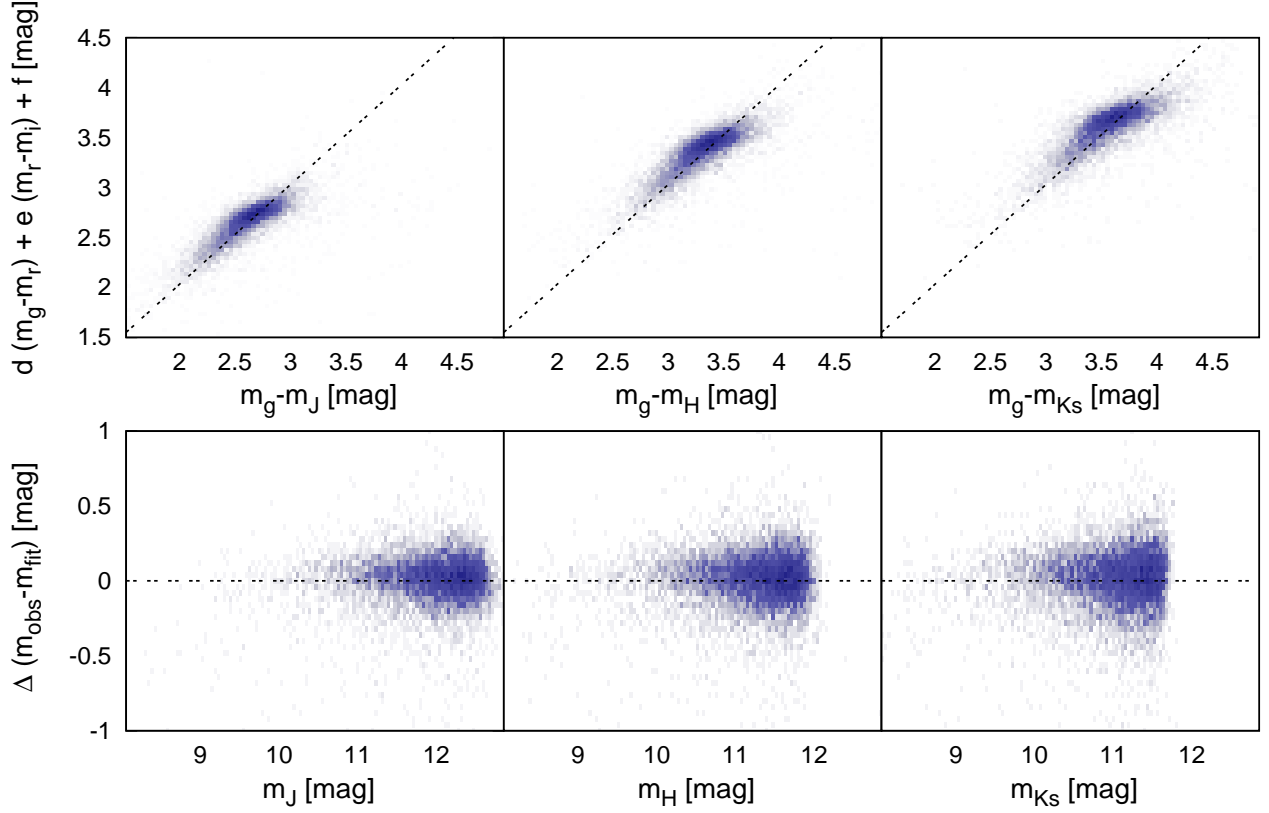}\\ 
\caption{Performance of our fit for the SDSS-2MASS transformation.  The top panels show an edge-on view on the colour-colour-colour plane for all 2MASS bands ($J$ in the top-left panel, $H$ in the top-middle panel, and $K_{s}$ in the top-right panel).  Our fits are always indicated by the dashed black lines.  The bottom panels show the residuals of the fit shown as the difference between the observed magnitude and the magnitude derived from the fit depending on the apparent magnitude in the 3 2MASS bands ($J$ in the bottom-left panel, $H$ in the bottom-middle panel, and $K_{s}$ in the bottom-right panel).}
\label{2mass_sdss_transform}
\end{center}
\end{figure*}
The root mean square (rms) increases with longer wavelength which was not surprising, because we performed an extrapolation from the SDSS wavelength range deeper into the infrared and the closer the central wavelength of a filter is located to the SDSS filters, the better is the fit.  In Figure \ref{2mass_sdss_transform}, we show edge-on projections on the fitting plane for all three 2MASS bands and also the residuals, which are displayed for the 2MASS filters themselves and not the colours.

\section{Mock K-correction}
\label{mockK}
\begin{figure*}[ht]
\begin{center}
\includegraphics[width=0.90\textwidth]{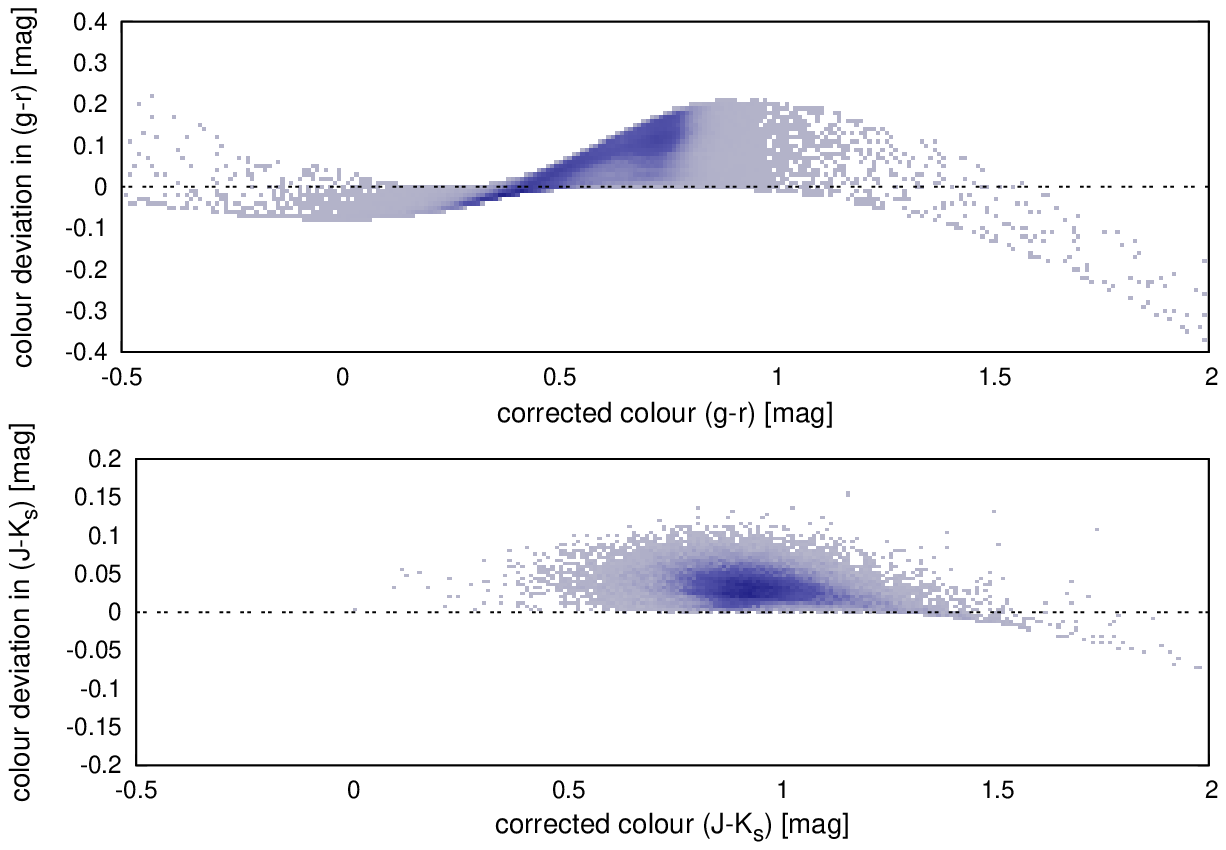}\\ 
\caption{Deviation of the uncorrected colour from the K-corrected colour using the real values for each survey.  Top panel: SDSS colours.  Bottom panel: 2MASS colours.}
\label{colour_deviation}
\end{center}
\end{figure*}

When creating a mock catalogue, one cannot simply change the sign of same K-corrections used to corrected the observed data, because the colours before applying the K-correction are, albeit similar, not the same as the colours after applying the K-correction.  Although, it is not a huge difference (see Figure \ref{colour_deviation}), it is one that can be taken into account with relatively small effort.  To this end, we used the K-corrections from \citet{Chilingarian:2010} with the updated coefficients for SDSS from \citet{Chilingarian:2012,Saulder:2013} and the new coefficients for the 2MASS filters taken directly from \url{http://kcor.sai.msu.ru/}.  To derive the new mock K-correction coefficients $\bar{B}_{ij}$, we fitted a two-dimensional polynomial function:

\begin{align}
K(z,m_{\textrm{uncor},f_{1}}-m_{\textrm{uncor},f_{2}})=\nonumber\\
\bar{K}(z,m_{\textrm{Kcor},f_{1}}-m_{\textrm{Kcor},f_{2}})=\\
\sum\limits_{i,j} \bar{B}_{ij} z^{i} (m_{\textrm{Kcor},f_{1}}-m_{\textrm{Kcor},f_{2}})^{j}\nonumber
\label{mockKcorrection}
\end{align}

which is similar to Equation \ref{Kcorrection} to the obtained K-corrections $K(z,m_{\textrm{uncor},f_{1}}-m_{\textrm{uncor},f_{2}})$, K-corrected colours $(m_{\textrm{Kcor},f_{1}}-m_{\textrm{Kcor},f_{2}})$ and redshifts $z$ of the galaxies from the SDSS and 2MASS galaxies and a grid of artificial values following the K-correction equation.  The wild cards $f_{1}$ and $f_{2}$ stand for two different filters.  By definition, the real K-corrections $K$ for the uncorrected colours $(m_{\textrm{uncor},f_{1}}-m_{\textrm{uncor},f_{2}})$ are the same as the mock K-corrections $\bar{K}$ for the K-corrected colours $(m_{\textrm{Kcor},f_{1}}-m_{\textrm{Kcor},f_{2}})$.  The coefficients $\bar{B}_{ij}$ of the mock K-correction are listed in Tables \ref{invKcorr_gband} to \ref{invKcorr_Ksband} for all colours and filters used in this paper.

\begin{table}[H]
\begin{center}
\begin{tabular}{c|cccc}
& ($g-r$)$^0$ & ($g-r$)$^1$ & ($g-r$)$^2$ & ($g-r$)$^3$ \\ \hline
$z^0$ &0& 0& 0& 0\\
$z^1$ &-0.230125 &1.76255 &6.30428 &-10.4609 \\
$z^2$ &-41.3522 &15.2747 &139.091 &23.9396 \\
$z^3$ &726.982 &-1337.47 &-443.452 &0\\
$z^4$ &-1827.06 &5009.87 &0 &0\\
$z^5$ &-5260.39 &0 &0 &0
\end{tabular}
\end{center}
\caption{Coefficients for the inverse K-correction in the $g$ band using $g-r$ colours.}
\label{invKcorr_gband}
\end{table}

\begin{table}[H]
\begin{center}
\begin{tabular}{c|cccc}
& ($g-r$)$^0$ & ($g-r$)$^1$ & ($g-r$)$^2$ & ($g-r$)$^3$ \\ \hline
$z^0$ &0& 0& 0& 0\\
$z^1$ &2.64192 &-3.63656 &3.87578 &-2.8507 \\
$z^2$ &-51.1976 &58.4272 &15.9944 &-0.19916 \\
$z^3$ &356.875 &-537.807 &31.3718 &0\\
$z^4$ &-554.669 &1091.06 &0 &0\\
$z^5$ &-2439.93 &0 &0 &0
\end{tabular}
\end{center}
\caption{Coefficients for the inverse K-correction in the $r$ band using $g-r$ colours.}
\label{invKcorr_rband}
\end{table}

\begin{table}[H]
\begin{center}
\begin{tabular}{c|cccc}
& ($J-K_{s}$)$^0$ & ($J-K_{s}$)$^1$ & ($J-K_{s}$)$^2$ & ($J-K_{s}$)$^3$ \\ \hline
$z^0$ &0& 0& 0& 0\\
$z^1$ &-2.90632 &1.84899 &0.687978 &-0.435851 \\
$z^2$ &28.7738 &-35.0671 &12.645 &0.814889 \\
$z^3$ &-124.868 &44.1619 &-33.6223 &0\\
$z^4$ &671.941 &123.024 &0 &0\\
$z^5$ &-1864.17 &0 &0 &0
\end{tabular}
\end{center}
\caption{Coefficients for the inverse K-correction in the $J$ band using $J-K_{s}$ colours.}
\label{invKcorr_Jband}
\end{table}

\begin{table}[H]
\begin{center}
\begin{tabular}{c|cccc}
& ($J-K_{s}$)$^0$ & ($J-K_{s}$)$^1$ & ($J-K_{s}$)$^2$ & ($J-K_{s}$)$^3$ \\ \hline
$z^0$ &0& 0& 0 &0\\
$z^1$ &-5.23228 &0.0686118 &4.15743 &-0.901711\\
$z^2$ &73.2808 &-63.8764 &-0.528324 &2.40482 \\
$z^3$ &-398.914 &197.991 &-27.4839 &0\\
$z^4$ &1726.93 &-30.6288 &0 &0\\
$z^5$ &-4240.1 &0 &0 &0
\end{tabular}
\end{center}
\caption{Coefficients for the inverse K-correction in the $K_{s}$ band using $J-K_{s}$ colours.}
\label{invKcorr_Ksband}
\end{table}
\FloatBarrier
\section{Additional figures}

\begin{figure*}[ht]
\begin{center}
\includegraphics[width=0.45\textwidth]{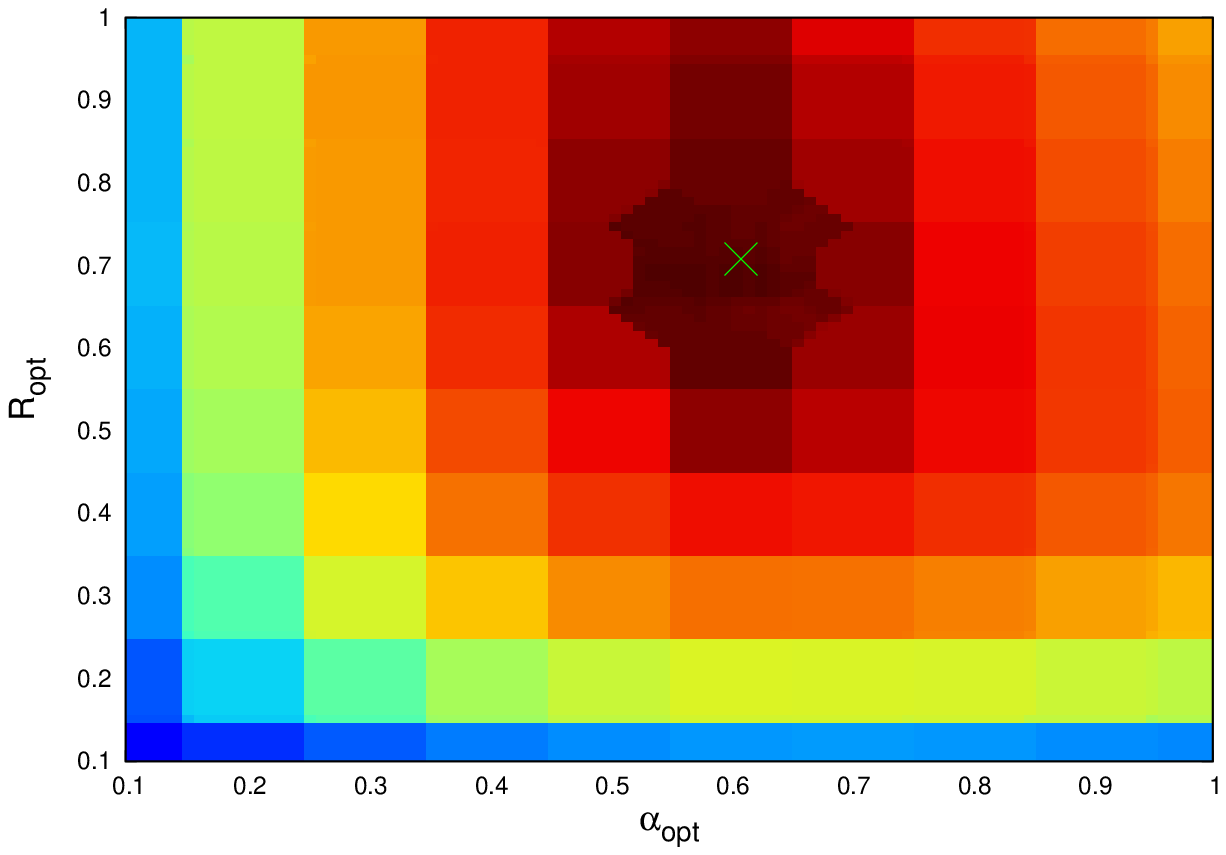}
\includegraphics[width=0.45\textwidth]{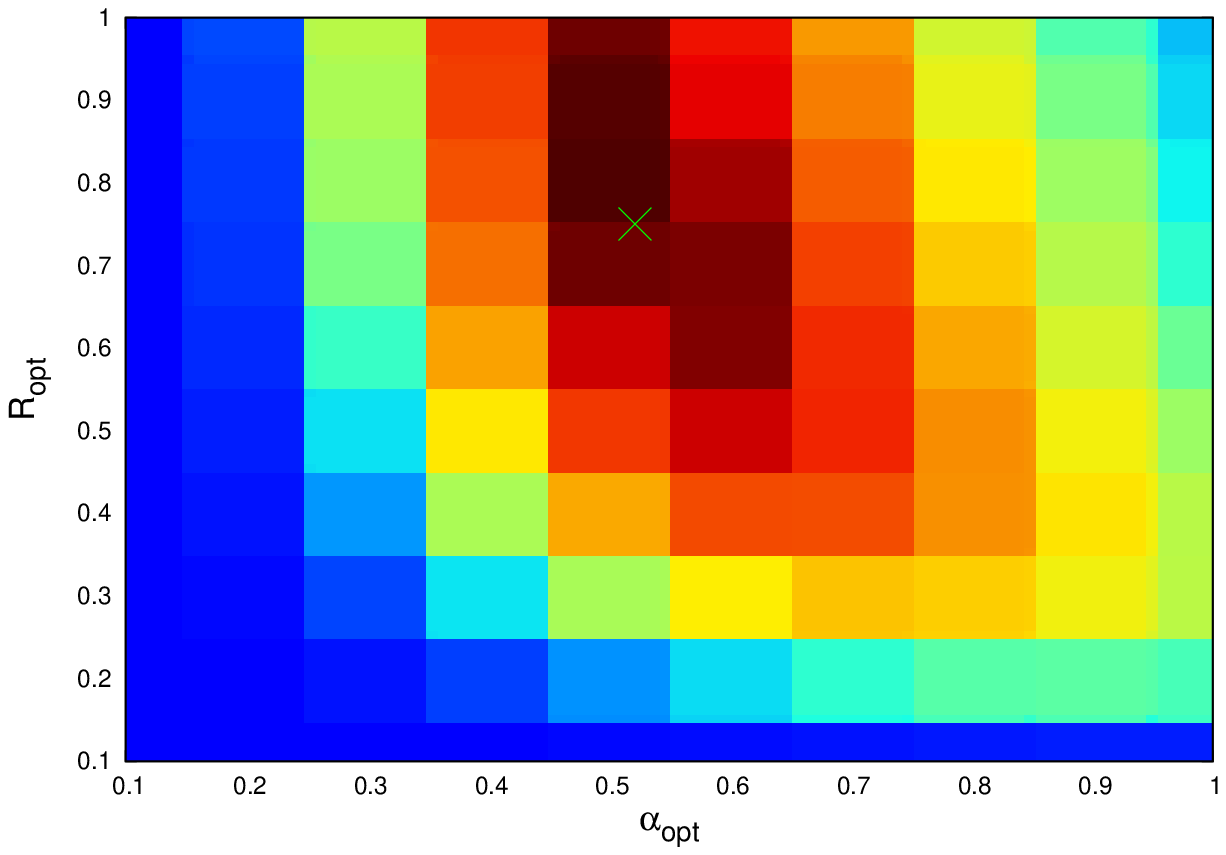}\\ 
\caption{Distribution of the values of the median group cost function $S_{\textrm{tot}}$ using the 2MRS mock catalogues (in the left panel) and the SDSS mock gatalogues (in the right panel) for different values of $\alpha_{\textrm{opt}}$ and $R_{\textrm{opt}}$.  High values of $S_{\textrm{tot}}$ are indicated by dark red, while low values are indicated by dark blue.  $\lambda_{\textrm{opt}}$ is fixed to its optimal values, which is listed in Table \ref{optimal_coefficients}.  The green X denotes result of the optimal values of $\alpha_{\textrm{opt}}$ and $R_{\textrm{opt}}$ according to our simplex fit.}
\label{optimal_2MRS_SDSS}
\end{center}
\end{figure*}

\begin{figure*}[ht]
\begin{center}
\includegraphics[width=0.90\textwidth]{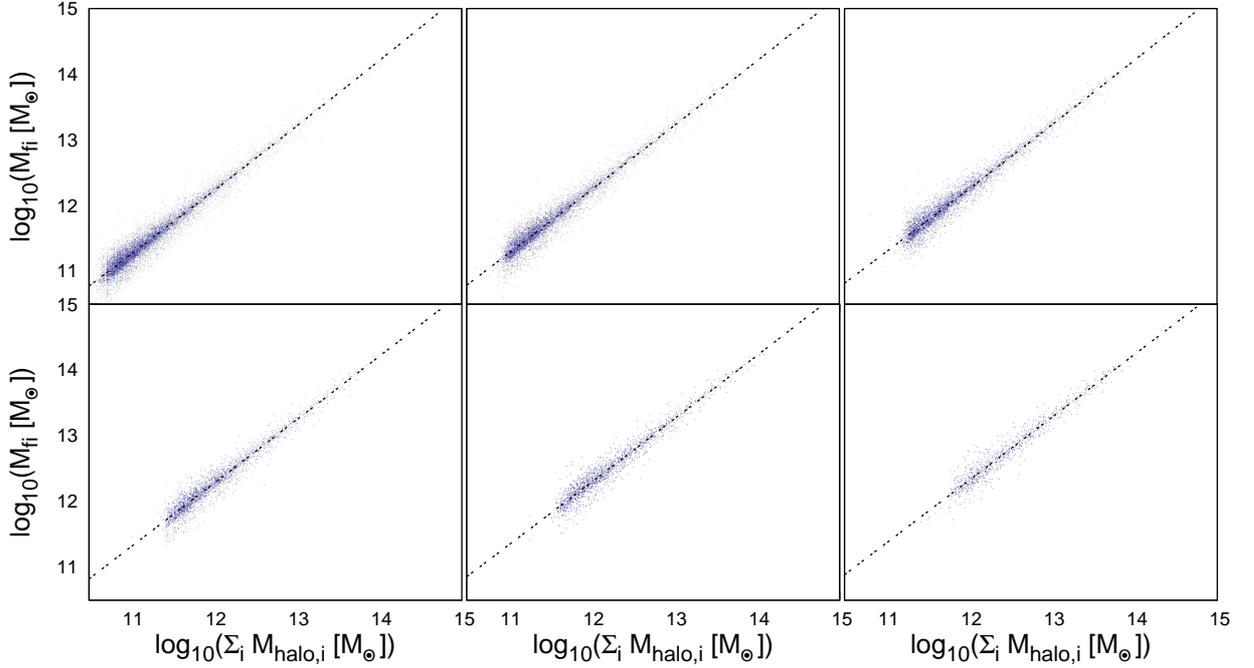}\\ 
\caption{Distribution of masses within the finite infinity regions before expanding them iteratively depending on the sum of the masses of the groups they are composed of.  The dotted black line indicates our fit on this relation.  The panels show the 6 snapshots used from the last one (number 63) in the top left panel to the earliest one (number 58) in the bottom right panel.}
\label{map_fi_first}
\end{center}
\end{figure*}

\begin{figure*}[ht]
\begin{center}
\includegraphics[width=0.90\textwidth]{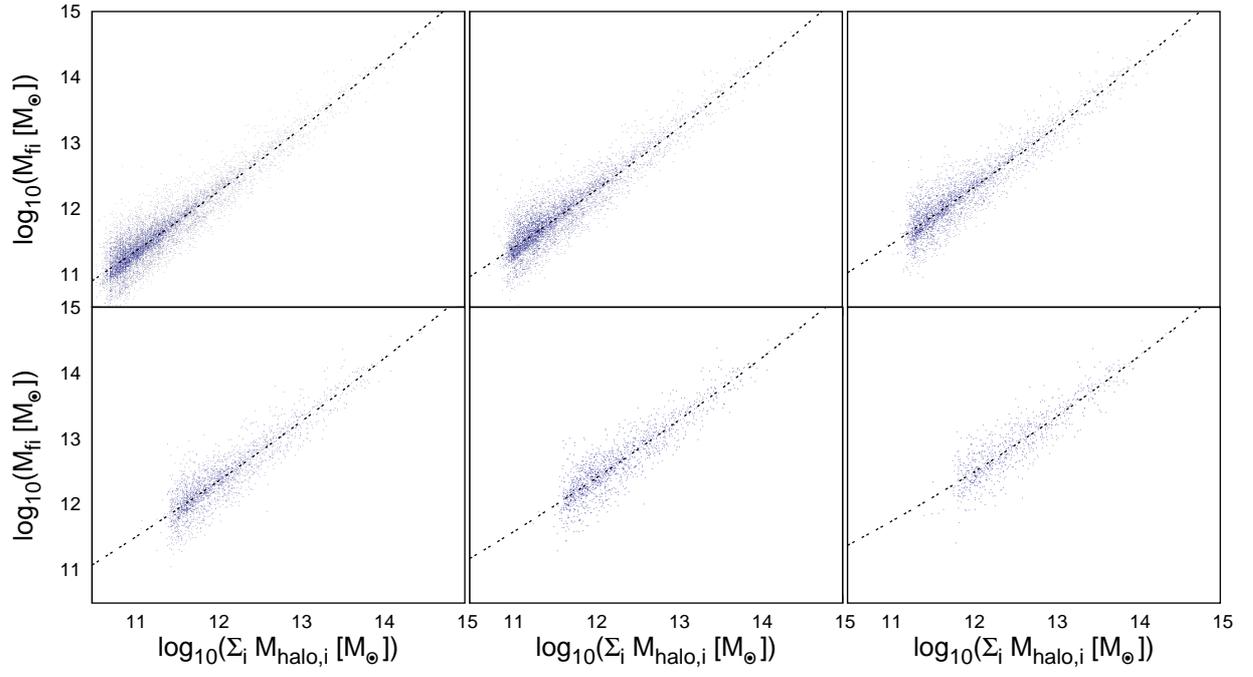}\\ 
\caption{Distribution of the final masses of the finite infinity regions depending on the sum of the masses of the groups they are composed of. Symbols and panels: as in Figure \ref{map_fi_first}.}
\label{map_fi_final}
\end{center}
\end{figure*}

\end{document}